\documentclass[twocolappendix]{emulateapj}
\usepackage{apjfonts}
\usepackage[colorlinks=true,urlcolor=blue,citecolor=blue,linkcolor=blue,breaklinks]{hyperref}
\usepackage{graphicx}
\usepackage{enumitem}
\newcommand{\kms}{km$\,$s$^{-1}$}
\newcommand{\sone}{045403}
\newcommand{\stwo}{050941}
\newcommand{\sthree}{051906}
\newcommand{\sfour}{052124}
\newcommand{\sfive}{053244}
\newcommand{\ssix}{053342}
\newcommand{\sseven}{053431}

\newcommand{\lsone}{045403.62--671618.5}
\newcommand{\lstwo}{050941.94--712742.1}
\newcommand{\lsthree}{051906.69--682137.4}
\newcommand{\lsfour}{052124.90--660412.9}
\newcommand{\lsfive}{053244.25--693005.7}
\newcommand{\lssix}{053342.15--684602.8}
\newcommand{\lsseven}{053431.46--683513.9}

\newcommand{\mecl}{$M_{\rm{ecl}}$}
\newcommand{\mmax}{$m_{\rm{max}}$}

\newcommand{\beq}{\begin{equation}}
\newcommand{\eeq}{\end{equation}}

\newcommand{\AmS}{{\protect\the\textfont2
  A\kern-.1667em\lower.5ex\hbox{M}\kern-.125emS}}
\newcommand{\lsim}{\ \raise
-2.truept\hbox{\rlap{\hbox{$\sim$}}\raise5.truept\hbox{$<$}\ }}
\newcommand{\gsim}{\ \raise
-2.truept\hbox{\rlap{\hbox{$\sim$}}\raise5.truept\hbox{$>$}\ }}
\newcommand{\simsim}{\ \raise
-2.truept\hbox{\rlap{\hbox{$\sim$}}\raise5.truept\hbox{$\sim$}\ }}

\def\Q{\ifmmode\mathcal{Q}\else$\mathcal{Q}$\fi}
\slugcomment{Version Dated \today}

\righthead{}
\lefthead{}

\begin{document}
 
\title{Stellar clusterings around ``Isolated'' Massive YSOs in the LMC}
%\title{Investigating Isolated Massive Star Formation in the Large Magellanic Cloud}

%\footnotemark[1]
%\footnotetext[1]{}


%\author{Dimitrios A. Gouliermis}
%\affil{University of Heidelberg, Centre for Astronomy, Institute for Theoretical Astrophysics, Albert-Ueberle-Str. 2, 69120 Heidelberg, Germany }
%\affil{Max Planck Institute for Astronomy, 
%         K\"onigstuhl 17, 69117 Heidelberg, Germany; dgoulier@mpia-hd.mpg.de}
%\email{dgoulier@mpia-hd.mpg.de}

\author{Ian W. Stephens\altaffilmark{1,2,15}, Dimitrios Gouliermis\altaffilmark{3,4}, Leslie W. Looney\altaffilmark{2}, Robert A. Gruendl\altaffilmark{2}, You-Hua Chu\altaffilmark{2,5}, \\ Daniel R. Weisz\altaffilmark{6,7}, Jonathan P. Seale\altaffilmark{8}, C.-H. Rosie Chen\altaffilmark{9}, Tony Wong\altaffilmark{2}, Annie Hughes\altaffilmark{10,11}, Jorge L. Pineda\altaffilmark{12}, J\"urgen Ott\altaffilmark{13}, Erik Muller\altaffilmark{14}}
\altaffiltext{1}{\itshape Institute for Astrophysical Research, Boston University, Boston, MA 02215, USA}
\altaffiltext{2}{\itshape Department of Astronomy, University of Illinois, 1002 West Green Street, Urbana, IL 61801, USA}
\altaffiltext{3}{\itshape Zentrum f\"{u}r Astronomie der Universit\"{a}t Heidelberg, Institut f\"{u}r Theoretische Astrophysik, Albert-Ueberle-Str. 2, 69120 Heidelberg, Germany; gouliermis@uni-heidelberg.de}
\altaffiltext{4}{\itshape Max Planck Institute for Astronomy, K\"onigstuhl 17, 69117 Heidelberg, Germany; dgoulier@mpia.de}
\altaffiltext{5}{\itshape Academia Sinica Institute of Astronomy and Astrophysics, P.O. Box 23-141, Taipei 106, Taiwan}
\altaffiltext{6}{\itshape Astronomy Department, Box 351580, University of Washington, Seattle, WA, USA}
\altaffiltext{7}{\itshape Hubble Fellow}
\altaffiltext{8}{\itshape Space Telescope Science Institute, 3700 San Martin Drive, Baltimore, MD 21218, USA}
\altaffiltext{9}{\itshape Max Planck Institute for Radio Astronomy, D-53121 Bonn, Germany}
\altaffiltext{10}{\itshape CNRS, IRAP, 9 Av. du Colonel Roche, BP 44346, F-31028 Toulouse cedex 4, France}
\altaffiltext{11}{\itshape Universit\'{e} de Toulouse, UPS-OMP, IRAP, F-31028 Toulouse cedex 4, France}
\altaffiltext{12}{\itshape Jet Propulsion Laboratory, California Institute of Technology, 4800 Oak Grove Drive, Pasadena, CA 91109-8099, USA}
\altaffiltext{13}{\itshape National Radio Astronomy Observatory, P.O. Box O, 1003 Lopezville Road, Socorro, NM 87801, USA}
\altaffiltext{14}{\itshape National Astronomical Observatory of Japan, Chile Observatory, 2-21-1 Osawa, Mitaka, Tokyo 181-8588, Japan}
\altaffiltext{15}{\itshape Current address: Harvard-Smithsonian Center for Astrophysics, Cambridge, MA 02138, USA; ian.stephens@cfa.harvard.edu}
\interfootnotelinepenalty=10000

\begin{abstract} 
Observations suggest that there is a significant fraction of O-stars in the field of the Milky Way that appear to have formed in isolation or in low mass clusters ($<$100~$M_\odot$). The existence of these high-mass stars that apparently formed in the field challenges the generally accepted paradigm, which requires star formation to occur in clustered environments. In order to understand the physical conditions for the formation of these stars, it is necessary to observe isolated high-mass stars while they are still forming. With the \emph{Hubble~Space~Telescope}, we observe the seven most isolated massive ($>$8~$M_\odot$) young stellar objects (MYSOs) in the Large~Magellanic~Cloud (LMC). The observations show that while these MYSOs are remote from other MYSOs, OB associations, and even from known giant molecular clouds, they are actually not isolated at all. Imaging reveals $\sim$100 to several hundred pre--main-sequence (PMS) stars in the vicinity of each MYSO. These previously undetected PMS stars form prominent compact clusters around the MYSOs, and in most cases they are also distributed sparsely across the observed regions. Contrary to what previous high-mass field star studies show, these observations suggest that high-mass stars may not be able to form in clusters with masses less than 100~$M_\odot$. If these MYSOs are indeed the best candidates for isolated high-mass star formation, then the lack of isolation is at odds with random sampling of the IMF. Moreover, while isolated MYSOs may not exist, we find evidence that isolated clusters containing O-stars can exist, which in itself is rare.
%We propose that the search for isolated high-mass star formation is in fact the search for high-mass stars forming in isolated stellar clusterings.
\end{abstract}

\subjectheadings{stars: formation -- stars: massive -- ISM: clouds -- Magellanic Clouds}

\section{Introduction}\label{sec:intro}
Approximately 20~percent of the Galactic main sequence O-stars are isolated field stars \citep[e.g.,][]{Mason1998}. After correcting for clustered environments and runaways, only \mbox{4\,--\,10}~percent of all O-stars appear to be truly isolated \citep[e.g.,][]{deWit2004,deWit2005,Zinnecker2007}. Isolated field O-stars are also suggested to account for 20\,to\,30\,percent of the high-mass stellar populations in star-forming galaxies \citep{Oey2004}. The existence of these stars is perplexing when one considers two theoretical expectations: 1) the relation between the maximum stellar mass and the hosting-cluster mass excludes O-stars from forming in clusters with masses $\leq$ 250~M$_\sun$ \citep[e.g.,][]{Weidner2006}, and  2) the maximum stellar mass is set by the high-mass end of a fully-populated stellar initial mass function \citep[IMF;][]{OeyClarke2005}. In favor of {\em in situ} formation, Monte Carlo simulations of a randomly sampled IMF suggest that ``isolated'' O-stars are likely formed in clusters with numerous unseen lower-mass stars \citep{Parker2007}, while contrary to being formed {\em in situ}, field O-type stars are proposed to be explained as runaway stars that are difficult to trace back to their original cluster or are remnants of clusters that have undergone significant dissolution \citep[e.g.,][]{Pflamm2010, Gvaramadze2012}.

%However, after formation in a cluster, O-stars can be expelled via a binary ejection event coupled with a subsequent sling-shot due to a supernova explosion, making it impossible to trace them back to clusters \citep{Pflamm2010}. 

According to the generally accepted paradigm of star formation, stars typically form in giant molecular clouds (GMCs). \citet{Lamb2010} presented a \emph{Hubble Space Telescope} (\emph{HST}) study on isolated high-mass stars for eight main sequence OB-stars in the Small Magellanic Cloud (SMC). With a detection limit of 1 M$_\sun$, these authors found that two stars are runaways, three are in small clusters, and the remaining three appear to be isolated. Furthermore, two of these isolated OB stars are in \ion{H}{2} regions without bow-shocks, increasing the likelihood that they are in their natal environment. \cite{Oey2013} identified in the SMC 14 additional field OB stars with symmetric dense \ion{H}{2} regions around them, minimizing the likelihood that these objects have transverse runaway velocities. All stars are confirmed spectroscopically to be strong candidates for field high-mass stars that formed in situ \citep{Lamb2015}. Given that the main sequence lifetime of these particular O-stars is about an order of magnitude shorter than that of a GMC \citep[$\sim$20--40 Myr in the Local Group,][]{Kawamura2009,Miura2012}, these observations suggest that high-mass star formation may not require GMCs. Therefore, the fact that some O-stars may form in isolation allows for a new and interesting probe of high-mass star formation. 

\begin{figure*}[t!]
\begin{center}
\includegraphics[scale=0.55,angle=90]{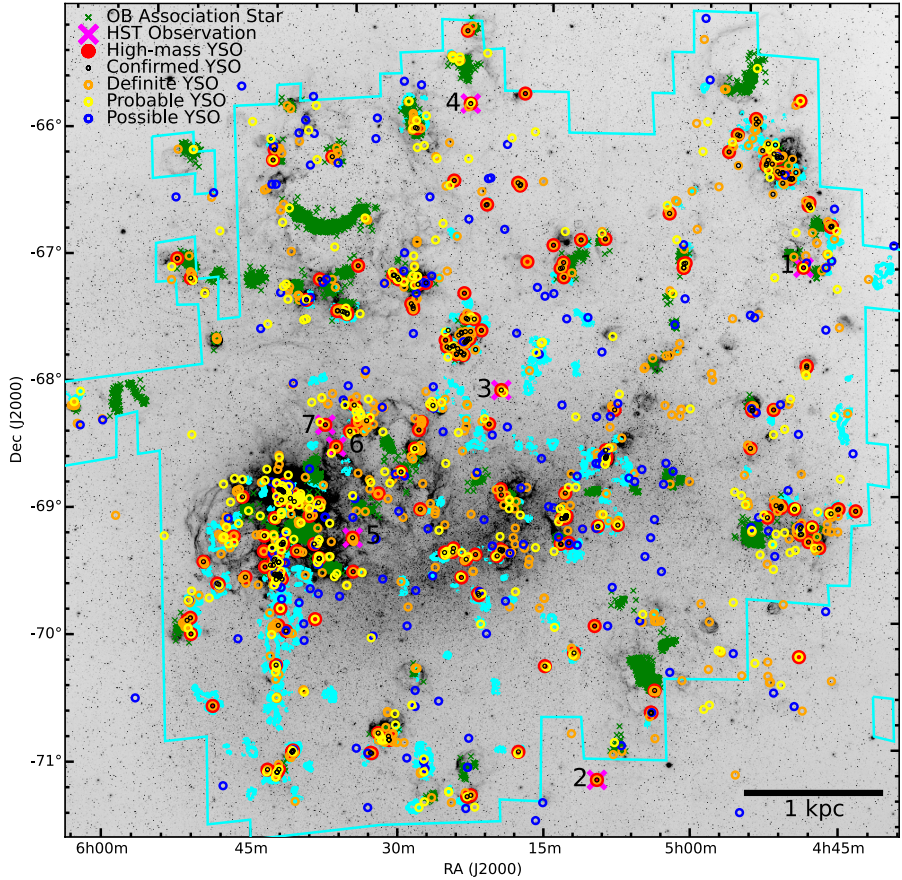}
\caption{H$\alpha$ image (with no background subtraction) of the LMC from the Magellanic Cloud Emission Line Survey \citep{SmithMCELS1999}. As indicated in the legend, small green $\times$'s are known stars in OB associations from \citet{Lucke1970}, magenta large $\times$'s are MYSOs observed with $HST$ presented in this paper, large red filled circles are MYSOs identified in GC09, small black empty black circles are sources which were confirmed to be MYSOs in \citet{Seale2009}, and orange, yellow, and blue empty circles are definite, probable, and possible YSOs, respectively, as categorized in GC09. The cyan contours show the CO(1--0) from MAGMA \citep{Wong2011} and indicate locations of GMCs. The cyan border indicates the entire LMC NANTEN \citep{Fukui2008} CO(1--0) survey. The MYSOs observed with $HST$ are numbered, corresponding to the following names: (1) 045403.62--671618.5, (2) 050941.94--712742.1, (3) 051906.69--682137.4, (4) 052124.90-660412.9, (5) 053244.25--693005.7, (6) 053342.15--684602.8, and (7) 053431.46--683513.9. \label{LMC_YSOs}
}
\end{center}
\end{figure*}

If high-mass field stars represent a population of stars that previously formed in isolation, there should be many high-mass stars that are currently forming in isolation. Specifically, considering that there are thousands of stars with $M>$10\,M$_\sun$ that are currently in the accretion phase in the Galaxy \citep{Zinnecker2007}, and it has been proposed that 4--10\% of O-stars are formed in isolation, the Galaxy should contain 100s of isolated high-mass stars under formation. However, convincing evidence of isolated field stars that are currently in the accretion phase is lacking.  The only investigation of such a candidate is that of the compact star-forming region N33 in the SMC, reported by  \cite{Selier2011}. These authors did not find any traces of a stellar clustering around the region on scales $\gsim$\,3\,pc, while on smaller scales a marginal concentration of faint stellar sources was discovered clustered around a high-mass O6.5-O7 main-sequence star.

As pointed out by \citet{Bressert2012}, the term ``isolated high-mass star formation'' can be unclear. Specifically, these authors suggested three possible criteria that may suggest a high-mass star is \emph{not} isolated: 1) a high-mass star is forming with other high-mass stars in a molecular cloud;  2) the formation of a high-mass star may be triggered by another high-mass star; and 3) a high-mass star was gravitationally bound (within $\sim$3 pc) with another high-mass star sometime in the past. \citet{Bressert2012} were specifically interested in criterion 3, the least restrictive of the criteria, and found 15 candidates in the 30~Doradus region that may satisfy this criterion. This study is more concerned with the most restrictive of the criteria, criterion 1, and therefore it is more akin to the investigations of field O-stars by \citet{deWit2004,deWit2005} and \citet{Zinnecker2007}, who suggested that 4--10\% of all Galactic O-stars are not runaways, but formed in isolation.  

Our analysis also focuses in particular on high-mass stars at early stages of their formation. During its formation, the high-mass star will typically reach the main-sequence (i.e., commencing hydrogen fusion) while still accreting \citep[e.g.,][]{YorkeSonnhalter2002,Zinnecker2007}. Since the term ``protostar" is typically reserved for pre--main-sequence (PMS) stars, we use the term young stellar object (YSO) for embedded sources.  Indeed, the massive YSOs (MYSOs) targeted in this study are associated with ionized gas and are embedded \citep{Seale2009}, and thus are on the main sequence and are likely still accreting.
%Moreover, it is possible that the MYSO was once in a GMC that has since been dissipated. There is much difficulty in determining the molecular gas history, particularly without supplementary observations, but selecting MYSOs that are not close to any other MYSO \emph{may} avoid this scenario. Moreover, as mentioned in the introduction, the lifetime of a MYSO is expected to be much shorter than the natal GMC.

While observations of high-mass star forming regions can be studied at the highest resolution in our Galaxy, surveys of these regions have complications. Distances are typically measured kinematically and have high uncertainties -- especially since there is an ambiguity of assigning the velocity to a ``near arm'' distance or a ``far arm'' distance. Moreover, the Galaxy has high extinction and confusion along the line of sight, which causes significant difficulty in assigning which emission is happening at which distance. Therefore, it is very challenging to analyze Galactic emission at GMC-scales around MYSOs and to create unbiased and uniform surveys for high-mass star formation in the Galaxy. The Large Magellanic Cloud (LMC), being one of the nearest galaxies to the Milky Way, mitigates most of these problems, and therefore it is an ideal laboratory for uniform surveys of high-mass star formation. Specifically, all sources are at a similar distance of about 50~kpc \citep[][$\sim$0.25~pc per arcsecond]{Feast1999} and the nearly face-on orientation and low extinction allows for large regions to be studied unambiguously. Due to the observational advantages over the Galaxy, the entire LMC has been targeted by large surveys (e.g., \emph{Spitzer}, \citealt{Meixner2006};  \emph{Herschel}, \citealt{Meixner2013}).

Based on the first criterion for isolation proposed by \citet{Bressert2012}, a high-mass star forms in isolation if it is not member of an OB association or of a runaway population.  
This criterion extended to MYSOs requests that the isolated source should be forming away from an OB association, as well as of any GMC. 
The close connection between GMCs and high-mass star formation was confirmed by \citet{Wong2011} in the LMC, where the more CO luminous GMCs are found more likely to contain MYSOs.  
Using $Spitzer$ observations, \citet[][hereafter GC09]{GC09} constructed one of the best, carefully-selected samples of MYSOs across the entirety of the LMC. Specifically, they compiled a catalog of 248 best MYSO 
candidates. \citet{CG08} found that 85\% of these MYSOs are in GMCs and 65\% are in OB associations. Only 7\% of the MYSOs are outside of both GMCs and OB associations, comparable to the amount of Galactic O-stars that appear to be isolated, non-runaway field stars.

\begin{figure}[t!]
\begin{center}
\includegraphics[scale=0.5]{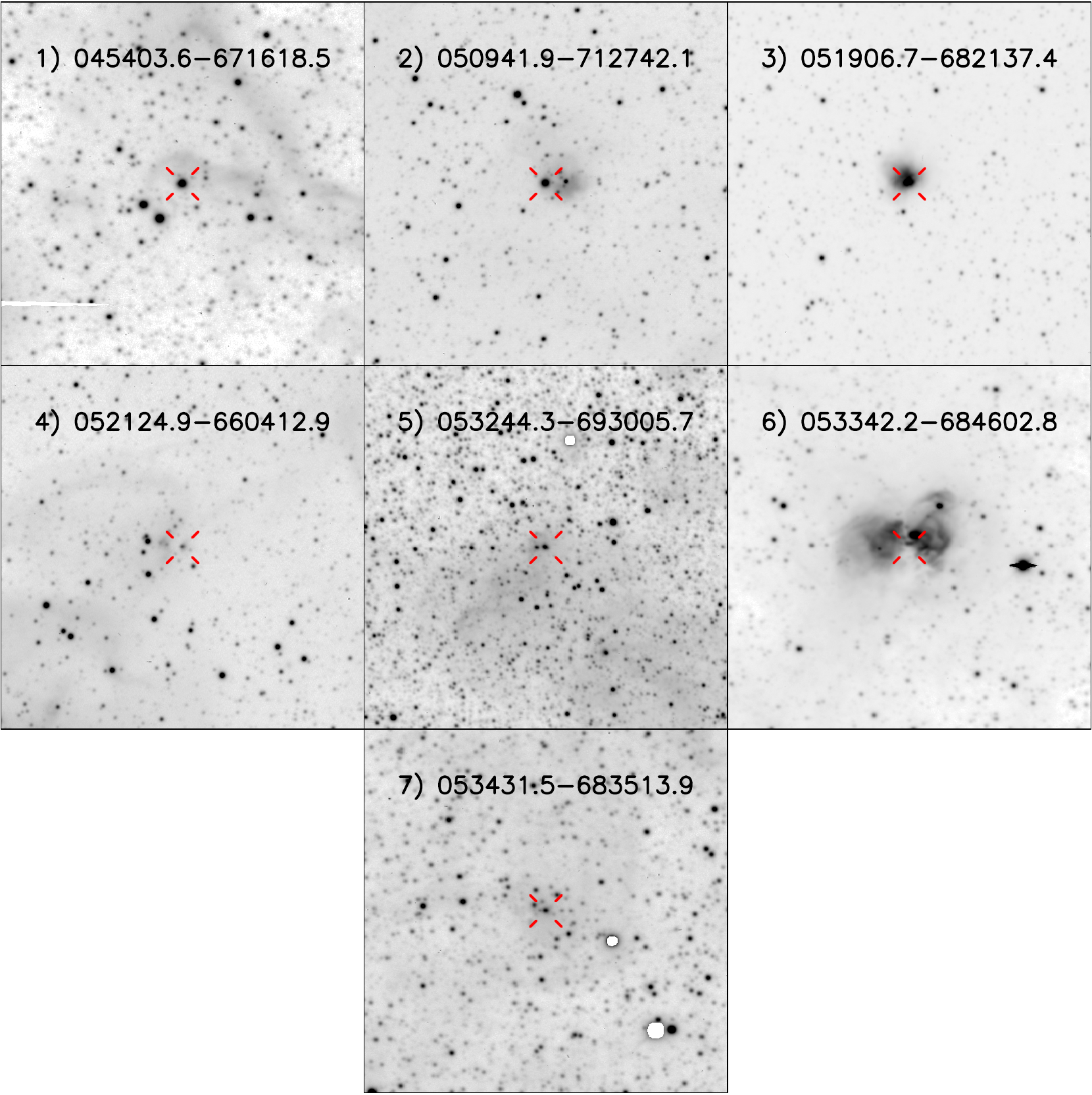}
\caption{ 
Ground-based H$\alpha$ images (no continuum subtraction) in inverted grayscale illustrates that the seven isolated MYSO targets have central H$\alpha$ regions. North is up and east is left. Observations were taken with the MOSAIC2 camera on the Blanco 4~m telescope; see \citet{Stephens2014} for details of the observations. The field-of-view for each panel is 150$\arcsec$ (37.5~pc) on a side. The \citet{GC09} position is marked with an open cross. MYSO numbering is the same as Figure\,\ref{LMC_YSOs}. \label{Halpha}
}
\end{center}
\end{figure}

We employed \emph{HST} to follow up on seven of the sources identified in \citet{CG08} since they are the best candidates for isolated MYSOs in the LMC. This sample is selected based upon the fact that within 80\,pc  (see Section\,\ref{ssel}), none of the sources are associated with (i) other MYSOs, (ii) OB associations, or (iii) any GMC. In all cases ground-based H$\alpha$ observations show that these MYSOs are affiliated with non-elongated, small \ion{H}{2} regions and therefore are unlikely to be part of a runaway population. We acquired WFC3 observations in the F656N, F555W, F814W, F110W, and F160W bands to examine the interstellar environment and determine the surrounding stellar populations down to $\sim$\,0.7\,M$_\sun$. The exquisite resolution of \emph{HST} immediately demonstrated in the reduced images that in fact none of the sources is single and therefore actually isolated. Instead, they are all associated with prominent stellar clusterings around them. %(Figures \ref{045403_2panel.png}\,-\,\ref{053431_2panel.png}). 

In this paper we present our observations for the search of ongoing isolated high-mass star formation in the LMC and describe the data reduction and point spread function (PSF) photometry applied.  We present the analysis of the data for these seven MYSOs in order to characterize in depth the natal environments of high-mass stars that appear as forming in isolation and to constrain more accurately the definition of isolated high-mass star formation. In Section\,\ref{ssandobs}, we describe our source selection of the seven MYSOs, the Mopra and \emph{HST} observations, and the \emph{HST} photometry. In Section\,\ref{section3}, we characterize the isolation of each target in our sample. In Section\,\ref{identPMS} and \ref{s:clusanl}, we identify the stellar populations associated to these seven MYSOs and characterize their clustering behavior through the entire \emph{HST} fields. Finally, in Section\,\ref{discussion} we discuss our findings in the general context of the phenomenon of isolated high-mass star formation.
%We further select three of the most prominent sources in our sample on the basis of their mass and isolation (sources marked with an asterisk in Table \ref{tab:MYSOmags}).

%\section{What is isolation?}\label{definingisolation}

%%%%%%%%
\section{Source Selection and Observations}\label{ssandobs}
\subsection{Source selection}\label{ssel}
As discussed in Section\,\ref{sec:intro}, the LMC provides a uniform, unbiased survey of high-mass star formation. GC09 identified 248 MYSOs with [8.0~$\mu$m]~$\leq$~8 mag.  Almost all of these MYSOs had follow-up $Spitzer$ IRS observations to confirm their MYSO-like spectral energy distributions \citep{Seale2009}. In Figure\,\ref{LMC_YSOs}, we show the distribution of all GC09 YSOs throughout the LMC. As per GC09, the figure includes: (1) \emph{Definite YSOs}, where the spectral features are very consistent with a YSO, (2) \emph{Probable YSOs}, which appear to be YSOs but have a suggestion of a feature of an alternate source, such as a background galaxy, and (3) \emph{Possible YSOs}, which are most likely other sources (i.e., stars, background galaxies, diffuse non-sources, and planetary nebulae) but cannot be ruled out as a YSO. Figure\,\ref{LMC_YSOs} also indicates the most massive definite and probable YSOs (8~$\mu$m magnitude, [8.0]~$\leq$~8, as per GC09), OB stars in known OB associations from \citet{Lucke1970}, and GMCs from the MAGMA CO(1--0) survey \citep{Wong2011}. The MAGMA survey shows the locations of LMC GMCs with molecular gas masses $M_{\rm{CO}} \gtrsim 2\times 10^4~M_\sun$. This survey used Mopra to re-observe GMCs identified in the LMC NANTEN CO(1--0) survey \citep{Fukui2008} with a higher resolution (11~pc) and approximately the same mass detection limit.

\begin{figure}[t!]
\begin{center}
\includegraphics[scale=0.335]{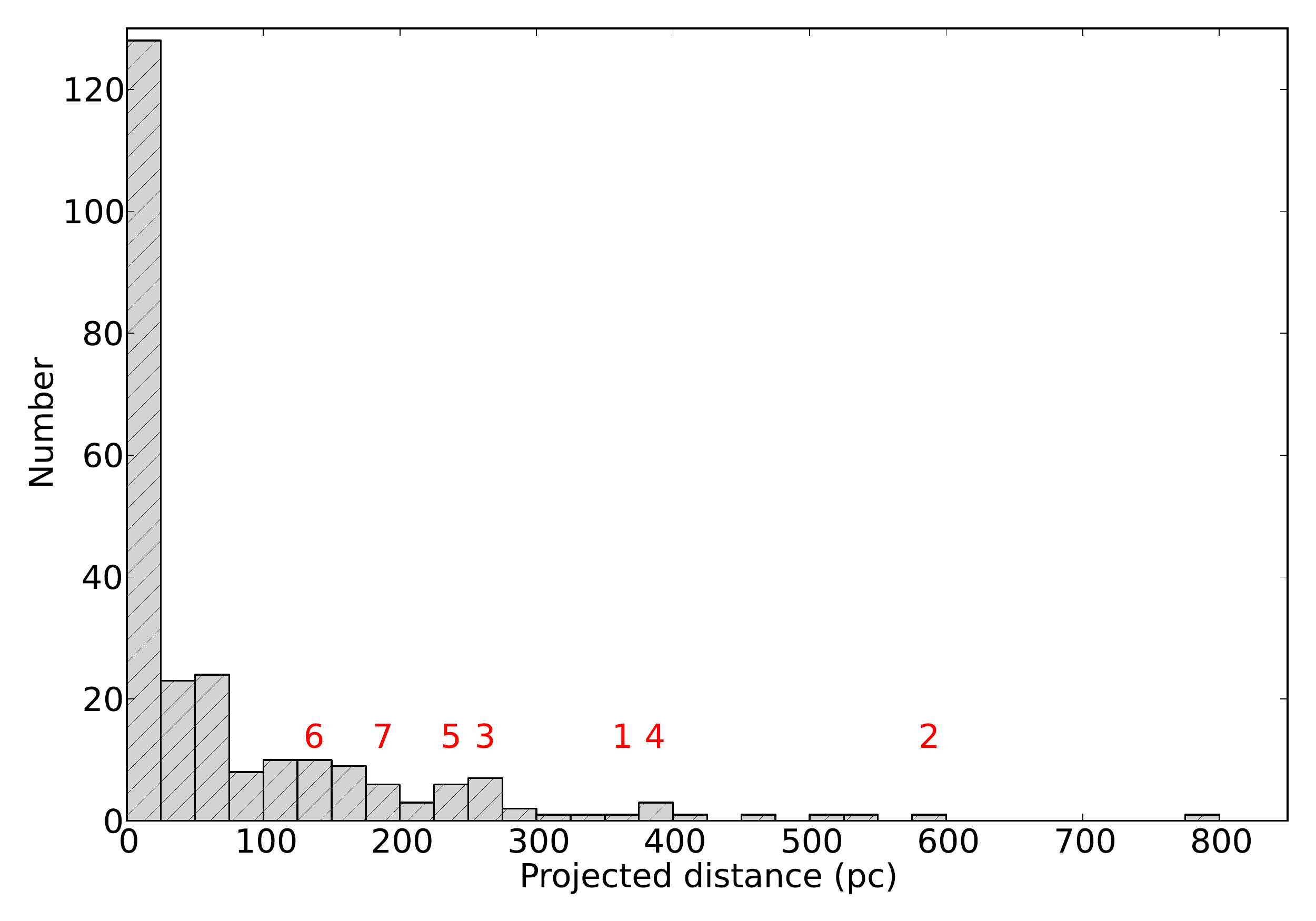}
\caption{Histogram of the projected distance (using 1$\arcsec$ = 0.25~pc) between each of the GC09 248 MYSO and its nearest neighbor. The red numbers indicate which MYSO lies within each bin, using the same MYSO numbering as Figure\,\ref{LMC_YSOs}. \label{nearestneighbor}
}
\end{center}
\end{figure}

\begin{deluxetable*}{crccccccccc}
%\tabletypesize{\tiny}
%\tablecolumns{4}
\setlength{\tabcolsep}{0pt}
\tablewidth{0pc}
\tablecaption{Magnitudes of the MYSOs in our sample in various bands.\label{tab:MYSOmags} }
\tablehead{
\colhead{No.} &
\colhead{Isolated}			&
\colhead{2MASS}   	&
\colhead{2MASS}     	&
\colhead{2MASS}     	&
\colhead{$Spitzer$}  	&
\colhead{$Spitzer$}       &
\colhead{$Spitzer$}  	&
\colhead{$Spitzer$}  	&
\colhead{$Spitzer$}  	&
\colhead{$Spitzer$}     	\\
\colhead{}  	&
\colhead{MYSO}  	&
\colhead{J}			&
\colhead{H} &
\colhead{K}&
\colhead{3.6~$\mu$m}  	&
\colhead{4.5~$\mu$m}&
\colhead{5.8~$\mu$m}&
\colhead{8.0~$\mu$m} &
\colhead{24~$\mu$m} &
\colhead{70~$\mu$m}     		 
}
\startdata
1 & \lsone & & & $14.72\pm0.18$ & $12.12\pm0.09$ & $11.19\pm0.07$ &  $9.37\pm0.08$ &  $7.69\pm0.09$\\
2 & \lstwo & $15.29\pm0.09$ & $14.77\pm0.15$  & $14.15\pm0.09$  &  $11.99\pm0.06$  &  $11.89\pm0.06$  &  $9.43\pm0.06$  &  $7.68\pm0.06$  &  $3.21\pm0.11$  &  $-1.62\pm0.22$ \\
3 & \lsthree & & & & $11.39\pm0.06$  & $10.99\pm0.06$  & $8.97\pm0.07$  &  $7.10\pm0.06$  &  $2.28\pm0.11$  &  $-2.42\pm0.22$ \\
4 & \lsfour & $16.66\pm0.18$ & $15.56\pm0.17$ & $14.63\pm0.12$ & $12.35\pm0.06$ & $11.96\pm0.06$ & $9.58\pm0.06$ & $7.82\pm0.06$ & $2.69\pm0.11$ & $-1.38\pm0.22$ \\
5 & \lsfive & $15.65\pm0.09$ & $15.14\pm0.10$ & $14.82\pm0.14$ & $12.25\pm0.06$ & $12.26\pm0.06$ & $9.47\pm0.06$ &  $7.73\pm0.06$ & $4.04\pm0.11$ & $-0.87\pm0.22$ \\
6 & \lssix & & & & $11.47\pm0.08$ & $10.75\pm0.07$ & $8.73\pm0.07$ & $6.83\pm0.07$ & $0.80\pm0.10$ &  $-3.95\pm0.22$ \\
7 & \lsseven & $15.32\pm0.08$ & $14.32\pm0.07$ & $13.14\pm0.04$ & $11.12\pm0.06$ & $10.56\pm0.05$ & $9.27\pm0.06$ & $7.74\pm0.06$ & $3.33\pm0.11$ & $-1.50\pm0.22$
\enddata
\tablecomments{All numbers are apparent magnitudes for the MYSO taken from GC09. All numbers are apparent magnitudes for the MYSO taken from GC09.}
\tablenotetext{a}{ The isolated MYSO name is based on the GC09 location of the right ascension ($\alpha$) and declination ($\delta$) of each MYSO; e.g., MYSO \lsone\ has the coordinates $\alpha$ = 4$^{\rm{h}}$54$^{\rm{m}}$3$\fs$62 and $\delta$ = --67$^\circ$16$\arcmin$18$\farcs$5.}
\end{deluxetable*}

To select for the most isolated MYSOs, we reduced the sample of isolated MYSOs from \citet{CG08} down to seven sources based on the following criteria:
\begin{enumerate}[itemsep=0.5mm]
\item The MYSO must be spectroscopically confirmed as a MYSO with Spitzer IRS \citep{Seale2009}.
\item The MYSO must have an H$\alpha$ region centered around it (see Figure\,\ref{Halpha}) in order to confirm that the star is massive enough to have significant hydrogen ionizing photons.
\item The MYSO must be far from any GMC, i.e., CO emission from the NANTEN/MAGMA surveys.
\item The MYSO must be far from known OB associations.
\item The MYSO must not be near another MYSO.
\end{enumerate}

The latter three criteria are particularly important because MYSOs can achieve high-velocities, and we do not want to include runaway YSOs that have been ejected from their natal environment (e.g., a GMC or OB association). Runaway stars are defined to have peculiar velocities larger than 40~\kms\ but can achieve velocities upward of 200~\kms\ \citep{Blaauw1961}, though the observed MYSOs likely have lower velocities since bow-shocks are not seen in the H$\alpha$ emission in Figure\,\ref{Halpha}. Assuming these MYSOs have velocities in the plane of the sky of 40~\kms\ and ages of 10$^6$~yr, a MYSO can travel a projected distance from its natal environment of $\sim$40~pc. We chose MYSOs with distances that were twice this value, i.e., sources that have projected distances of more than 80~pc from any GMC, OB association, or other MYSOs. This reduces our sample of sources to seven isolated MYSOs, listed\footnote{The sources listed in the table are referred to by their full GC09 names (based on their central positions). Throughout the paper for simplicity we refer to them by their first six digits of the right ascension, i.e., MYSOs \sone, \stwo, \sthree, \sfour, \sfive, \ssix, and \sseven.}  in Table\,\ref{tab:MYSOmags}.

Given these criteria, the members of our sample are found particularly far from other MYSOs, with projected distances ranging from $\sim$\,150\,pc to 600\,pc. The distribution of the projected distances between all MYSOs and their nearest neighbor is shown in Figure\,\ref{nearestneighbor}. Seventy-seven (31\%) of the 248 MYSOs in the GC09 catalog do not have another MYSO within a projected distance of 80\,pc. Of these, 14 sources have brighter 8\,$\mu$m emission than the  brightest (and most likely the most massive) of our seven sources (MYSO \ssix). While these sources could also be appropriate candidates for isolated high-mass star formation, they did not satisfy all of our isolation criteria, and therefore they were not considered for the sample. These criteria are set to assure that our sample observes some of the most isolated MYSOs in the LMC. %The detailed analysis presented later for three of these MYSOs focuses on the two most massive objects (MYSOs \sthree\ and \ssix) and the one in the apparently most isolated environment (MYSO \stwo). %. Initial analysis of the other four isolated YSOs show similar results.

%13 of these 14 sources had similar 8~$\mu$m magnitudes ranging from 6.8 to 6.2~mag, and the remaining source, 052646.61--684847.2, is much brighter ([8~$\mu$m] = 4.25). The MYSO 052646.61-684847.2 is located in the high-mass LMC star-forming complex N44 \citep{Henize1956} near OB association LH49 \citep{Lucke1970} and contained within a GMC \citep{Fukui2008}. Since \ssix\ is an late-type O-star YSO (see Section X), this analysis suggests that early O-star MYSOs cannot form without having other MYSOs forming nearby.

%actual distances were 111 and 563~pc, but let's not get too specific!

%052646.61-684847.2 4.25
%053252.41-694620.1 6.21
%051351.51-672721.9 6.25
%053931.19-701216.8 6.31
%050520.28-665506.5 6.31
%052504.10-682824.5 6.33
%054330.33-692446.6 6.37
%051024.09-701406.5 6.42
%052249.88-664056.1 6.43
%045709.73-684448.2 6.44
%053630.81-691817.2 6.5
%051324.50-691048.3 6.51
%045111.39-692646.7 6.67
%045550.62-663434.6 6.74
%052343.48-680033.9 6.76
%050051.45-662359.9 6.79

\subsection{Observations}\label{s:obs}

\subsubsection{\emph{HST} Observations and Photometry}\label{s:obsphot}

We acquired \emph{HST} images of the seven MYSOs in our sample in five different filters using the Wide Field Camera 3 (WFC3). The observations were performed during Cycle 20 for project GO-12941 (PI: I. Stephens).
Two broad-band filters (F555W and F814W) and one narrow-band filter (F656N) were used with the WFC3/UVIS imager, and the broad-band filters F110W and F160W were used with the WFC3/IR imager. Filters F555W, F814W, F110W, and F160W roughly correspond to 
standard $V$, $I$, $J$, and $H$ bands respectively, while the F656N filter corresponds to H$\alpha$. We simultaneously used ACS/WFC for parallel observations in the filters F555W, F658N, and F814W ($\sim$\,$V$, H$\alpha$, and $I$). The angular resolution of these observations can be calculated as $R=\sqrt{R^{\prime2}+p^{\prime2}}$, $p^\prime$ being the pixel size in seconds of arc and $R^\prime = 0\farcs21  \lambda/ D$, where wavelength $\lambda$ is in $\mu$m and the diameter of the telescope $D$ is in meters (2.4\,m). The pixel sizes are 0\farcs04 and 0\farcs13 for WFC3/UVIS and WFC3/IR respectively and 0\farcs05 for ACS/WFC. The corresponding resolutions are given in Table\,\ref{tab:obs}, where the \emph{HST} observations are summarized. The integration times for \sone\ are lower than the rest of the MYSOs due to observing restrictions enforced for \emph{HST} Cycle 20 observations.
The field-of-view of both WFC3/UVIS and WFC3/IR is 162\arcsec\,$\times$\,162\arcsec and 123\arcsec\,$\times$\,136\arcsec, respectively, and that of ACS/WFC is  202\arcsec\,$\times$\,202\arcsec.

%The field of view for WFC3/UVIS is 162$\arcsec$~$\times$~162$\arcsec$ and for WFC3/IR is 123$\arcsec$~$\times$~136$\arcsec$. The angular resolution can be calculated using $R=\sqrt{R^{\prime2}+p^{\prime2}}$, where $p^\prime$ is the pixel size in arcseconds and $R^\prime = 0.21\arcsec  \lambda/ D$, where $\lambda$ is the wavelength in $\mu$m and D is the diameter of the telescopes (2.4~m). Given pixel sizes of 0.04$\arcsec$ and 0.13$\arcsec$ for WFC3/UVIS and WFC3/IR respectively, we calculate resolutions of 0\farcs06, 0\farcs07, 0\farcs08, 0\farcs16, and 0\farcs19 for F555W, F656N, F814W, F110W, and F160W respectively. Summary of the observations are seen in Table \ref{tab:obs}.
%\begin{deluxetable}{@{}cc@{}c@{}c@{}c@{}cccc@{}c@{}c@{}c@{}c@{}c@{}c@{}}
\begin{deluxetable*}{@{}c@{}c@{}cc@{}c@{}c@{}c@{}c@{}c@{}c@{}}
%\tabletypesize{\tiny}
%\tablecolumns{4}
%\rotate
\tablewidth{0pc}
\tablecaption{Summary of the \emph{HST} observations \label{tab:obs} }

\tablehead{
\colhead{}			&
\colhead{WFC3/UVIS}    	&
\colhead{WFC3/UVIS}    	&
\colhead{WFC3/UVIS}     	&
\colhead{WFC3/IR}      	&
\colhead{WFC3/IR}     	&
\colhead{ACS/WFC\tablenotemark{a}}     	&
\colhead{ACS/WFC\tablenotemark{a}}  	&
\colhead{ACS/WFC\tablenotemark{a}}  	&      	\\
\colhead{}		&
\colhead{F555W ($V$)}&
\colhead{F656N (H$\alpha$)\tablenotemark{b}}&
\colhead{F814W ($I$)}     	&
\colhead{F110W ($J$)}     	&
\colhead{F160W ($H$)}      	&
\colhead{F555W ($V$)}\ &
\colhead{F658N (H$\alpha$)} &
\colhead{F814W ($I$)}   	 
}
\startdata
%&     	8 & \nodata &  \nodata &  \nodata &  \nodata &  \\
%Field of View &    \multicolumn{3}{c}{162$\arcsec$~$\times$~162$\arcsec$} &  \multicolumn{2}{c}{123$\arcsec$~$\times$~136$\arcsec$}&  \multicolumn{3}{c}{202$\arcsec$~$\times$~202$\arcsec$}\\ 
%Field of View (\arcsec) &   162~$\times$~162 & 162~$\times$~162 & 162~$\times$~162 & 162~$\times$~162 & 162~$\times$~162 &  202~$\times$~202 &  202~$\times$~202 &  202~$\times$~202\\ 
Effective $\lambda$ (nm) & 530.8 & 656.1 &  802.4 &  1153.4 &  1536.9  & 536.1  & 658.4 & 805.7 & \\ 
Resolution (\arcsec) & 0.06 & 0.07 &  0.08 &  0.16 &  0.19  & 0.07  & 0.08 & 0.09 & \\ 
%Exposure Time 1 (s) &  $3\times400$ &  $1\times400$, $1\times440$, $1\times692$  &  $2\times1000$, $1\times693$ &  $3\times299$ &  $3\times499$, $1\times599$ &   $1\times858$, $1\times980$ & $3\times750$ & $1\times1692$, $1\times2091$, $1\times460$ &  \\ 
\hline
\multicolumn{9}{c}{Exposures A (s)}\\
\hline
 &  $3\times400$ &  $1\times400$  &  $2\times1000$ &  $3\times299$ &  $3\times499$ &   $1\times858$ & $3\times750$ & $1\times1692$&  \\ 
 &   & $1\times440$  &  $1\times693$ &   &  $1\times599$ &  $1\times980$ & & $1\times2091$  &  \\ 
                              &                              &   $1\times692$  &   &   &   &   & &  $1\times460$&  \\ 
Total time (s) & 1200 & 1532 & 2693 & 898 & 2097 & 1838 & 2250 & 4243 & \\
\hline
\multicolumn{9}{c}{Exposures B (s)}\\
\hline
 &  $3\times400$ &  $2\times400$  &  $1\times1000$ &  $3\times299$ &  $2\times499$ &   $1\times280$ & $2\times750$ & $1\times1690$ &  \\ 
                              &                            & $1\times429$&  $1\times600$&   &  $1\times599$ &  $1\times820$ & $1\times833$ & $1\times1353$ &  \\ 
                              &                              &     & $1\times404$   &                                 &                            &                            &                              &                                &  \\  
Total time (s)  &   1200 & 1229 & 2004 & 898 &  1598 &  1100 &  2333 & 3043
\enddata
\tablecomments{
%The character(s) in parentheses next to the filter name indicates the approximate band. 
``Exposures A'' describe the exposure times applied for all sources except for \sone. ``Exposures B'' describe the exposure times applied for source \sone.
}
\tablenotetext{a}{Observations taken parallel to WFC3 and thus are not focused on our targets.}
\tablenotetext{b}{We included a short 10-second exposure with the F656N filter to provide measurements for the saturated sources.} %However, this exposure was not used for the science in this paper.}
\end{deluxetable*}

%ACTUAL BREAKDOWNS FROM RETRIEVING PHASE II APT
%WFC3
%F555W: 400+400+400=1200
%F656N: 400+440+692=1532 also + 10
%F814W: 1000+1000+693=2693
%F110W: 897.694=897.694
%F160W: 1497.696+599.232=2096.928
%ACS WFC1:
%F555W: 858+980=1838
%F658N: 2250=2250
%F814W: 1692+2091+460=4243
%Total: 16290.612

%045403.6-671618.5 BREAKDOWN
%WFC3
%F555W: 400+400+400=1200
%F656N: 400+429+400=1229 NO + 10
%F814W: 1000+600+404=2004
%F110W: 598.463+299.231=897.694 (first is 2x)
%F160W: 998.464+599.232=1597.696
%ACS WFC1:
%F555W: 280+820=1100
%F658N: 1500+833=2333
%F814W: 1690+1353=3043
%Total: 13404.39

\begin{figure}[t!]
\begin{center}
\includegraphics[scale=0.5]{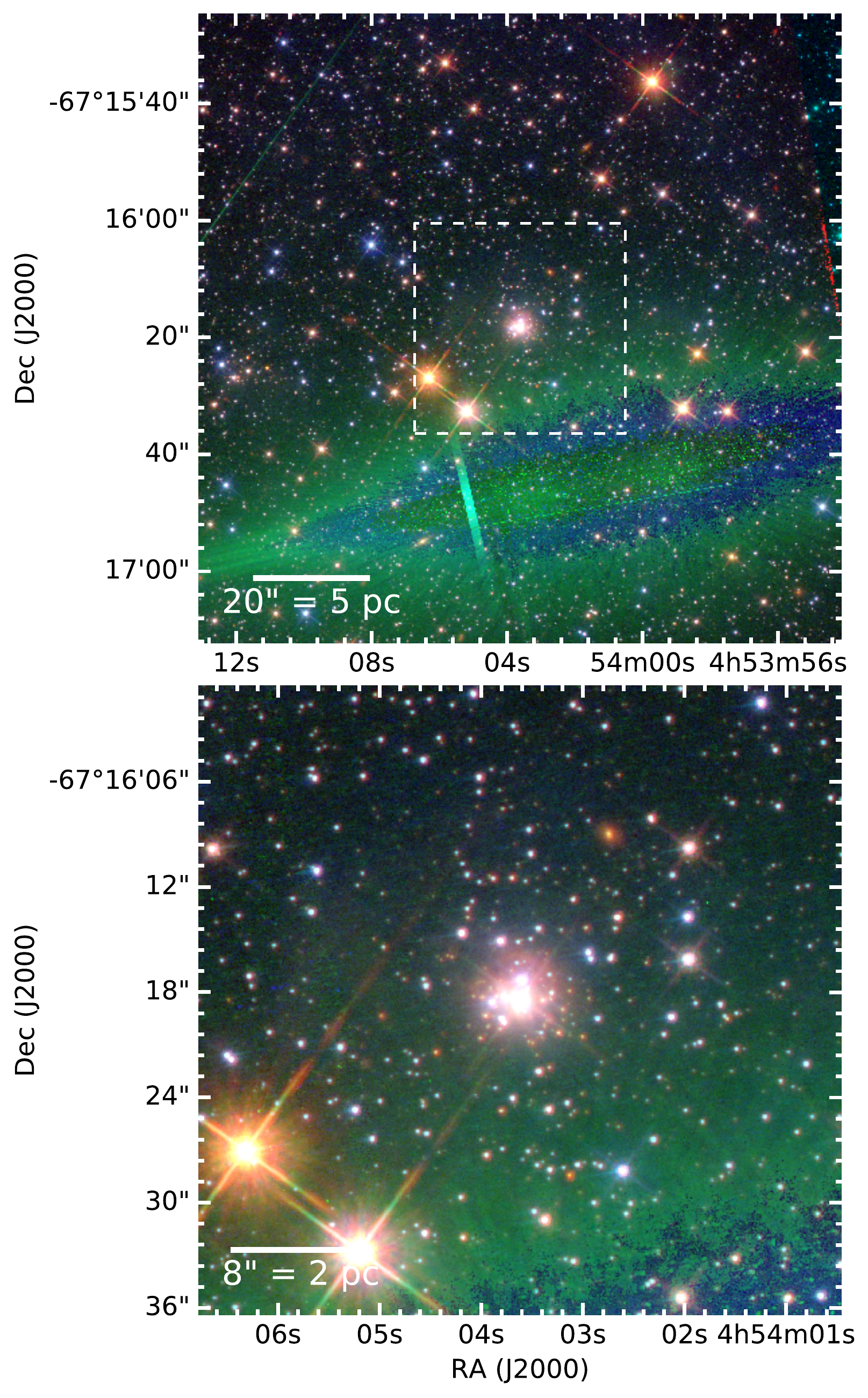}
\caption{Three-color image of MYSO 045403.6--671618.5. The colors in red, green, and blue are F160W, F814W, and F555W ($\sim$1.5, 0.80, and 0.53\,$\mu$m), respectively. Both panels are centered on the brightest photometric source of the high-mass star forming region.The top panel shows a large field of view, with a white box indicating the zoom-in shown in the bottom panel.  Each color is on an arcsinh scale and colors were adjusted in each panel to best show the stellar content. For this image, the large green/blue spike toward the south 
of the YSO is due to a bright foreground star located to the east, outside the field of view. \label{045403_2panel.png}
}
\end{center}
\end{figure}

\begin{figure}[t!]
\begin{center}
\includegraphics[scale=0.5]{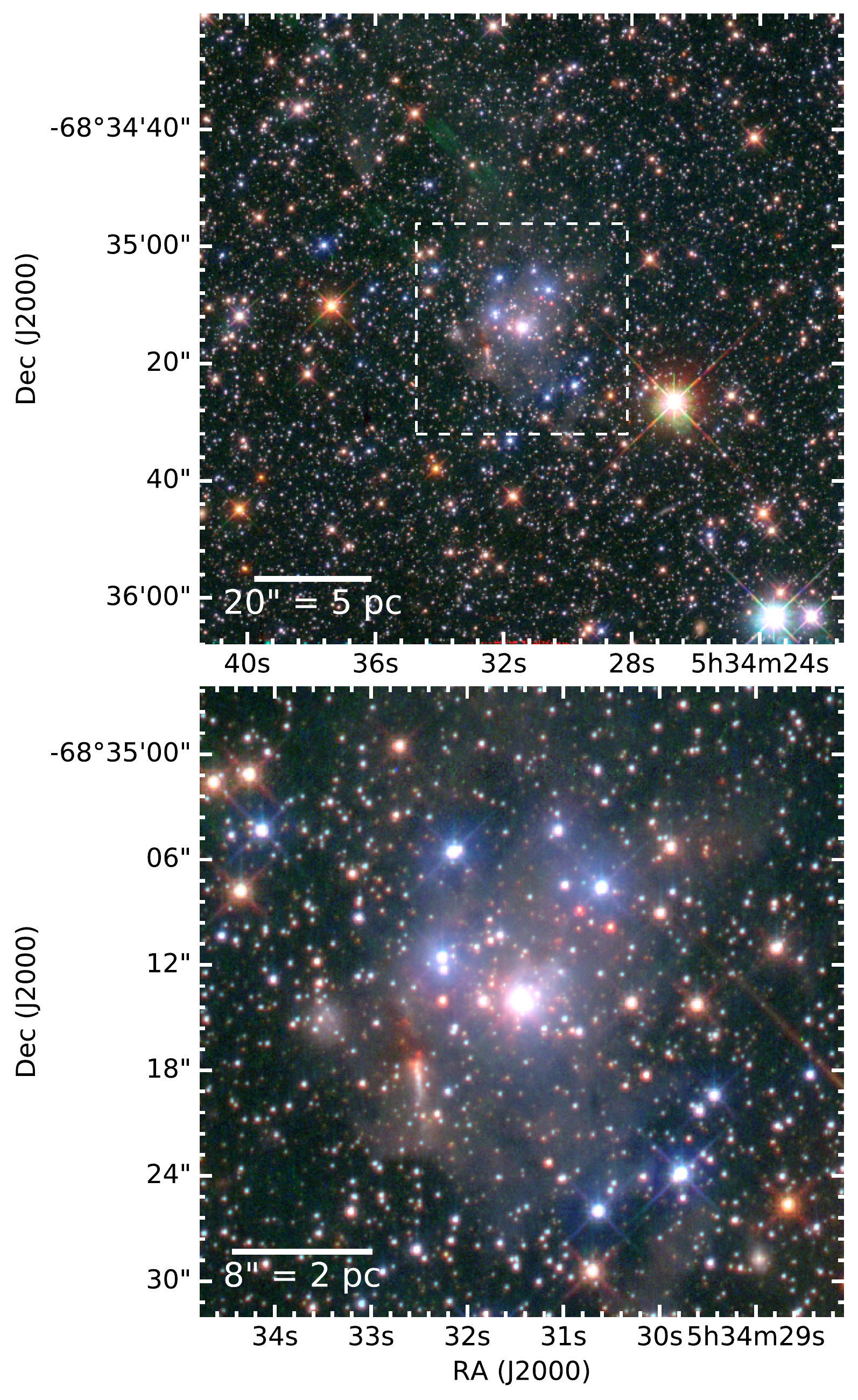}
\caption{Three-color image of MYSO 053431.5--683513.9. Image description as in Figure\,\ref{045403_2panel.png}. \label{053431_2panel.png}
}
\end{center}
\end{figure}

The observation strategy is primarily focused on deep imaging with two filters, F814W and F160W, in order to create color-magnitude diagrams (CMDs) that are more sensitive to low-mass protostars. Deep observations were chosen for the F814W over F555W because the F555W filter is usually more contaminated by diffuse nebular emission. F160W was chosen over F110W in order to detect the more embedded protostars. Our observations also include both F555W and F110W filters because they provide accurate identification of less embedded stellar populations. These filters also allow for the search for possible areas of extinction through color-color diagrams. %They can also be used for extending our band coverage, allowing the modeling of spectral energy distributions (SEDs) of the stars. 
Moreover, at low extinction \emph{HST} WFC3 can reach significantly deeper magnitudes in F110W than in F160W. F656N was included in order to determine locations of classical and compact \ion{H}{2} regions, Herbig Ae/Be stars, classical T\,Tauri stars, and bow shocks that can indicate possible runaway stars. The H$\alpha$ observations are also used to estimate the spectral types of the MYSOs (see Section\,\ref{s:spclass}).  Since in Cycle 20 WFC3 narrow-bands suffer from Charge Transfer Efficiency (CTE) loss, a post-flash was incorporated for F656N (using the parameter \texttt{Flash=12}). The WFC3 3-color images (using F160W, F555W, and F814W) for each MYSO are seen in Figures \ref{045403_2panel.png}\,-\,\ref{053431_2panel.png}. In these images it is immediately evident that \emph{none of these MYSOs are forming in complete isolation}. We investigate the clustering properties of stars in the vicinity of these MYSOs in Sections \ref{identPMS} and \ref{s:clusanl}. %These targets are specifically the two most massive in our sample (\sthree\ and \ssix) and that, which appears to be the most isolated (\stwo, located 600\,pc from any GC09 MYSO).

\begin{figure}[t!]
\begin{center}
\includegraphics[scale=0.5]{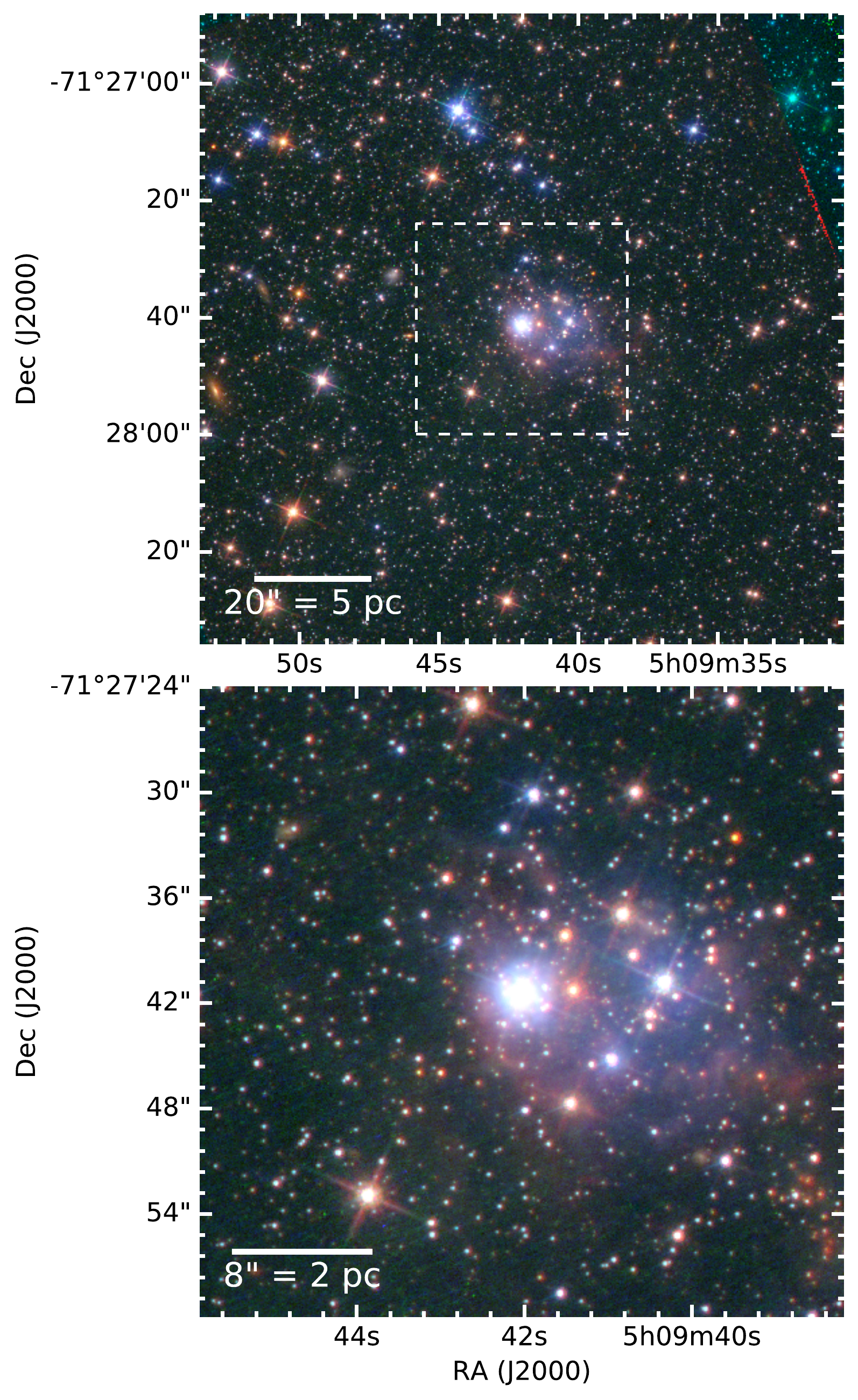}
\caption{Three-color image of MYSO 050941.9--712742.1. Image description as in Figure\,\ref{045403_2panel.png}.  \label{050941_2panel.png}}
%For this figure, the top-right corner of the left panel is missing F160W emission due to the orientation of \emph{HST} for this observation.
\end{center}
\end{figure}

\begin{figure}[t!]
\begin{center}
\includegraphics[scale=0.5]{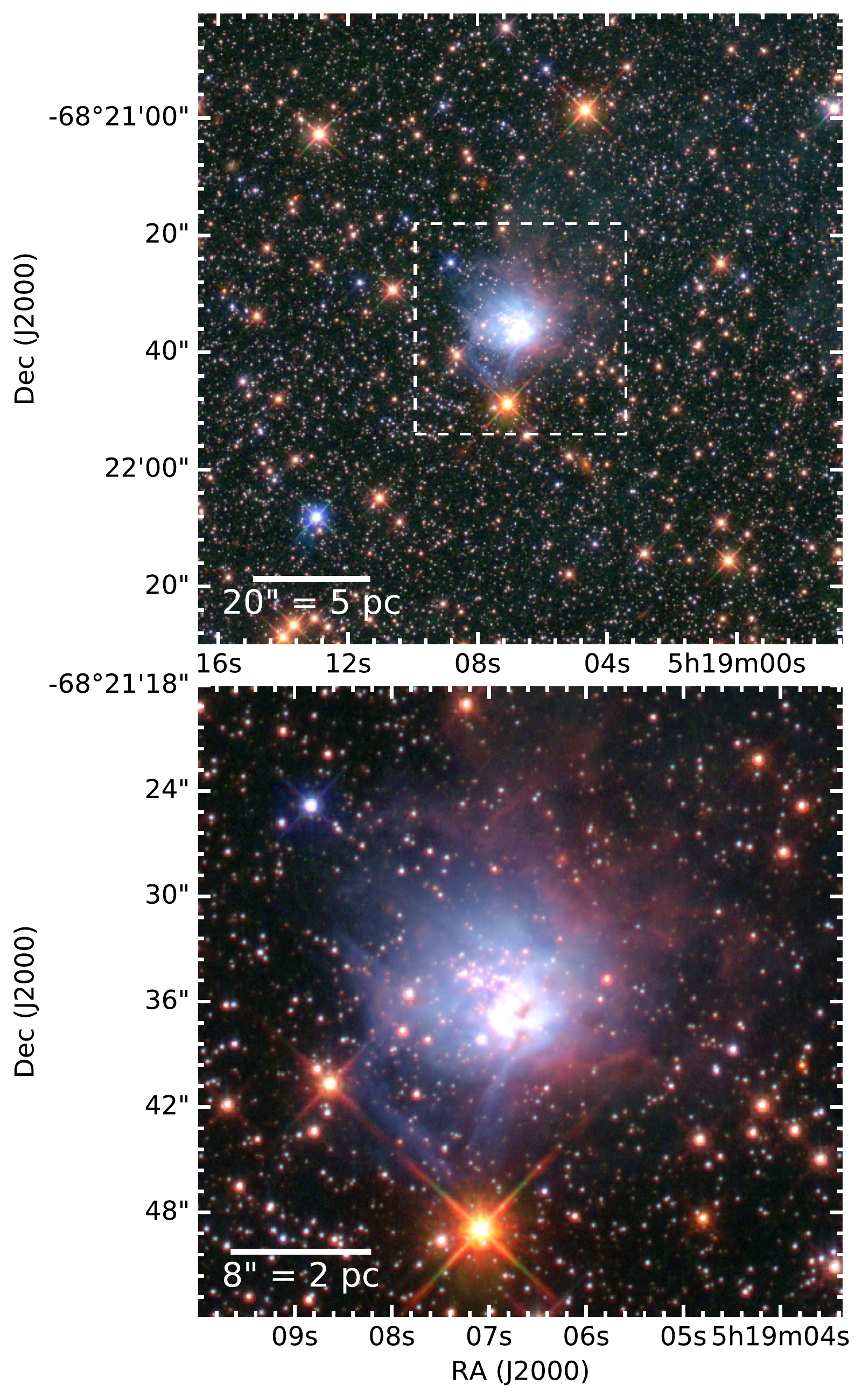}
\caption{Three-color image of MYSO 051906.7--682137.4. Image description as in Figure\,\ref{045403_2panel.png}. \label{051906_2panel.png}
}
\end{center}
\end{figure}

\begin{figure}[t!]
\begin{center}
\includegraphics[scale=0.5]{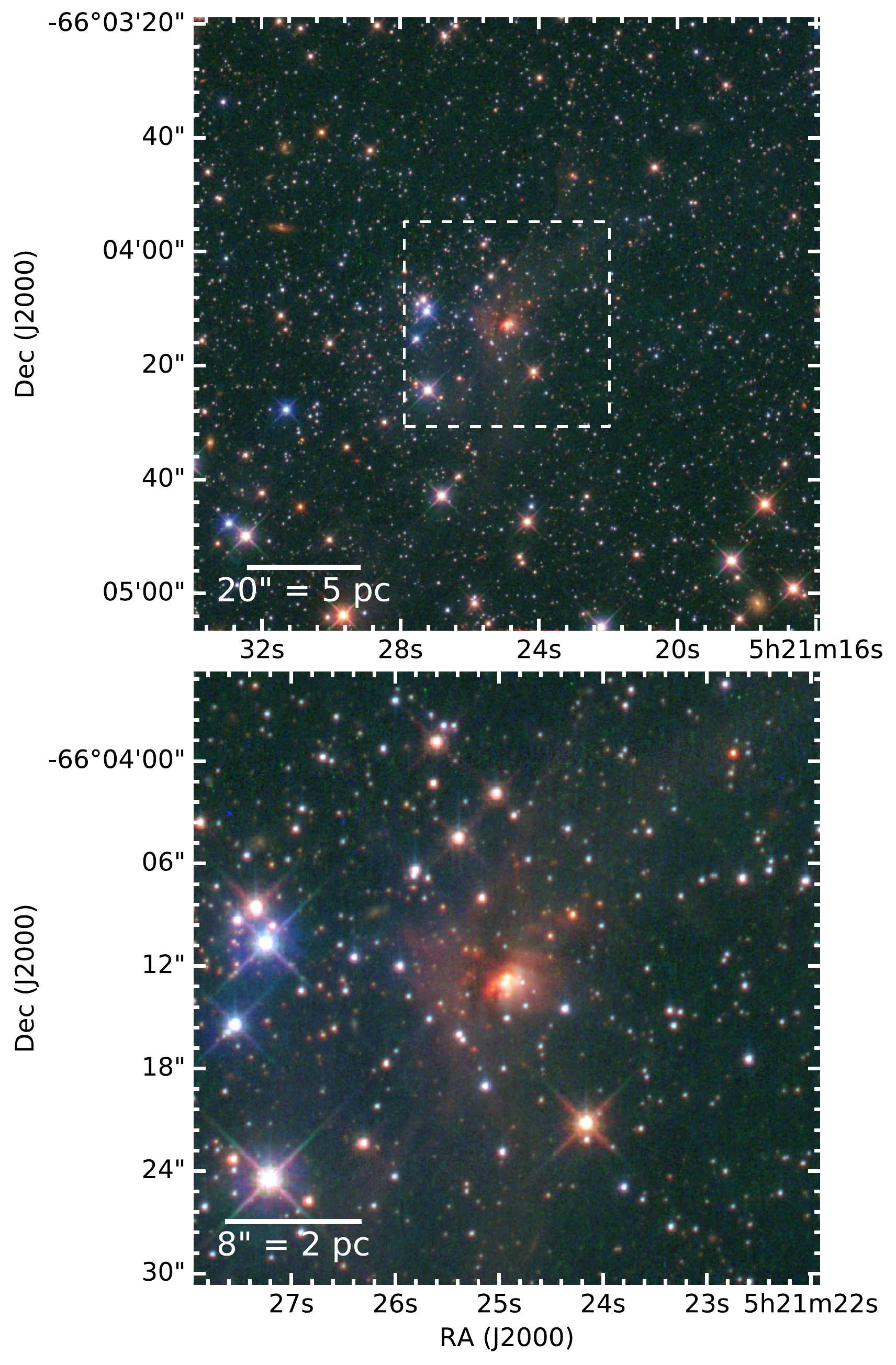}
\caption{Three-color image of MYSO 052124.9--660412.9. Image description as in Figure\,\ref{045403_2panel.png}. \label{052124_2panel.png}
}
\end{center}
\end{figure}

\begin{figure}[t!]
\begin{center}
\includegraphics[scale=0.5]{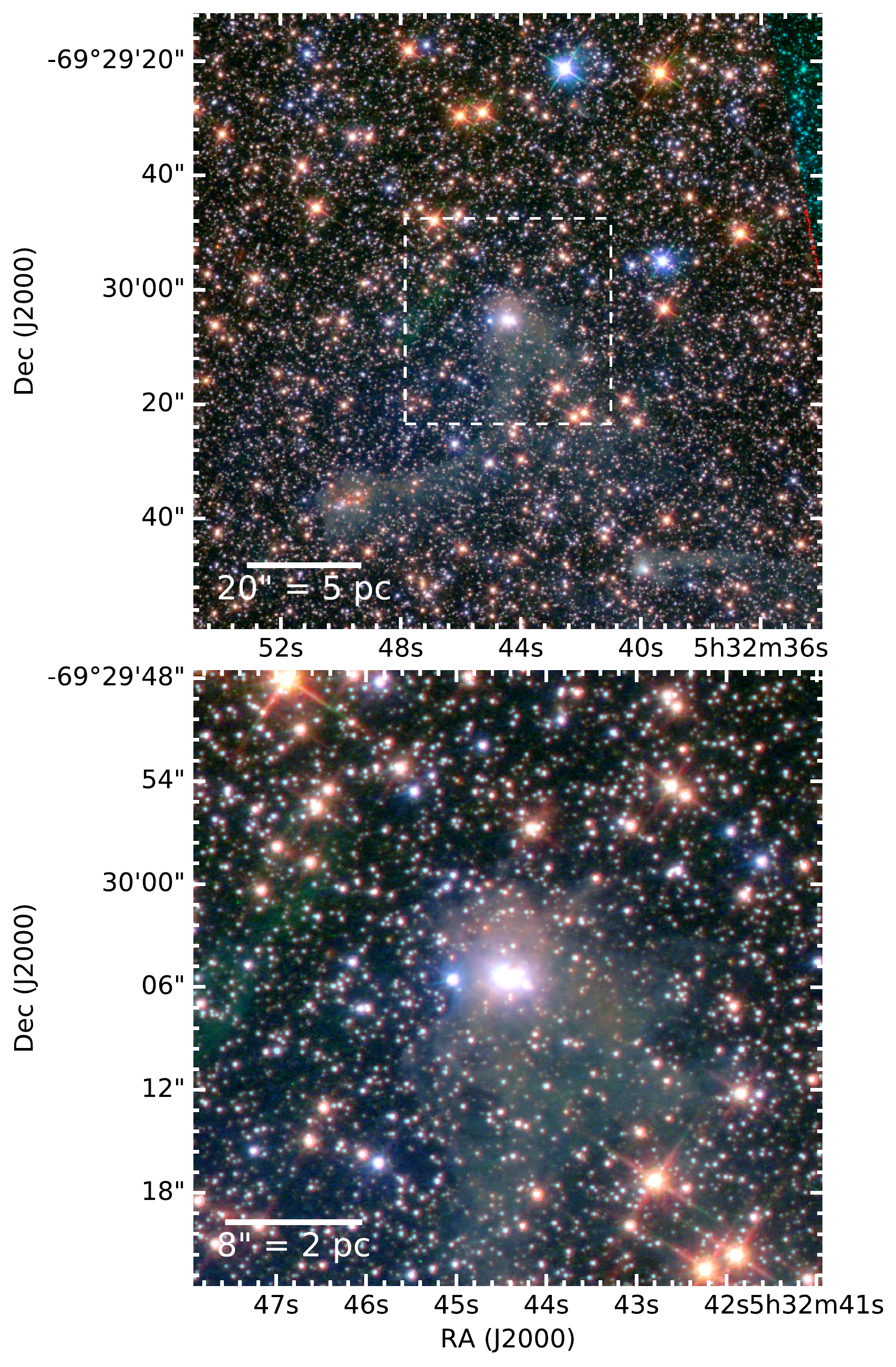}
\caption{Three-color image of MYSO 053244.3--693005.7. Image description as in Figure\,\ref{045403_2panel.png}. \label{053244_2panel.png}
%For this figure, the top-right corner of the left panel is missing F160W emission due to the orientation of \emph{HST} for this observation. 
}
\end{center}
\end{figure}

\begin{figure}[t!]
\begin{center}
\includegraphics[scale=0.5]{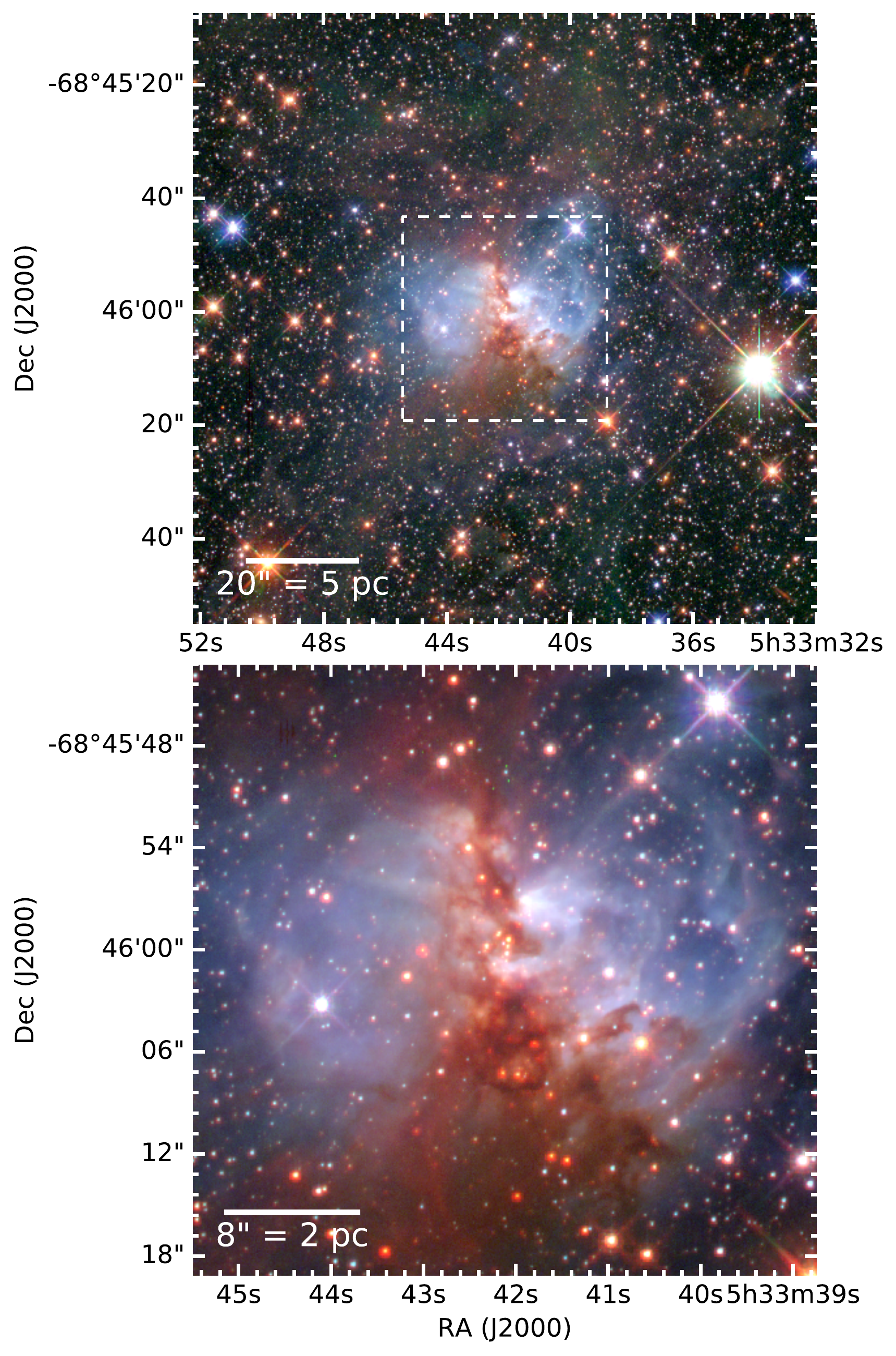}
\caption{MYSO 053342.2--684602.8. Image description as in Figure\,\ref{045403_2panel.png}. This image is centered on the star-forming region rather than the brightest source. The brightest embedded source is located slightly northwest of the image center. \label{053342_2panel.png}
}
\end{center}
\end{figure}

Photometry was performed with the {\sc dolphot} package\footnote{{\sc dolphot} is available online 
at \href{http://purcell.as.arizona.edu/dolphot}{http://purcell.as.arizona.edu/dolphot}} \citep{Dolphin2000}.
This package performs PSF fitting tailored to \emph{HST} cameras. The images were first prepared 
with the {\sc dolphot} routines {\tt acsmask} and {\tt splitgroups}, which respectively apply the image defect mask 
and split the original HST \_FLT FITS files into a single FITS file per chip. The main {\sc dolphot} routine was then 
used to make photometric measurements on the pre-processed images relative
to the coordinate system of the drizzled F555W image, which was used as a reference. The output 
photometry from {\sc dolphot} is on the calibrated VEGAmag scale based 
on the zeropoints provided on the WFC3 page.\footnote{\href{http://www.stsci.edu/hst/wfc3}{http://www.stsci.edu/hst/wfc3}} The VEGAmag zeropoints for F555W, F656N, F814W, F110W, and F160W are 
25.8160, 19.8215, 24.6803, 26.0628, and 24.6949 mag, respectively. The $HST$ magnitudes of the MYSOs are indicated in Table\,\ref{t:bright}.

For the photometric analysis presented in this paper, only the sources with the best photometric quality parameters were kept. Specifically, the 
sources in the original {\sc dolphot} output file should meet the following criteria: {\tt Object Type}\,=\,1 (i.e., a PSF consistent with stellar, non-extended objects), {\tt signal-to-noise}$\,>$\,5, {\tt sharp}$^2<$\,0.3, {\tt crowd}\,$<$\,2, and {\tt round}$^2 <$\,1. The final stellar photometric catalog is referred to as our {\em clean photometric sample}. %{\bf Do I mention F814W and F160W here or later? Include table of centers?}

%Source, F160W, F814W, RA, DEC
%\sone  14.456 99.999 4:54:03.613 -67:16:18.44
%\stwo 15.653 99.999 5:09:42.028 -71:27:41.55
%\sthree 15.622 99.999 5:19:06.746 -68:21:36.26
%\sfour 16.862 20.072 5:21:24.927 -66:04:12.72
%\sfive 16.815 99.999 5:32:44.418 -69:30:05.48
%\ssix  17.456 18.359 5:33:41.900 -68:45:57.16
%\sseven 15.205 18.318 5:34:31.471 -68:35:14.19
%16.792 17.877 5:34:31.436 -68:35:14.09
%Black circles show the locations the brightest sources (F160W magnitudes less than 18 mag) within 1$\arcsec$ of the central MYSO}

\begin{deluxetable*}{cccccccc}
\tablecaption{$HST$ Magnitudes of MYSOs \label{t:bright}}
\tablewidth{0pc}
\tablehead{
\colhead{MYSO} & \colhead{$\alpha$ (2000)} & \colhead{$\delta$ (2000)} & \colhead{F555W}  & \colhead{F814W} & \colhead{F110W} & \colhead{F160W} & \colhead{$A_V$\tablenotemark{a}} \\
\colhead{} &  \colhead{(h:m:s)} & \colhead{(d:m:s)} & \colhead{(mag)} & \colhead{(mag)} & \colhead{(mag)} & \colhead{(mag)} & \colhead{(mag)}
 }
\startdata
\sone & 04:54:03.61 & --67:16:18.4 &   &   & 15.562 & 14.456 & 0.58  \\
\stwo  & 05:09:42.03 & --71:27:41.6 &   &   & 15.923 & 15.653 & 0.20 \\
\sthree & 05:19:06.75 & --68:21:36.3 &   &   & 16.076 & 15.622 & 0.29 \\
\sfour & 05:21:24.93 & --66:04:12.7 & 22.562 & 20.072 & 18.175 & 16.862 & 0.69 \\
\sfive & 05:32:44.42 & --69:30:05.5 &   &   & 17.149 & 16.815 & 0.23 \\
\ssix & 05:33:41.90 & --68:45:57.2 & 19.093 & 18.359 & 17.816 & 17.456 & 0.24 \\
\sseven & 05:34:31.47 & --68:35:14.2 & 19.717 & 18.318 & 16.868 & 15.205 & 0.85
\enddata
\tablecomments{MYSO is defined as the central source with the brightest F160W magnitude. The right ascension and declination in the table indicate the photometric location of the MYSO. MYSOs without F555W or F814W magnitudes did not have valid photometric fits. Uncertainties in the photometric magnitudes were typically $\sim$0.001 mag.}
\tablenotetext{a}{$A_V$ was calculated based on the Padova isochrones using F110W and F160W magnitudes and assuming a zero-age main sequence star.}
\end{deluxetable*}

%sqrt((0.21*.555/2.4)^2+0.040^2) = 0.06291515243
%sqrt((0.21*.656/2.4)^2+0.040^2) = 0.06996256141
%sqrt((0.21*.814/2.4)^2+0.040^2) = 0.08168843629
%sqrt((0.21*1.11/2.4)^2+0.13^2) = 0.1622752773
%sqrt((0.21*1.6/2.4)^2+0.13^2) = 0.19104973174

%sqrt((0.21*.555/2.4)^2+0.05^2) = 0.06291515243
%sqrt((0.21*.658/2.4)^2+0.05^2) = 0.07625536456
%sqrt((0.21*.814/2.4)^2+0.05^2) = 0.08702298906

\subsubsection{Mopra Observations}
\begin{figure}[t!]
\begin{center}
\includegraphics[scale=0.45]{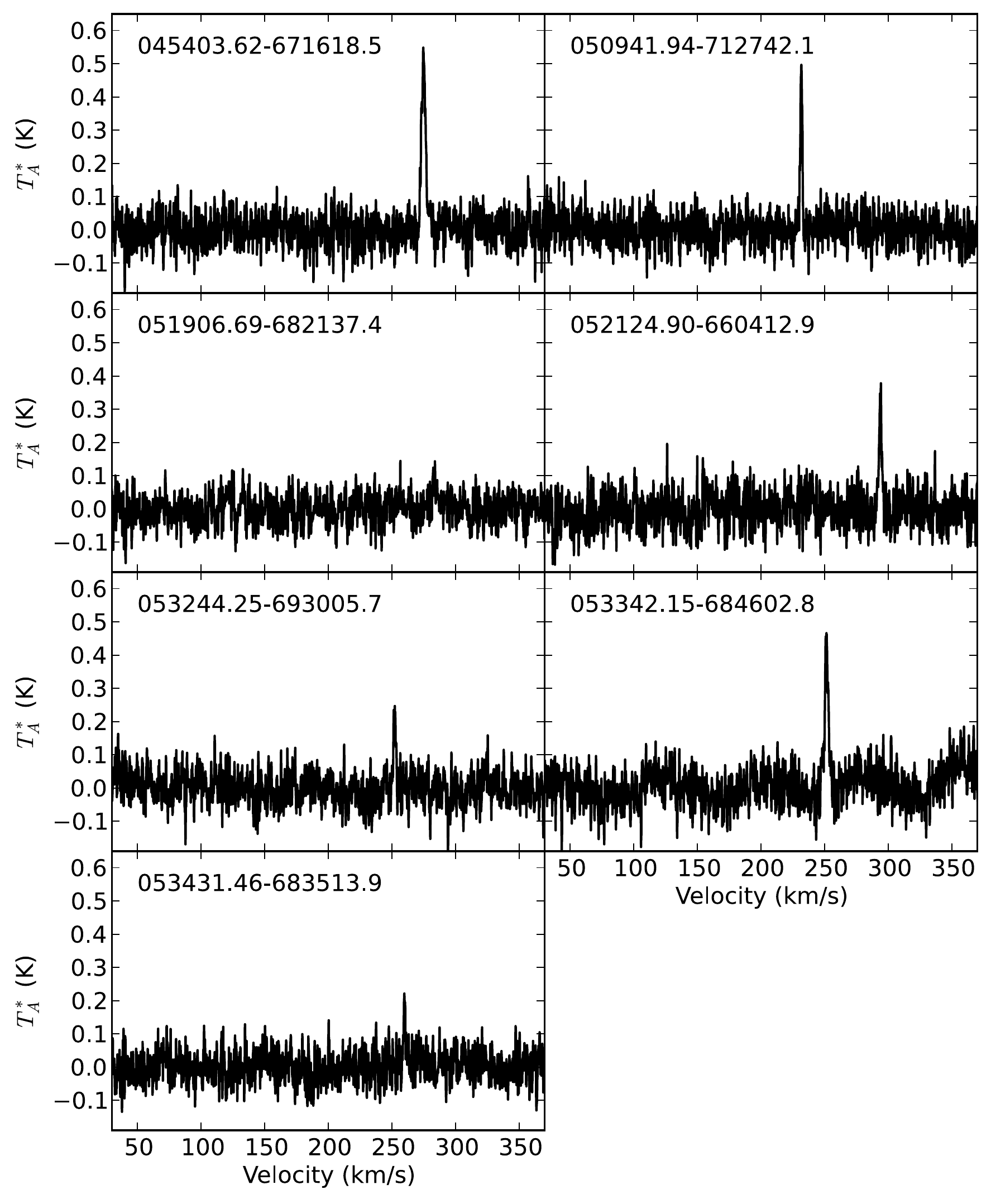}
\caption{Mopra CO(1--0) spectra centered on the MYSO based on the coordinates given in GC09. Velocities are given the radio kinematic local standard of rest (LSRK) frame.
Spectra were smoothed with a Gaussian kernel with a width of four channels (0.36~\kms) after deleting the baseline. \label{COspec}
}
\end{center}
\end{figure}

After our \emph{HST} observations, Mopra spectra on our sources became available. The MAGMA survey \citep{Wong2011} is a CO(1--0) survey of the LMC that targeted only the locations with NANTEN CO(1--0) emission \citep{Fukui2008}. While all seven of the isolated MYSO candidates except \stwo\ were covered by the NANTEN survey (Figure\,\ref{LMC_YSOs}), none were covered by the MAGMA survey. MAGMA mapped the LMC GMCs at the higher resolution of $\sim$\,11\,pc (NANTEN had a resolution of $\sim$\,40\,pc), but it covers just $\sim$80\% of the total emission detected by NANTEN. Of the 248 MYSOs identified in GC09, 76 MYSOs were not covered in the MAGMA survey, among which are our seven selected sources. In 2011 June and July, members of our team (PI: T. Wong) performed follow-up single pointing Mopra observations of all these 76 sources, integrating for 10 minutes on each source. The sensitivity is approximately a factor of 2 times better than that of the MAGMA survey. Our seven selected MYSOs were included in these runs, and we present here their Mopra spectra, which are shown in Figure\,\ref{COspec}.

%GMC: open LMC_MAGMA_DR2b.CO.45as.mmom0.fits.gz and use calculatedistance.py. Use ds9 ruler to figure out which pixel is closest. Then use a pixel sized box to place over that pixel and get the center RA/dec. Plug into calculatedistance.py
%OB: nearest_LH.py
%Spitzer MYSO:  nearestneighborhisto.py
%Spitzer IMYSO: nearestneighbors_intermediate.py plus Seale2014, search_seale2014.py   bica_distances.xlsx
%Herschel: search_seale2014.py   bica_distances.xlsx
%Cluster/Association/SNR: search_bica2008_append.py   bica_distances.xlsx
\begin{deluxetable*}{@{}c@{}c@{}c@{}c@{}c@{}c@{}c@{}c@{}c}
\tabletypesize{\scriptsize}
%\tablecolumns{4}
%\rotate
\tablewidth{0pc}
\tablecaption{Closest Objects to Isolated Sources in Parsecs \label{tab:isolation} }
\setlength{\tabcolsep}{1.5mm}
\tablehead{
\colhead{Isolated}			&
\colhead{GMC\tablenotemark{a}}   	&
\colhead{Star in OB}     	&
\colhead{{\em Spitzer}}     	&
\colhead{{\em Spitzer} Intermediate-}  	&
%\colhead{$Herschel$\tablenotemark{d}}       &
\colhead{{\em Herschel}\tablenotemark{d}}       &
\colhead{{\em Herschel}\tablenotemark{d}}  	&
\colhead{Cluster or }&
\colhead{Known}     	\\
\colhead{MYSO}  	&
\colhead{}			&
\colhead{Association\tablenotemark{b}}&
\colhead{MYSO\tablenotemark{c}}&
\colhead{Mass YSO\tablenotemark{c}   }  	&
\colhead{$L_{\rm{FIR}} > 3000~L_\odot$ }&
\colhead{$L_{\rm{FIR}} > 1000~L_\odot$ }&
\colhead{Association\tablenotemark{e}}&
\colhead{SNR\tablenotemark{e}}     		 
}
\startdata
%&     	8 & \nodata &  \nodata &  \nodata &  \nodata &  \\
%Field of View &    \multicolumn{3}{c}{162$\arcsec$~$\times$~162$\arcsec$} &  \multicolumn{2}{c}{123$\arcsec$~$\times$~136$\arcsec$}&  \multicolumn{3}{c}{202$\arcsec$~$\times$~202$\arcsec$}\\ 
\sone & 140 & 43 & 350 & 12 & 170 & 88 & 60 & 67 \\
\stwo & 530 & 230 &  600 &  220 &  510  & 380 & 64 & 1400\\ 
\sthree & 170 & 360 &  260 &  25 &  130 & 26 & 66 & 460 \\ 
\sfour & 320 & 160 & 400 & 7.1 & 320 & 310 & 27 & 330 \\
\sfive & 210 & 170 & 240 & 10 & 240 &  10 & 41 & 300 \\
\ssix  & 83 & 160  &  150 &  9.3 & 130  & 130 & 48 & 430 \\
\sseven & 140 & 230 & 180 & 44 & 180 & 180 & 28 & 560

\enddata
\tablecomments{All numbers are projected distances in pc, using 1$\arcsec = 0.25$~pc  and rounding to two significant figures. The distances are lower limits since we consider projected distances. YSOs, clusters, and non-OB associations are not included if they are associated with the nearby star-forming region of the MYSO.}
\tablenotetext{a}{CO(1--0) data from MAGMA \citep[][Data Release 2]{Wong2011}, measuring the distance to the center of the nearest pixel in the masked CO integrated intensity map. \stwo\ actually lies slightly outside the MAGMA and NANTEN survey region \citep[Figure\,\ref{LMC_YSOs},][]{Fukui2008}, but $Spitzer$ 8~$\mu$m emission does not indicate strong emission expected from a GMC.}
\tablenotetext{b}{This is the distance to the closest OB star in the associations from \citet{Lucke1970}. Distances to the center of the OB associations can be much larger.}
\tablenotetext{c}{MYSOs have [8.0] $\leq$ 8 mag and are considered ``definite'' or ``probable'' YSOs in \citet{GC09}. Intermediate YSOs are ``definite'' or ``probable'' YSOs with lower magnitudes. For \sone, a source identified as a galaxy in \citet{GC09} is declared as a YSO here (see Section\,\ref{sec:sone}).}
\tablenotetext{d}{Dust clumps that have a $Herschel$ derived far-infrared luminosity $L_{\rm{FIR}} > 1000~L_\odot$ should have an embedded source. $Herschel$ sources are from \citet{Seale2014}. Many YSOs and clumps have failed graybody fits \citep{Seale2014}, suggesting that some of these distances may be upper limits.}
\tablenotetext{e}{ The stellar clusters, associations, and supernova remnants (SNRs) are from \citet{Bica2008} and distances are measured to the center of these sources. }

\end{deluxetable*}

In these spectra, CO(1--0) is detected in every MYSO except for perhaps MYSO \sthree.  The spectrum of this source has two possible CO(1--0) peaks between 250 and 300 \kms, but since we do not have velocity information for this MYSO, we cannot confirm whether these are true detections. It should be noted that while there are Mopra detections for almost all considered sources, there are no NANTEN CO(1--0) detections on these MYSOs, which was one of the reasons for selecting them. The lack of NANTEN detections on them limits the molecular gas mass of any potentially associated GMCs to $M_{\rm{CO}} \lsim 2\times 10^4 M_\sun$. A possible exception is source \stwo\ because it was not covered by the NANTEN survey (see Figure\,\ref{LMC_YSOs}). However, based on the lack of very bright emission of this source in the $Spitzer$ 8\,$\mu$m band \citep{Meixner2006}, which CO would typically correlate with, we deduce that there is not likely a nearby GMC.
%2*1.11*2.35*0.768*0.495 = 1.98329472 for ~integrated intensity of 050941. this adjusts for mopra efficiency (assumed to be a factor of 2). i'm not sure that's the efficiency. apparently they vary from ~0.38 to 0.54

%\section{Analysis On The Three Prominent MYSOs in Isolation}\label{section3}
\section{Isolation Analysis of each MYSO Based on Existing Data}\label{section3}

%In Table \ref{tab:isolation} we summarize the approximate distance (in pc) for each isolated MYSO and a variety of source types. To measure the distance to each GMC, MAGMA moment maps and isolated MYSOs were loaded in ALADIN \citep{Bonnarel2000} and the distance tool was used to connect the source and the \emph{closest edge} (not the center) of the CO(1--0) emission. All other distances were calculated using the positions of each source given in the references contained in Table \ref{tab:isolation}.

We characterize the isolation of the seven MYSOs observed by $HST$ by calculating the distances to known astronomical sources. We particularly used the catalogs by \citet{Bica2008} and \citet{Seale2014}, which are constructed from previous known LMC catalogs. The \citet{Bica2008} catalog contains a list of known emission nebulae, star clusters, associations, and \ion{H}{1} shells and supershells, while \citet{Seale2014} used new \emph{Herschel} observations and existing YSO catalogs to find locations of active star-forming regions. \citet{Seale2014} classified {\em Herschel} sources as YSOs, dust clumps (which may or may not have cores), galaxies, or unclassified sources. \citet{Seale2014} classified sources as YSOs if they are not identified as galaxies or other sources (e.g., supernova remnants), are detected in 3 $Herschel$ bands, and have bright 24\,$\mu$m point-like emission. Dust clumps meet the same criteria, except they are not associated with a 24\,$\mu$m point source. The unclassified sources are all dim and may be very faint YSOs or dust clumps, but they may also be fluctuations in the interstellar medium (ISM). This catalog also provides far infrared (FIR) luminosities based on {\em Herschel} graybody fits. We also used Aladin \citep{Bonnarel2000} to provide a pictorial view of the environment encompassing each MYSO. MAGMA \citep{Wong2011} was used to find the closest known GMCs, and we confirmed that these GMCs are indeed the closest based on the lower resolution NANTEN survey \citep{Fukui2008}. The catalog of \citet{Lucke1970} was used to find the locations of the largest OB associations. All distances were measured using the GC09 positions of the MYSOs.

In Table\,\ref{tab:isolation} we summarize the distances (with $1\arcsec \simeq 0.25$\,pc) to the MYSOs. We also summarize each individual MYSO in the Appendix. It should be noted that in the table and the paper in general we are referring to the \emph{projected} distances; thus these are minimum distances to each of the objects. Indeed, the LMC is not entirely face-on, with an inclination of approximately 35$^\circ$ \citep[e.g.,][]{vanderMarel2001}, which implies we are typically underestimating the distances by approximately (1 -- cos 35$^\circ$) = 17\%. The summary in the appendix characterizes the isolation for each of the MYSOs, based primarily on the \citet{Bica2008} and \citet{Seale2014} catalogs. As is shown, all these sources are mostly isolated with no nearby high-mass star formation.

%\clearpage 
%%%%%%%%%%%%%%%%%%%%%%%%%%%% FIGURE %%%%%%%%%%%%%
%\begin{figure*}[t!]
%\begin{center}
%%\centerline{
%\includegraphics[clip=true, trim =  0 0 -0.5cm 0, width=0.33\textwidth]{fig12a.pdf}
%\includegraphics[clip=true, trim =  0 0 -0.5cm 0, width=0.33\textwidth]{fig12b.pdf}
%\includegraphics[clip=true, trim =  0 0 -0.5cm 0, width=0.33\textwidth]{fig12c.pdf}
%%}
%\end{center}
%\caption{ Color-magnitude diagrams for \stwo, \sthree, and \ssix\ using the F814W and F160W filters of the entire clean photometric sample. Typical stellar populations of the general LMC field are plotted with blue symbols. The young PMS stellar sources of every region, determined by statistically decontaminating the complete observed CMDs from the field contribution, are plotted in red. They represent the recent star formation events for each region. An indicative reddening vector for $A_{V} = 2$ mag is shown in the CMDs only to demonstrate the effect of extinction; the length of the vector does not correspond to the actual interstellar extinction in the regions. }
%\label{f:fs.cmds}
%\end{figure*}
%%%%%%%%%%%%%%%%%%%%%%%%%%%%%%%%%%%%%%%%%%%%%%

%\clearpage 
%%%%%%%%%%%%%%%%%%%%%%%%%%%% FIGURE %%%%%%%%%%%%%
\begin{figure*}[t!]
\begin{center}
%\centerline{
\includegraphics[clip=true, trim =  0 0 -0.5cm 0, width=0.247\textwidth]{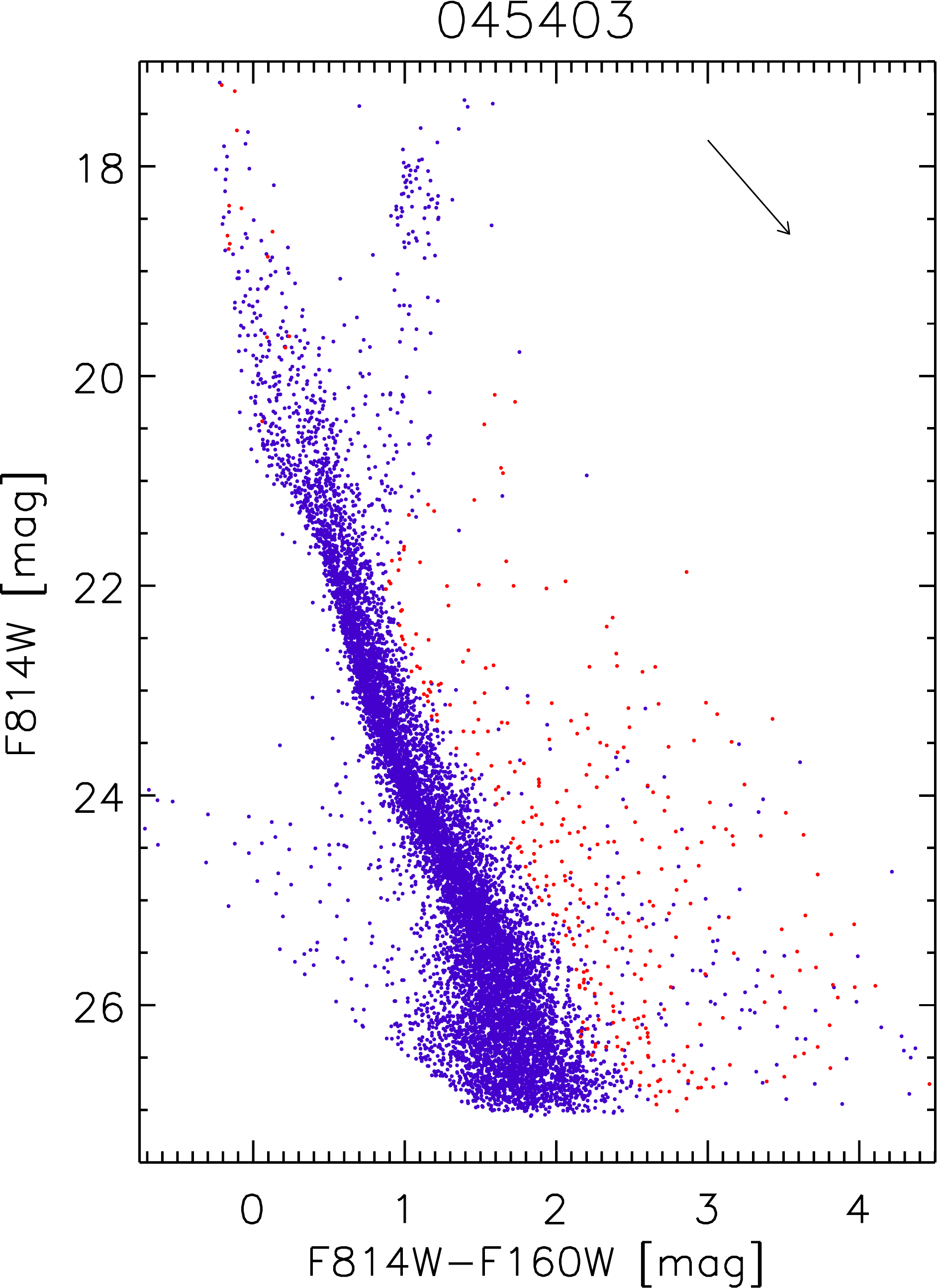}
\includegraphics[clip=true, trim =  0 0 -0.5cm 0, width=0.247\textwidth]{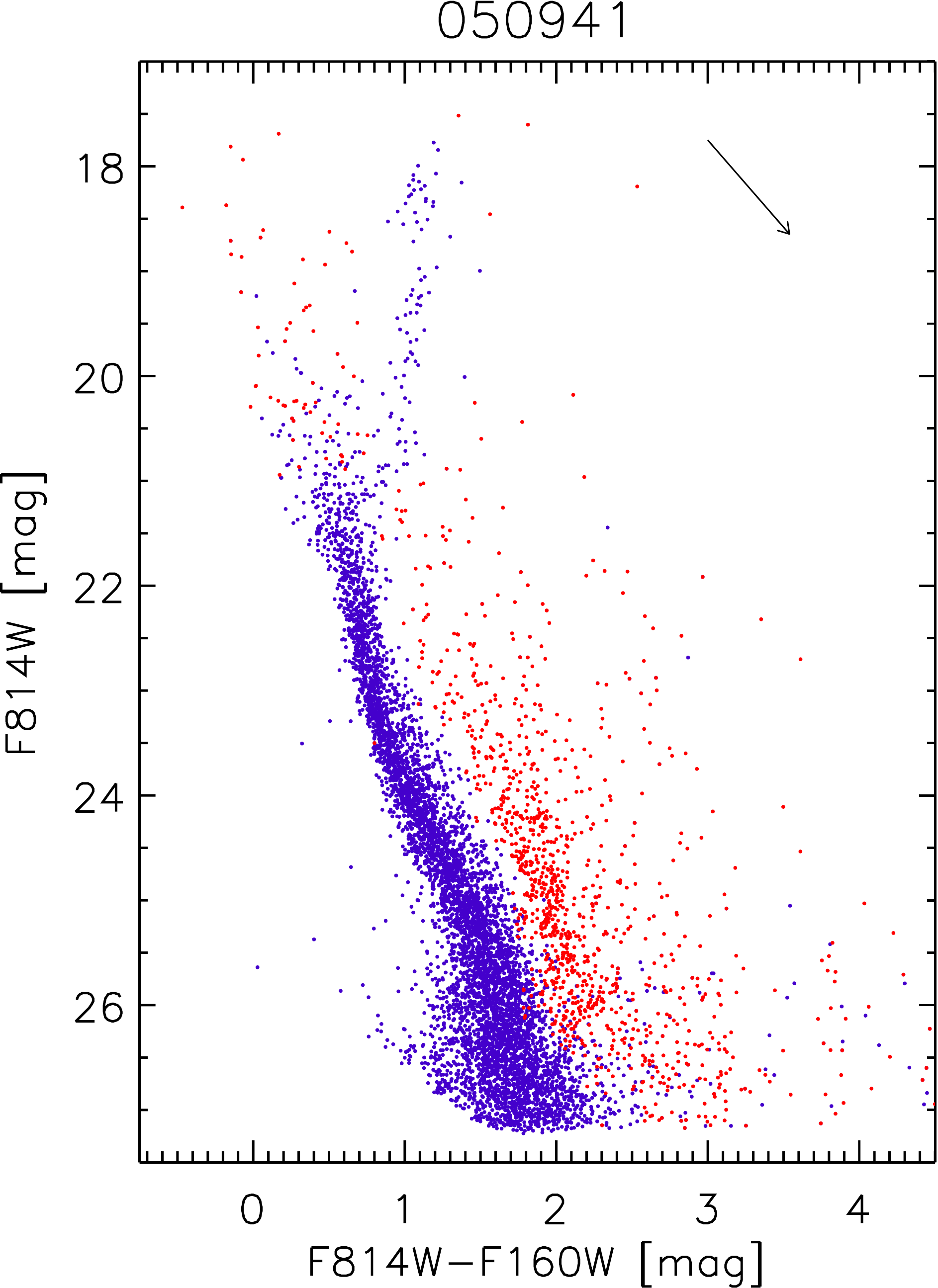}
\includegraphics[clip=true, trim =  0 0 -0.5cm 0, width=0.247\textwidth]{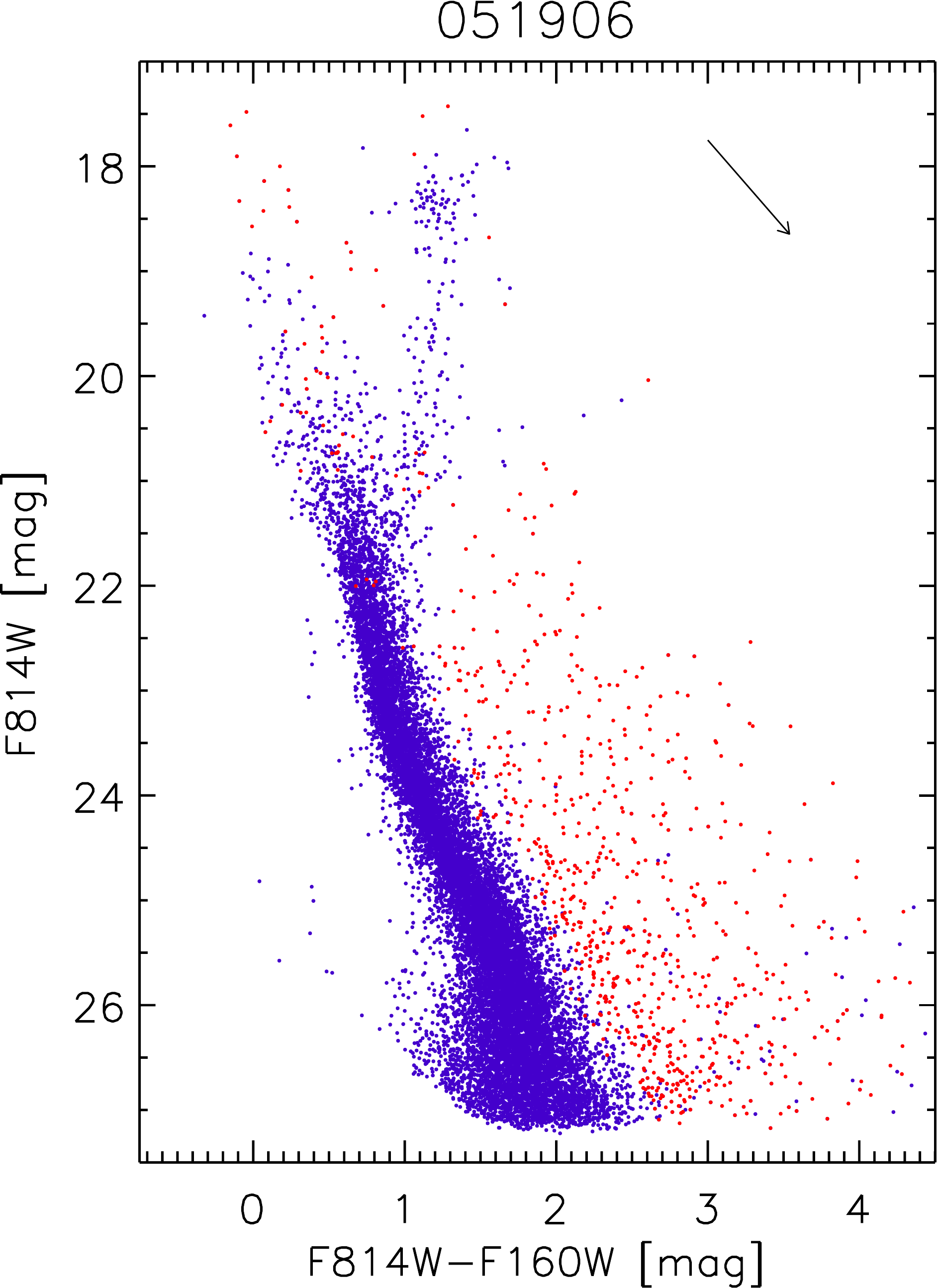}
\includegraphics[clip=true, trim =  0 0 -0.5cm 0, width=0.247\textwidth]{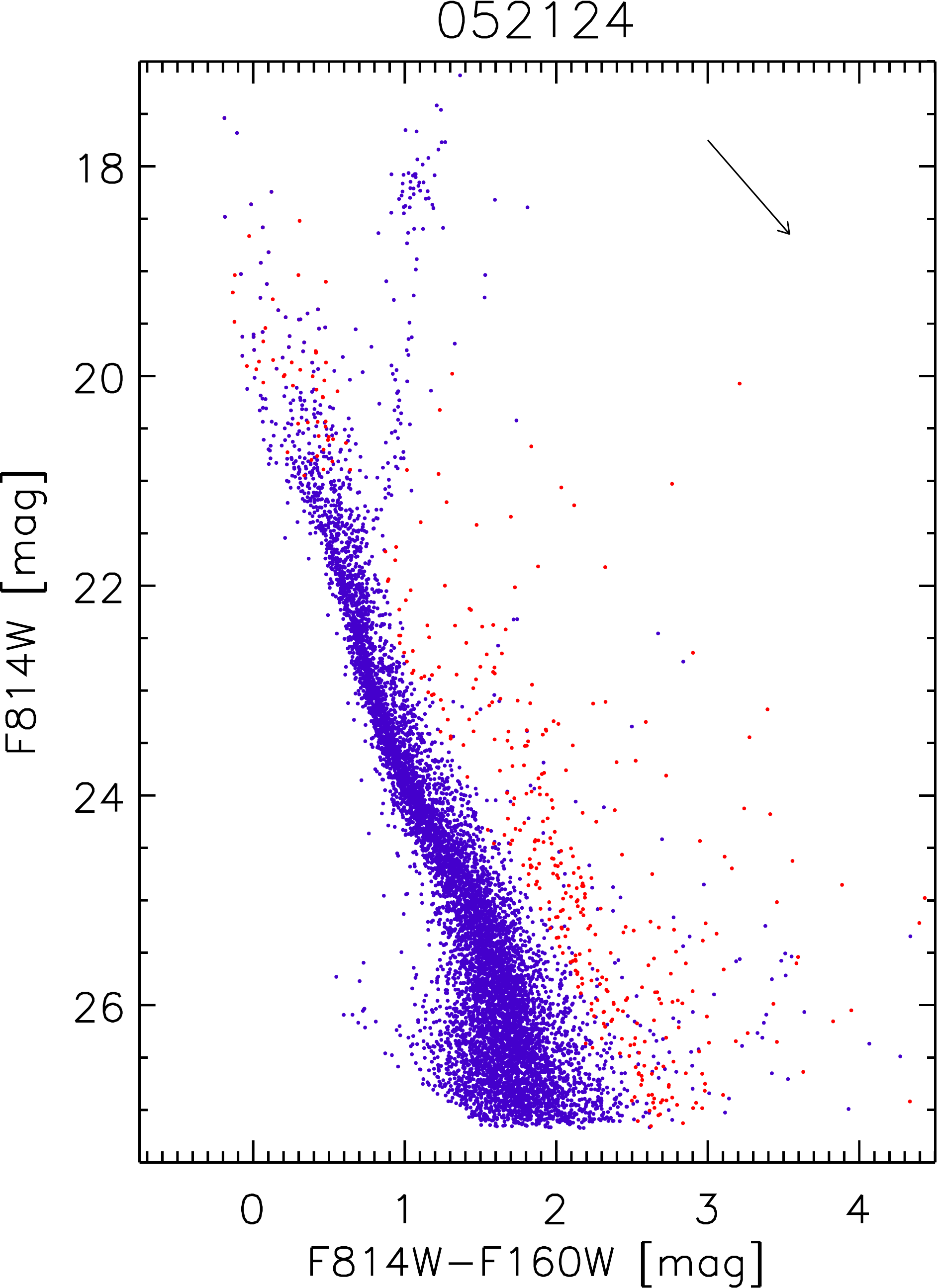}
\includegraphics[clip=true, trim =  0 0 -0.5cm 0, width=0.247\textwidth]{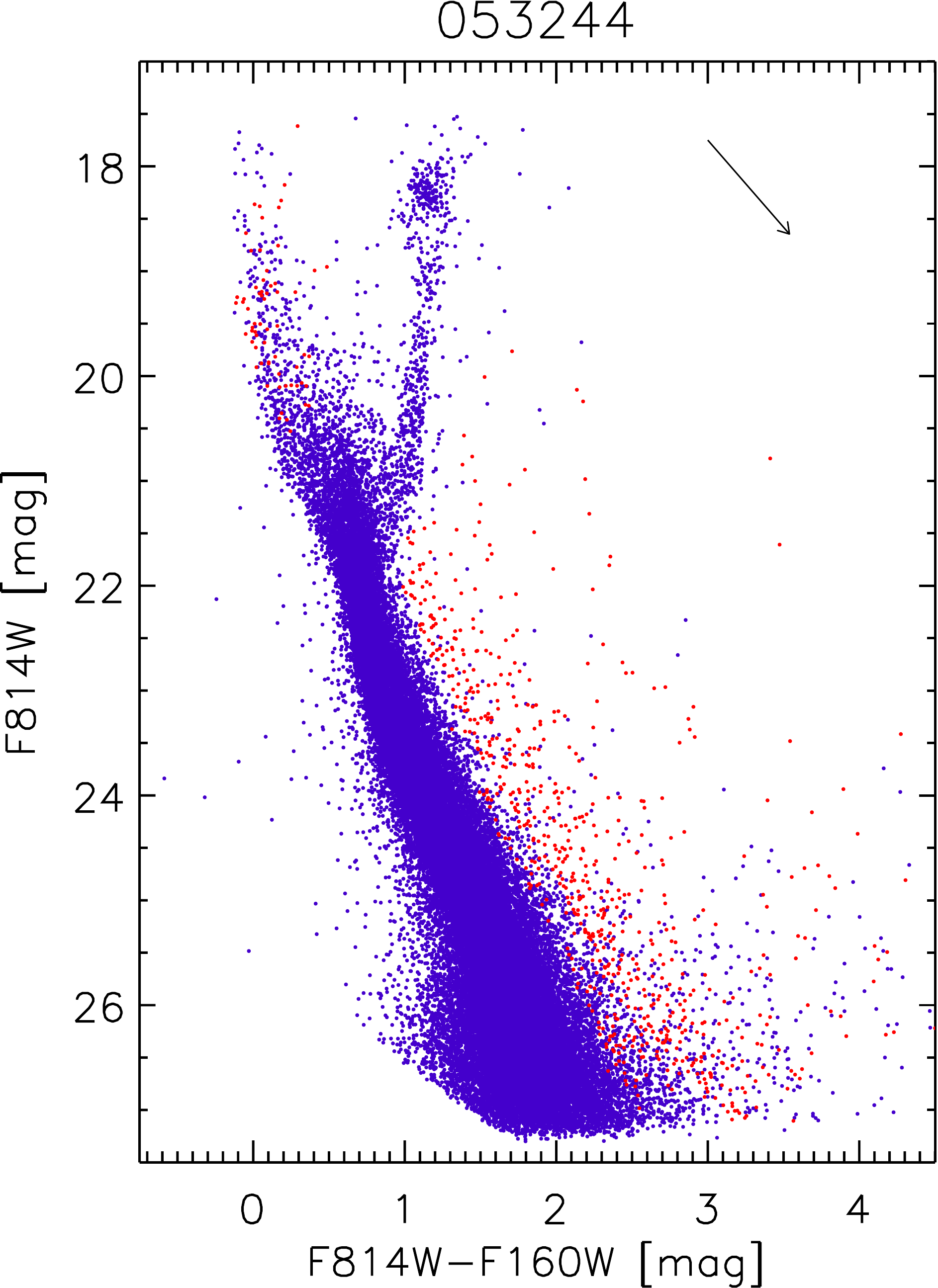}
\includegraphics[clip=true, trim =  0 0 -0.5cm 0, width=0.247\textwidth]{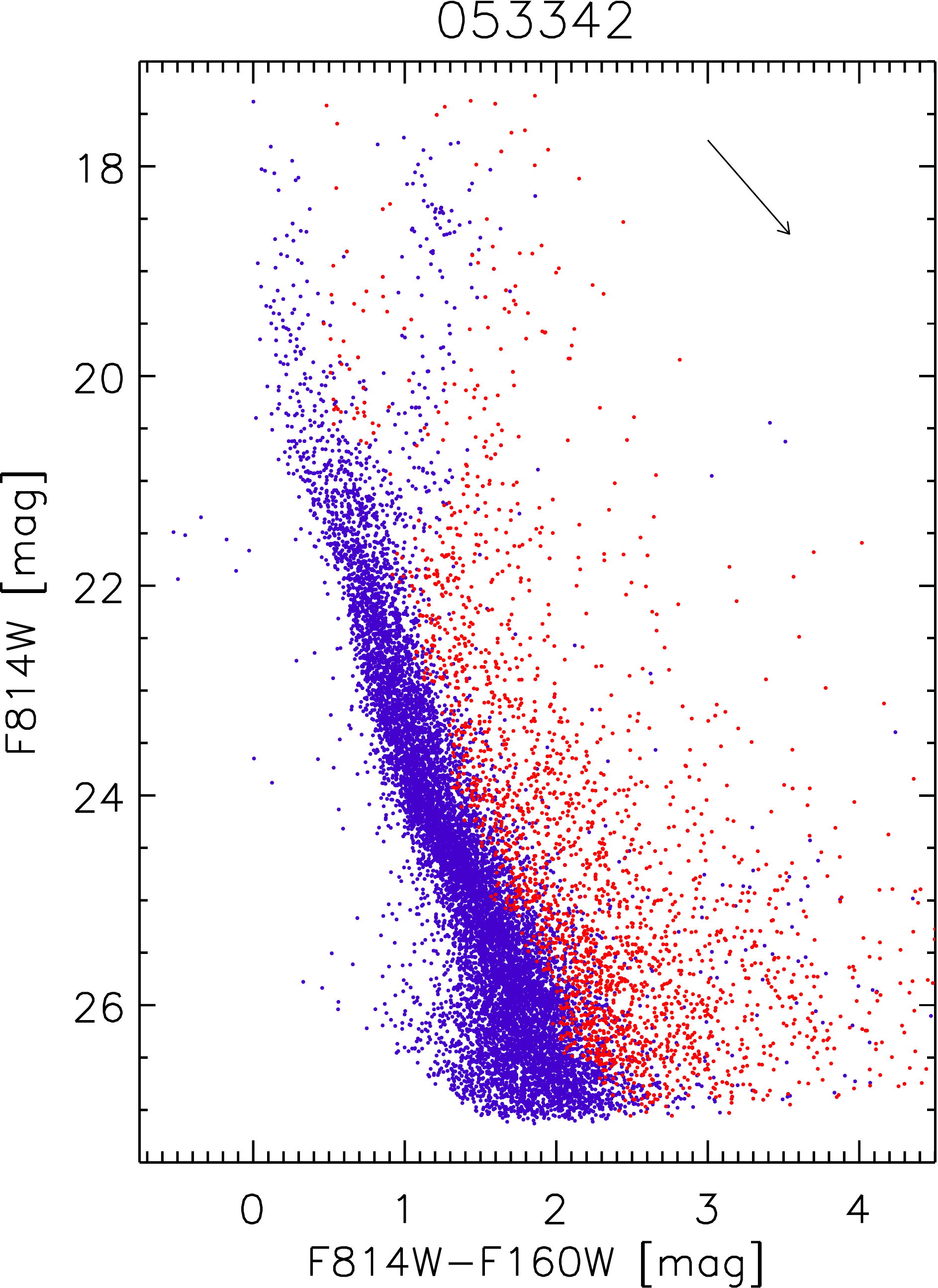}
\includegraphics[clip=true, trim =  0 0 -0.5cm 0, width=0.247\textwidth]{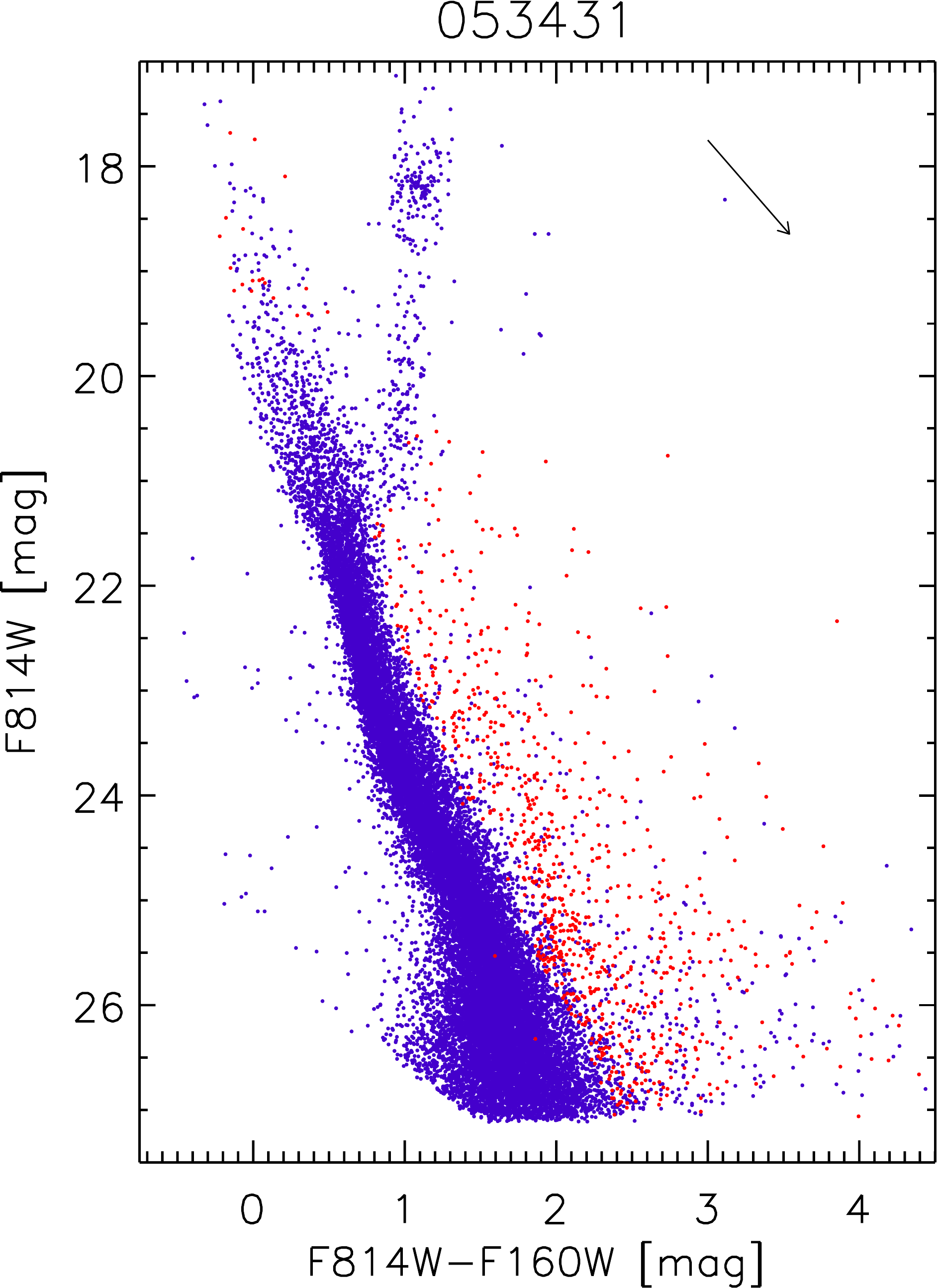}
%}
\end{center}
\caption{ Color-magnitude diagrams for the seven MYSOs using the F814W and F160W filters of the entire clean photometric sample. Typical stellar populations of the nearby LMC field are plotted with blue symbols. The young PMS stellar sources of each region, determined by statistically decontaminating the complete observed CMDs from the field contribution, are plotted in red. They represent the recent star formation events for each region. An indicative reddening vector for $A_{V} = 2$ mag is shown in the CMDs only to demonstrate the effect of extinction \citep{Fitzpatrick1999}; the length of the vector does not correspond to the actual interstellar extinction in the regions. }
\label{f:fs.cmds}
\end{figure*}
%%%%%%%%%%%%%%%%%%%%%%%%%%%%%%%%%%%%%%%%%%%%%%

%\clearpage 
%%%%%%%%%%%%%%%%%%%%%%%%%%%% FIGURE %%%%%%%%%%%%%
\begin{figure*}[t!]
\begin{center}
%\centerline{
\includegraphics[clip=true, trim =  0 0 -0.5cm 0, width=0.247\textwidth]{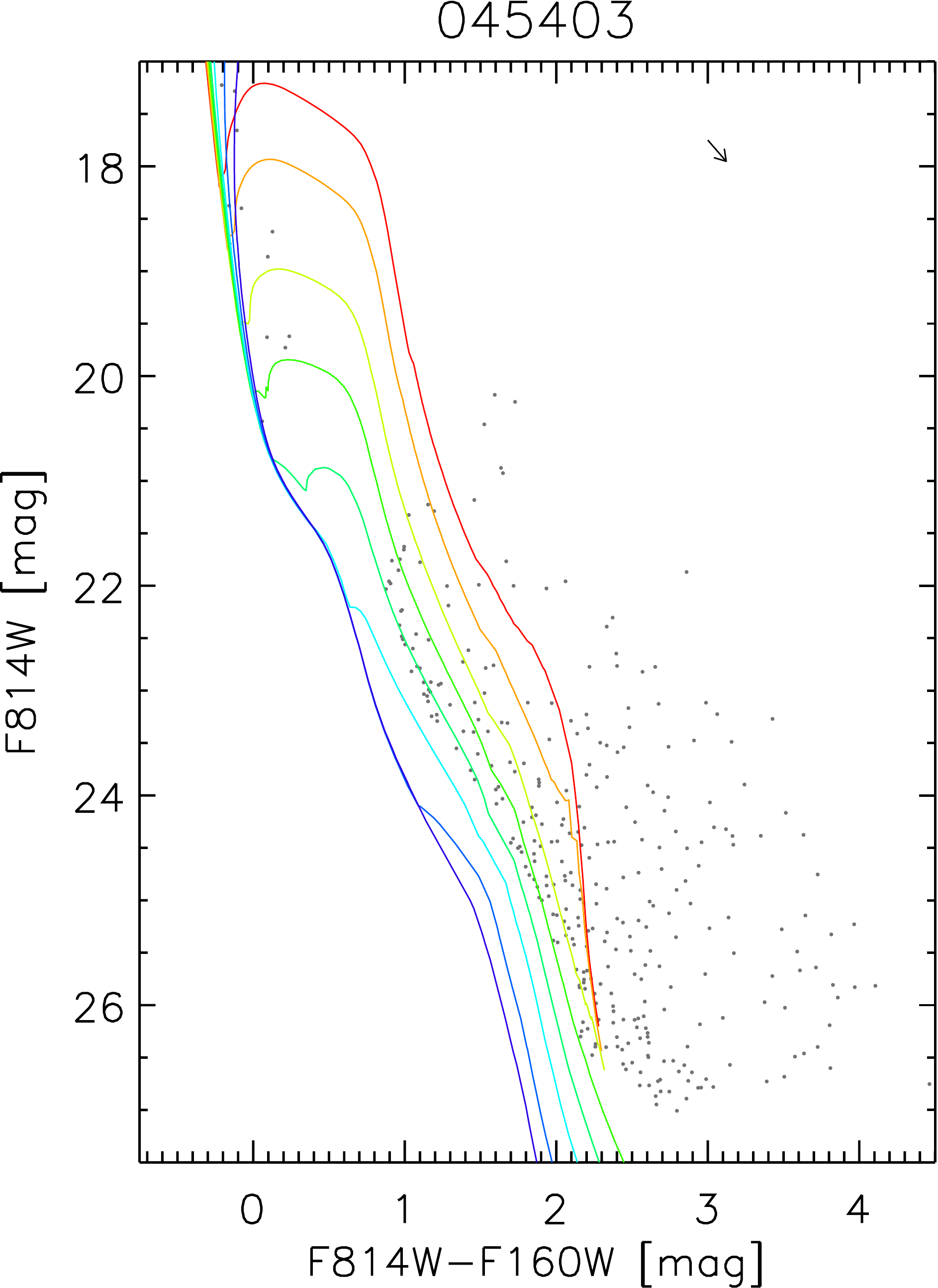}
\includegraphics[clip=true, trim =  0 0 -0.5cm 0, width=0.247\textwidth]{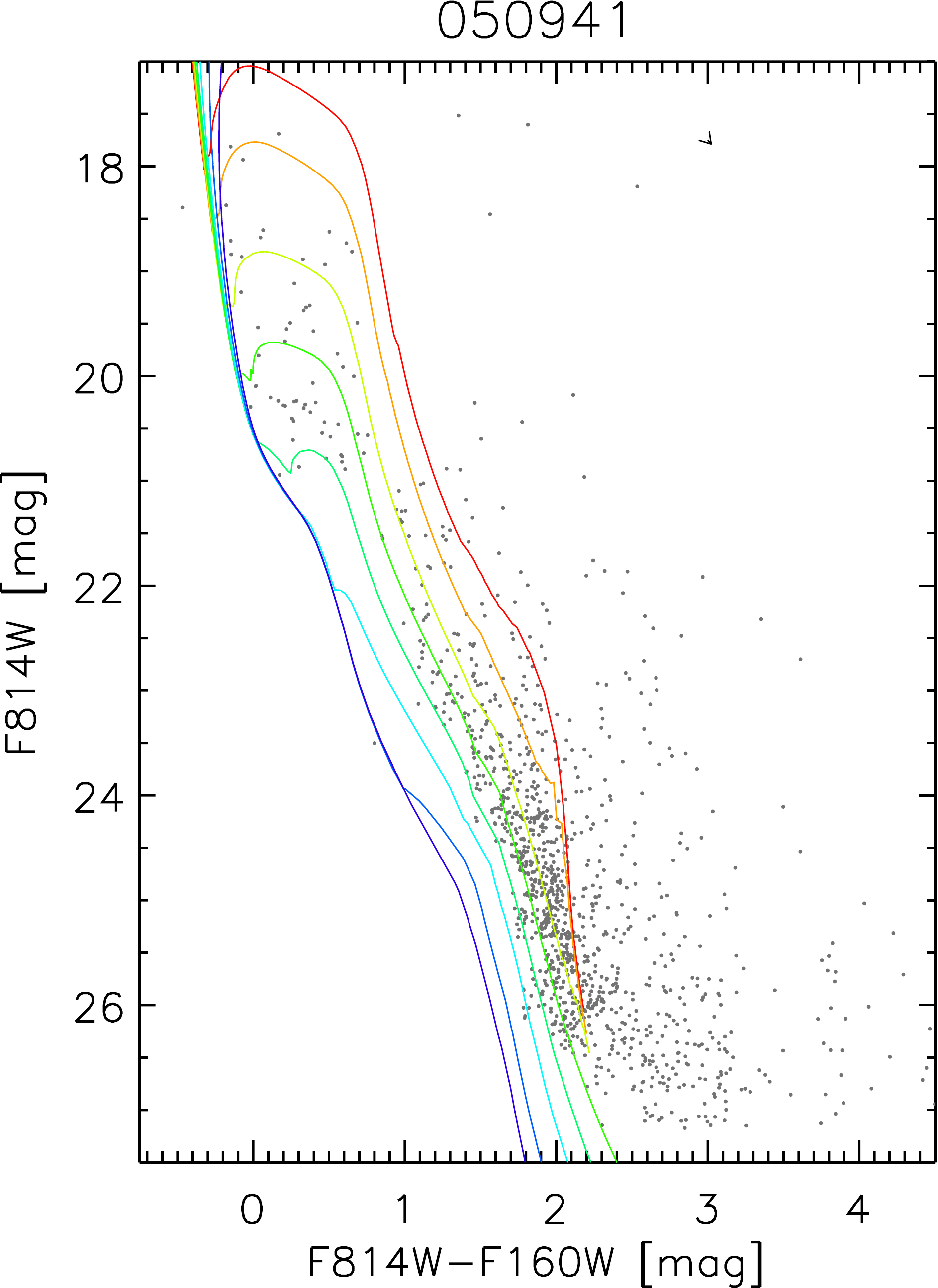}
\includegraphics[clip=true, trim =  0 0 -0.5cm 0, width=0.247\textwidth]{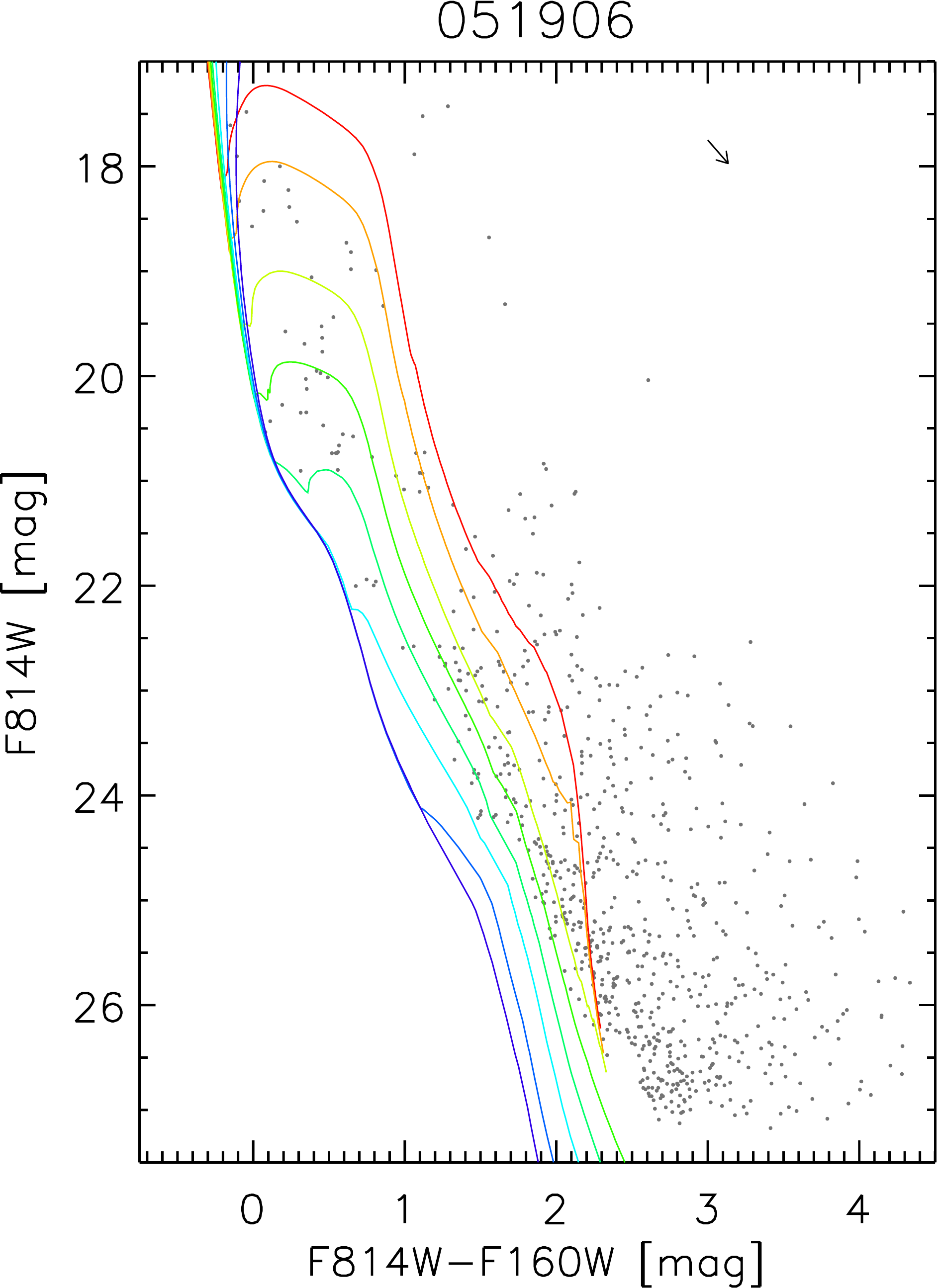}
\includegraphics[clip=true, trim =  0 0 -0.5cm 0, width=0.247\textwidth]{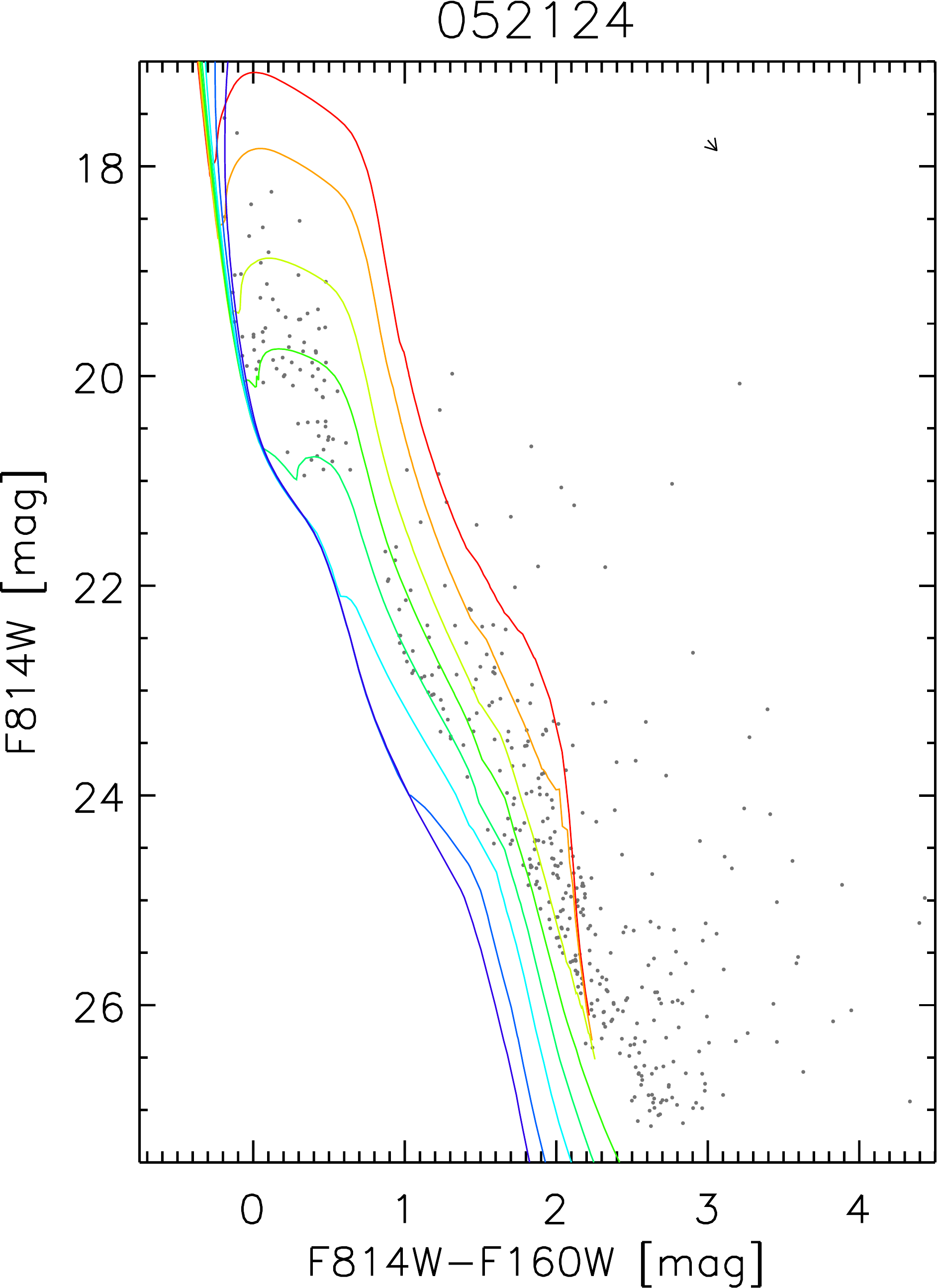}
\includegraphics[clip=true, trim =  0 0 -0.5cm 0, width=0.247\textwidth]{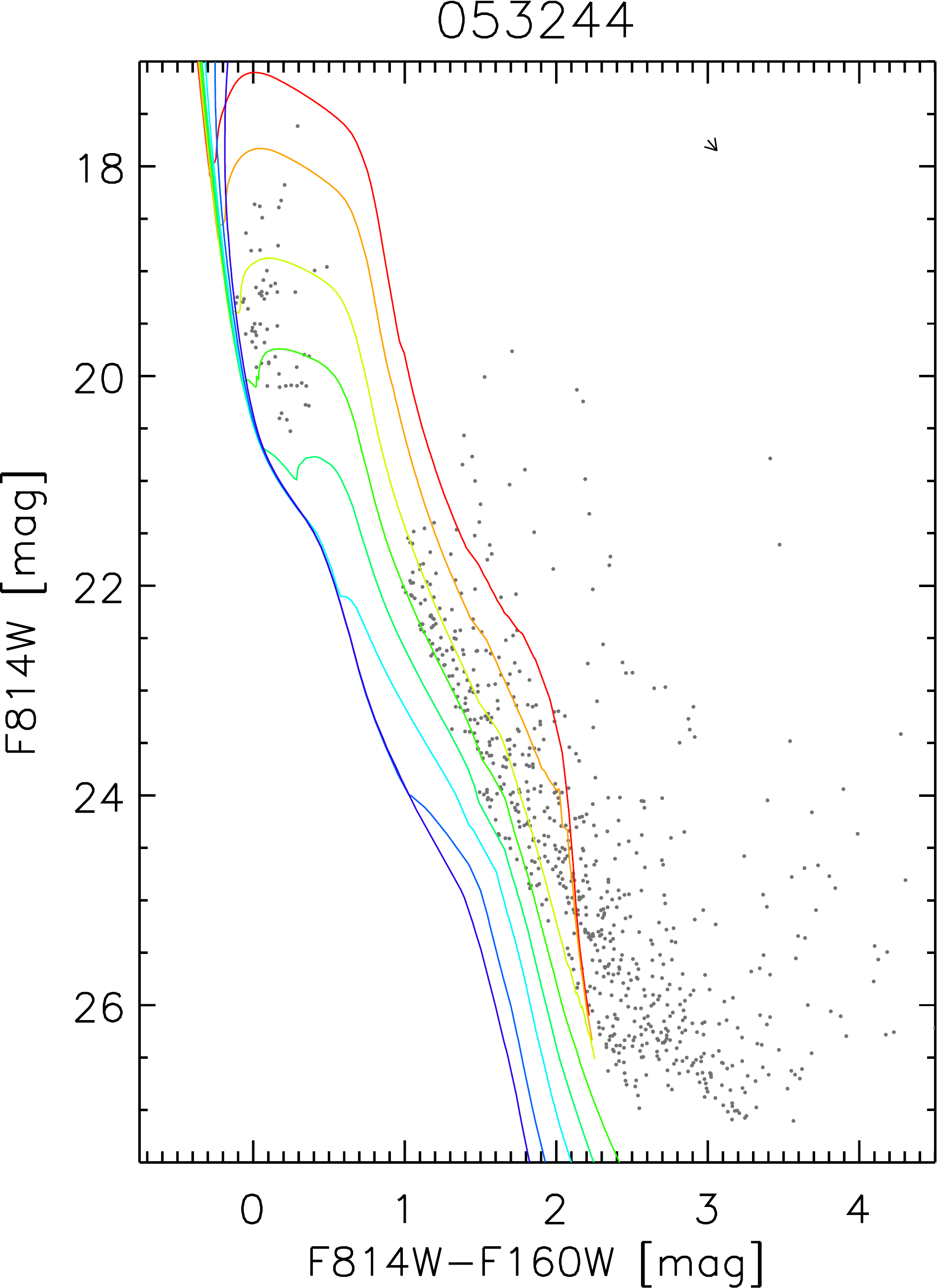}
\includegraphics[clip=true, trim =  0 0 -0.5cm 0, width=0.247\textwidth]{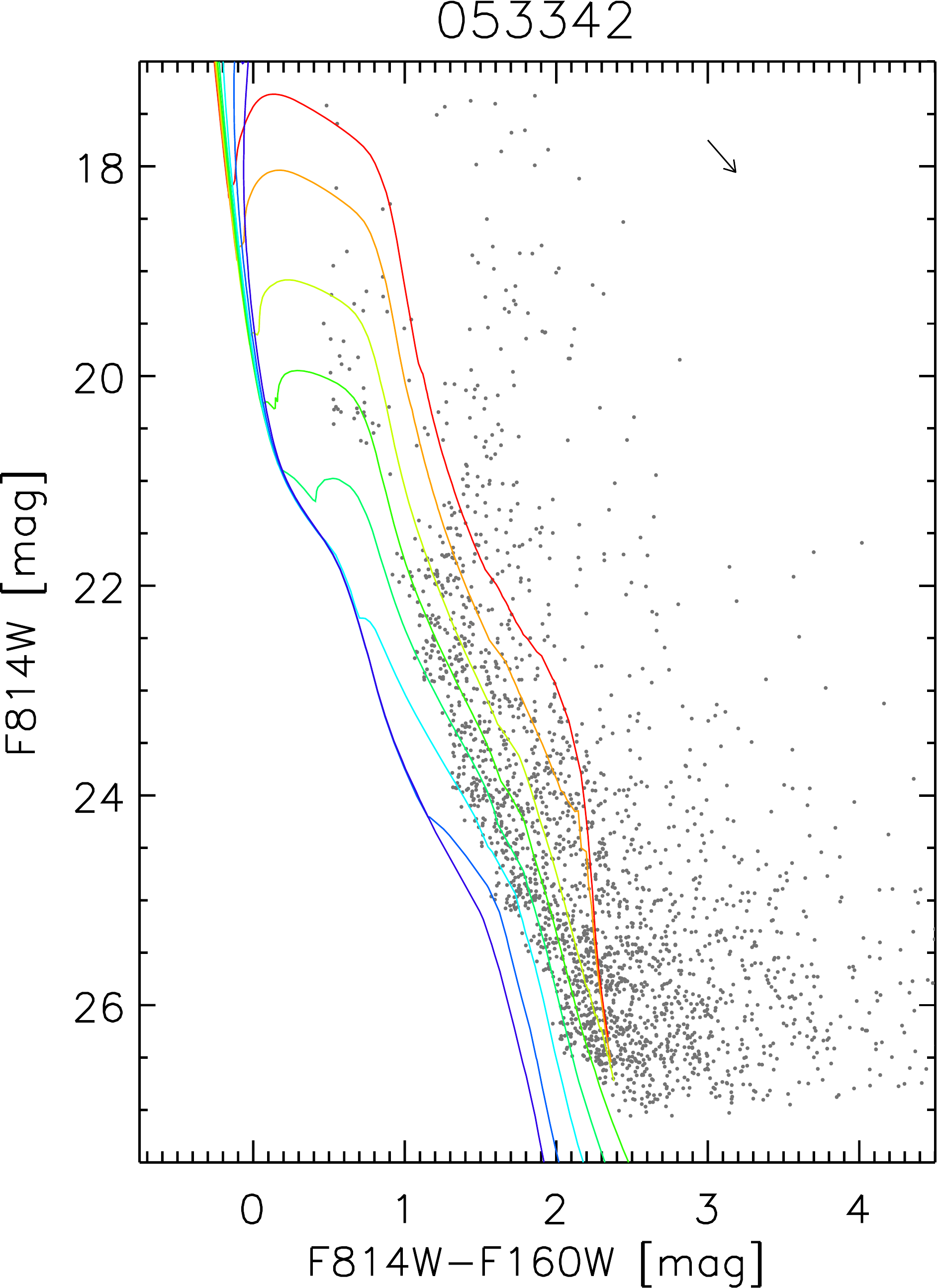}
\includegraphics[clip=true, trim =  0 0 -0.5cm 0, width=0.247\textwidth]{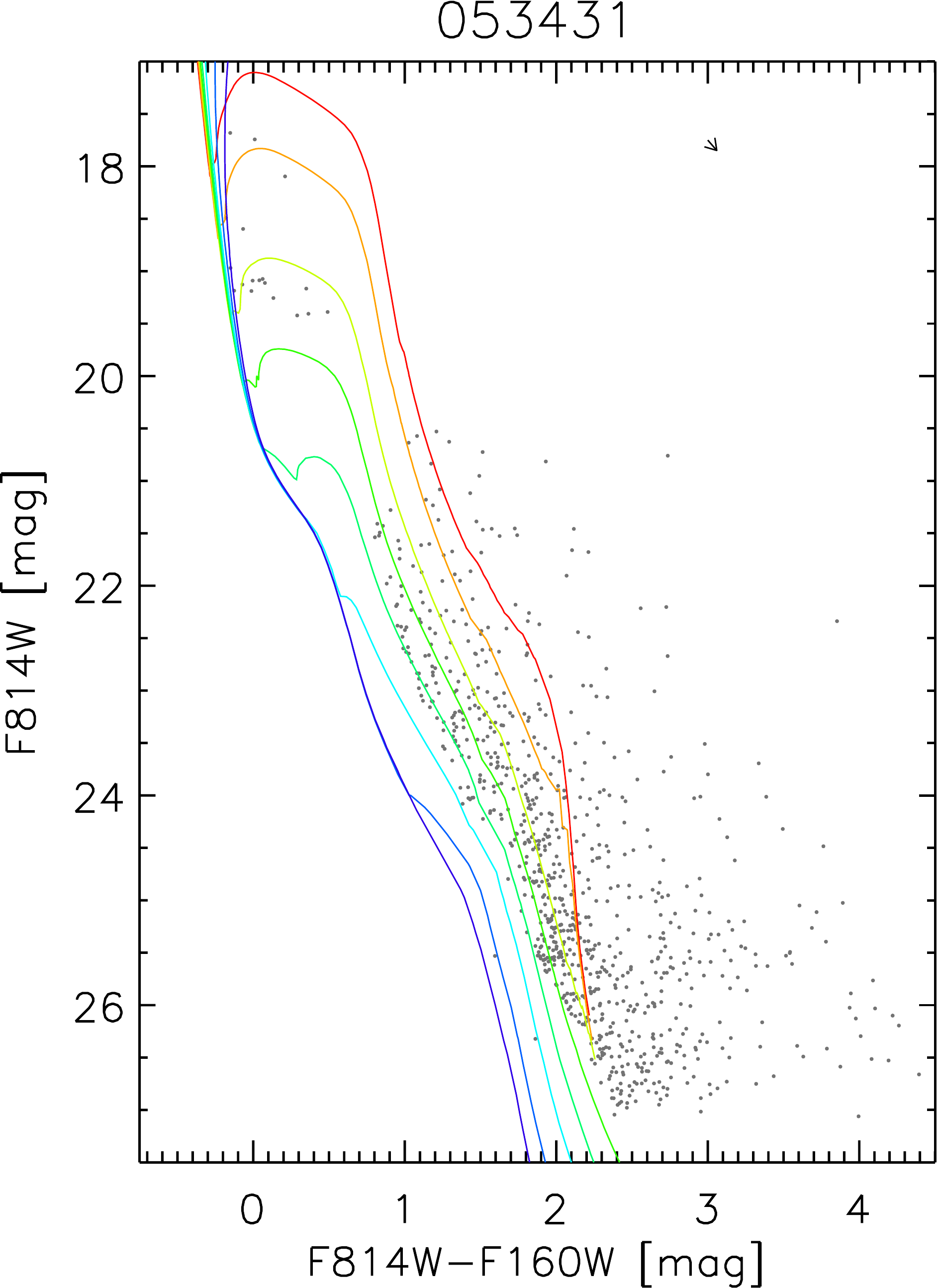}

%}
\end{center}
\caption{F814W, F160W color-magnitude diagrams of the most probable young PMS stellar populations around the seven MYSOs, as determined by statistically decontaminating the complete CMDs from the field contribution. Isochrones from the Padova family of models are shown corresponding to ages 0.5 (red), 1 (orange), 2.5 (yellow-green), 5 and 10 (green), 20 (cyan), 50 (blue), and 100~Myr (violet). While the masses probed by the isochrone strongly depends on the age, the typical isochrone shows a stellar mass range between $\sim$0.3 and 7\,$M_\odot$. Reddening vectors (top-right) represent the minimum correction applied to the evolutionary models in order to fit the observations. This correction corresponds to an $A_{V}$ of 0.50, 0.10, 0.55, 0.35, 0.50, 0.75, and 0.30 mag for \sone, \stwo, \sthree, \sfour, \sfive, \ssix, and \sseven, respectively. These corrections were determined so that the models fit the blue part of the main-sequence and red-giant branch of the {\em complete} CMDs of the regions. They represent thus the {\em minimum} $A_V$ in every stellar sample.}
\label{f:s.cmds}
\end{figure*}
%%%%%%%%%%%%%%%%%%%%%%%%%%%%%%%%%%%%%%%%%%%%%%

\section{PMS stars in the Observed Regions}\label{identPMS}

\subsection{Identification of the Pre--Main-Sequence Populations}\label{PMSstars}

We investigate the young stellar populations detected with our observations within the WFC3/UVIS field-of-view of the regions around the MYSOs. From the clean photometry (see Section\,\ref{s:obsphot}), we use the CMDs of the stars detected in the filters F814W and F160W, equivalent to standard $I$ and $H$ photometric bands. The choice of these bands is based on the fact that any young currently-forming stellar population will be visible at longer wavelengths. Stellar photometric measurements using F814W are not affected by diffuse hydrogen emission as in the F555W ($\sim$\,$V$) band. In addition, the F160W ($\sim$\,$H$) filter is less sensitive to extinction and thus allows for the selection of more embedded sources than the F110W ($\sim$\,$J$) filter. 
In the CMDs of the observed regions around our target MYSOs, shown in Figure\,\ref{f:fs.cmds}, it is seen that the observed fields cover a large variety of stellar populations. The observed stellar samples comprise the evolved populations of the surrounding LMC field designated by the prominent red giant branch and low--main-sequence features of the CMDs. The majority of the latter populations are essentially faint objects still in their PMS evolutionary stage, i.e., they have not started their lives on the main-sequence yet. They are located at the red part of the observed CMDs, almost parallel to the low--main-sequence. 

In order to identify these PMS stars and distinguish them from the evolved main-sequence stars of the nearby LMC field, we decontaminate the observed CMDs from the contribution of the local LMC field with the application of a statistical field-subtraction technique based on the Monte Carlo method.  Specifically, we construct the CMD of the most empty area in each observed field, which we consider to be the best-representative of the local LMC field population (we refer to it as the {\sl field CMD}). We then consider an circular subregion on the CMD around every star in the total CMD, and we subtract from the stars included in this region the corresponding number of randomly selected stars that belong in the same CMD-subregion of the field CMD. Since the area selected for the field CMD is only a portion of the complete observed area, the number 
of expected field stars in every CMD-subregion was scaled according to the fraction of the surface of the total area over that of the field area. We construct thus the `clean' CMD of each observed field, which contains only the most probable PMS stars in the region.

After statistically subtracting the field stars from the CMD of every area, each of the remaining red sources was visually inspected in the F814W and F160W images to ensure that they indeed correspond to real stellar sources in at least one of the filters. Note that while the source might not be visually confirmed as stellar in one filter, the {\sc dolphot} algorithm may still be able to fit an accurate PSF to the source; as discussed in Section \ref{s:obsphot}, the analyzed photometry all have a {\tt signal-to-noise$\,>$\,5}. The visual inspection was primarily used to remove sources that confused the {\sc dolphot} algorithm due to diffuse emission and bright halos and spikes emanating from bright stars. Moreover, this visual inspection removed a few sources that were obvious galaxies (e.g., extended sources with some structure) and artifacts in the observations. For \sone, we removed all photometric sources lying within the ``blue" glow of the diffraction spike seen in Figure\,\ref{045403_2panel.png} since the photometry here was found to be unreliable. In the CMDs of Figure\,\ref{f:fs.cmds}, field stars are plotted in blue and the {most probable} PMS stellar populations (derived from our field-subtraction technique and visual inspection) are shown in red. A small fraction of the red PMS stellar sample is expected to be still contaminated by some main sequence stars, but to a very small degree. Therefore we treat all these sources as true PMS stars. In Section\,\ref{s:clusanl} we investigate the clustering behavior of these stellar populations in the surroundings of our MYSOs.

Typically, the positions of the PMS stars in the CMD do not overlap with those of the main-sequence stars, and therefore it is quite straightforward for our field-subtraction  technique to eliminate completely features that are typical of old populations from the original CMD \citep[e.g.,][]{gouliermisetal11, gouliermisetal12}. However, our field decontamination method is not optimized for regions of high differential extinction because evolved stars in such regions strongly contaminate the CMD positions of the PMS stars due to reddening. This contamination also `hides' the main-sequence {\sl turn-on}, i.e., the position in the CMD where the PMS stars ignite hydrogen and reach the main-sequence, and which thus determines the youngest age of the PMS populations. As seen in the CMDs of Figure\,\ref{f:fs.cmds}, this issue is quite prominent in the case of the observed field around MYSO\,053342, where the main-sequence {\sl turn-off} contaminates strongly the {\sl turn-on} and therefore PMS stars are not easily distinguishable from the old field stars. While this method is not optimized for differential extinction across the field, it is sufficient at identifying young clusters in the field, which we discuss in more detail in the following sections.

%\clearpage 

\subsection{PMS Color-Magnitude Diagrams}\label{s:pmss}

The CMDs of the PMS stellar sources remaining after field-subtraction and visual inspection in the regions around each of the considered MYSOs are shown in Figure\,\ref{f:s.cmds}. In these CMDs, stellar evolutionary models for various ages are also plotted. These isochrones are taken from the {\sl Padova} grid of models \citep{Bressan2012, Chen2014, Tang2014} and range from 0.5 to 100\,Myr. They are used for guidance on the evolutionary stage of the observed PMS stars in the CMDs. Isochrones younger than $\sim$\,5\,Myr are generally considered not as well-determined as the older ones. These models also cover the PMS evolutionary phase for which they are qualitatively indistinguishable from the {\sl Pisa} family of PMS models \citep[FRANEC;][]{Tognelli2011} for the LMC metallicity ($Z=0.008$). While these isochrones provide an approximation of the age and age-spread of PMS stars in the ensembles, they cannot be used at face-value due to several physical characteristics of these PMS stars. In particular, a large fraction of these PMS stars are T\,Tauri-type stars, which are often dislocated from their theoretical CMD-positions due to, for example, rotational variability, accretion excess, and unresolved binarity \citep{Gouliermis2012, Jeffries2012, Preibisch2012}. Evolutionary models are also known to be inconsistent with each other \citep{Hillenbrand2008} at such a degree so that the choice of the appropriate grid of models practically depends on the specific dataset. %As a consequence we use the shown models tentatively in order to (a) determine an age-limit for the PMS populations, (b) assess any evolutionary difference between the young populations in different regions and (c) distinguish PMS populations from residual contaminating evolved field stars.

%The CMDs of both areas encompassing 050941 and 051906 show a clear sequence of PMS stars with ages younger than 5~Myr. 
The CMDs of the areas encompassing these YSOs typically have PMS stars with ages younger than $\sim$ 5~Myr. This age-limit is more prominent for the fainter stars, while the brighter PMS stars and those of the turn-on are shown to also fit ages of up to $\sim$10~Myr. Therefore, both regions host PMS stars at very similar evolutionary stages with ages younger than 10~Myr. However, as mentioned above no actual age can be assigned by a simple fit on the CMD. On the Padova isochrones shown in Figure\,\ref{f:s.cmds}, we applied extinction corrections on the basis of their fit to the blue part of the upper main--sequence (above the turn-off) of the complete CMDs of Figure\,\ref{f:fs.cmds}. The extinction corrections are indicated in the figure caption of Figure\,\ref{f:s.cmds}, with $A_V$ varying from 0.10 to 0.75\,mag. These measurements are based on the extinction law by \citet{Schlafly2011} for a coefficient $R_V = 2.1$, which was found to fit best the two-color diagrams of the populations. Based on these isochrones, we also give the $A_V$ values for the central MYSOs in Table\,\ref{t:bright}.

Of the seven observed MYSO fields, the region encompassing 053342 is extincted the most, experiencing a strong differential (i.e.,  spatially-variable) reddening. This is shown in the complete CMD of Figure\,\ref{f:s.cmds}, where it can be seen that our selection of the young PMS populations was not entirely successful, including several evolved highly extincted giants seen in the bright-red part of the CMD. The isochrones plotted on this CMD are corrected for a minimum extinction of $A_V \simeq 0.75$\,mag, determined so that the upper--main-sequence fits the blue part of the total observed CMD. As a consequence, the bright main-sequence stars seen on the right of the models in Figure\,\ref{f:s.cmds} are not poor fits, but highly reddened main-sequence stars, corresponding to a maximum reddening of $A_V \sim 3.25$\,mag. We can assume that the PMS stars in the observed area also suffer from the same differential extinction and therefore their CMD-positions are shifted in a variable manner, in addition to their ``intrinsic'' dislocation due to their characteristics, as discussed above. Consequently, it is quite difficult to separate the complete PMS population in the star-forming region around MYSO\,053342 from its surrounding field population. Nevertheless, for the purpose of this paper we are only interested in clusters surrounding the MYSOs, in the area of which we expect a minimal contamination by MS stars.

It should also be noted that while we cannot evaluate accurately the age of these clusters, for the purposes of estimating their masses (Section\,\ref{s:clusterest}) we assume ages of $\sim$\,1\,Myr and 2.5\,Myr. While the ages are uncertain, based on the isochrones these values represent the best estimate of the ages of the PMS stars within the field.
%This is a reasonable assumption, considering that the isochrones indicate that the PMS stars are generally younger than $\sim$\,5\,Myr.

%\appendix %\label{s:appendix}
%\begin{center}
%SPECTRAL TYPE CALCULATION
%\end{center}
\subsection{Spectral Types of the MYSOs}\label{s:spclass}

%%%%%%%%%%%%%%%%%%%%%%%%%%%% FIGURE %%%%%%%%%%%%%
\begin{figure*}[t!]
\begin{center}
%\centerline{
\includegraphics[clip=true, trim =  0 0 -0.5cm 0, width=0.33\textwidth]{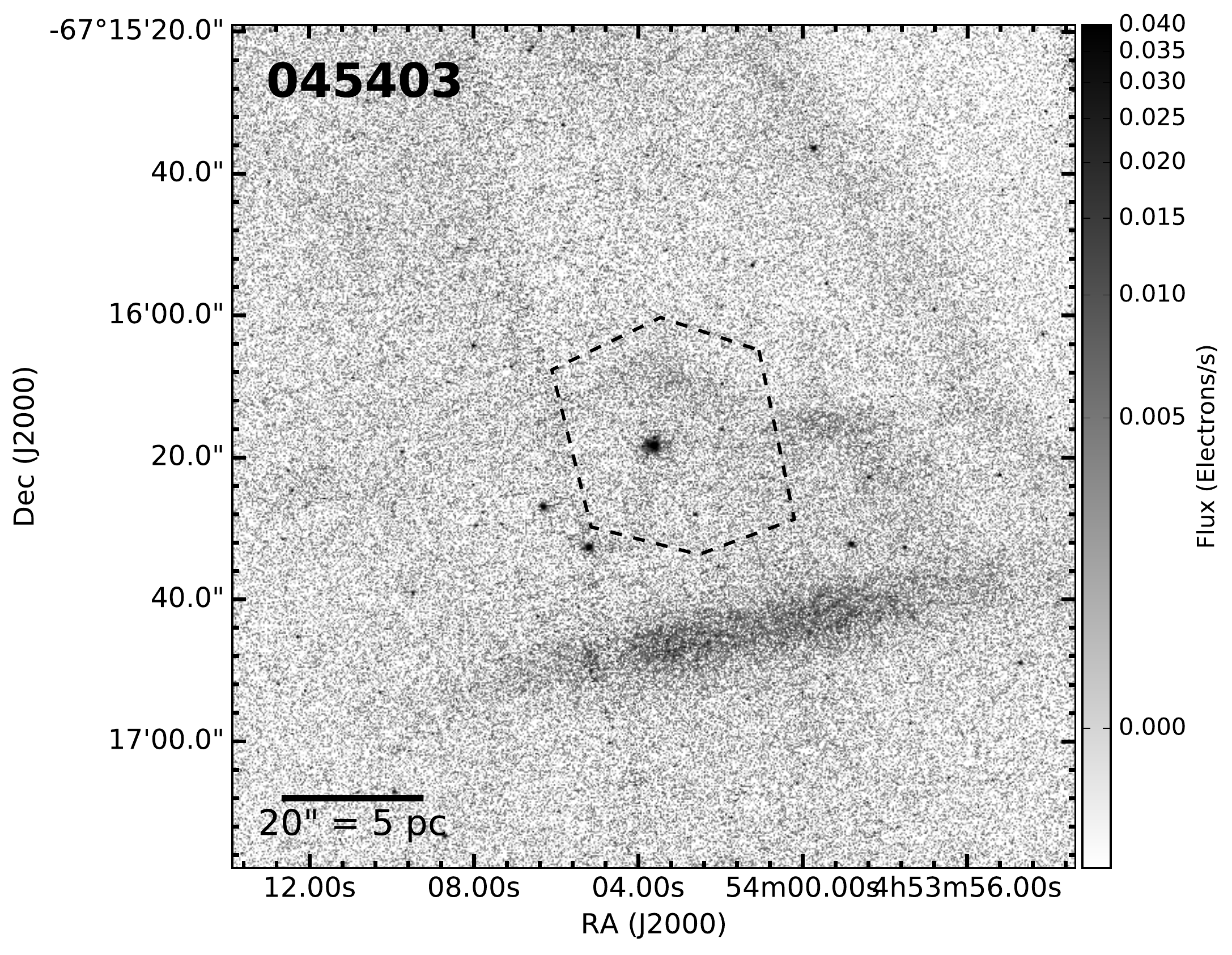}
\includegraphics[clip=true, trim =  0 0 -0.5cm 0, width=0.33\textwidth]{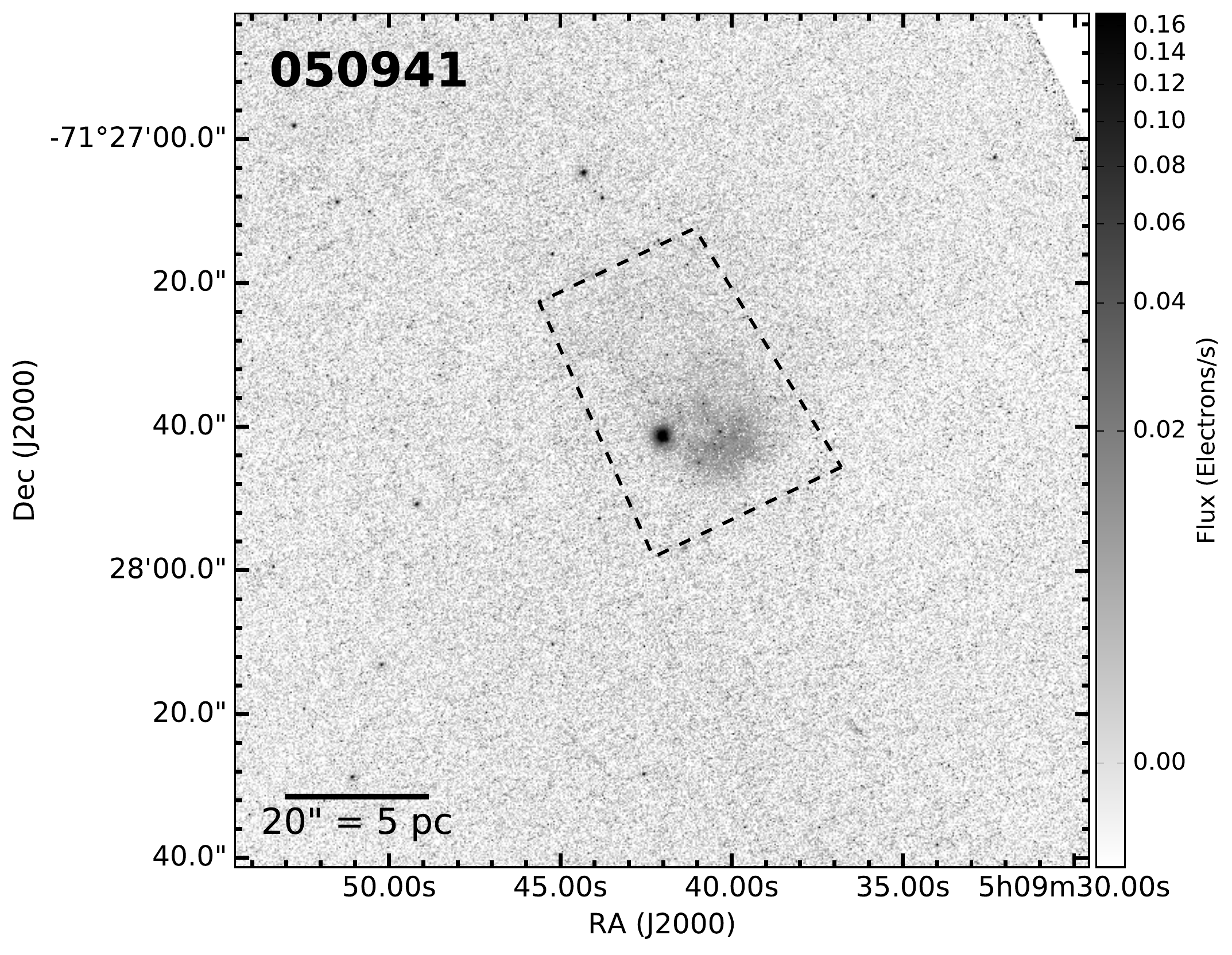}
\includegraphics[clip=true, trim =  0 0 -0.5cm 0, width=0.33\textwidth]{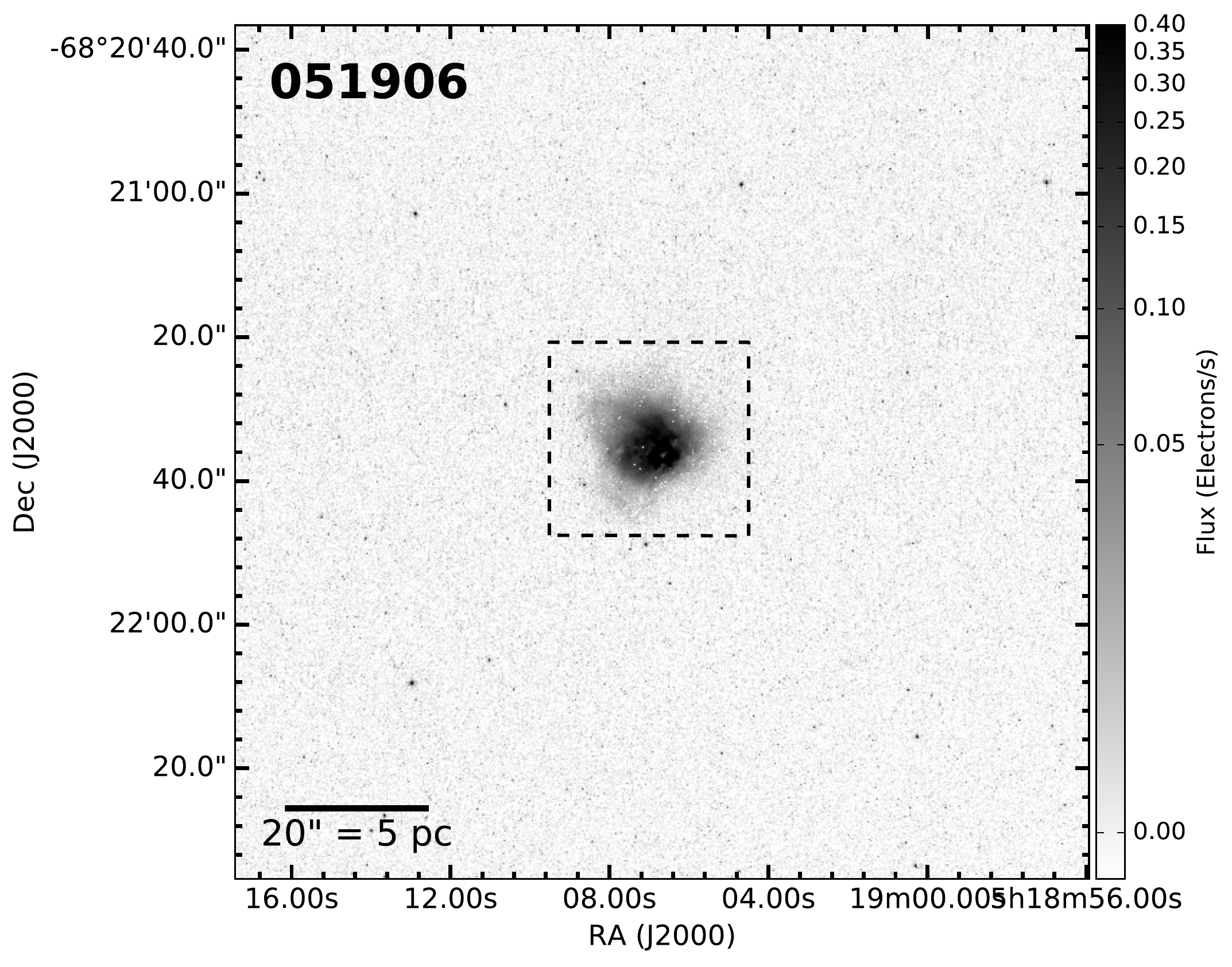}
\includegraphics[clip=true, trim =  0 0 -0.5cm 0, width=0.33\textwidth]{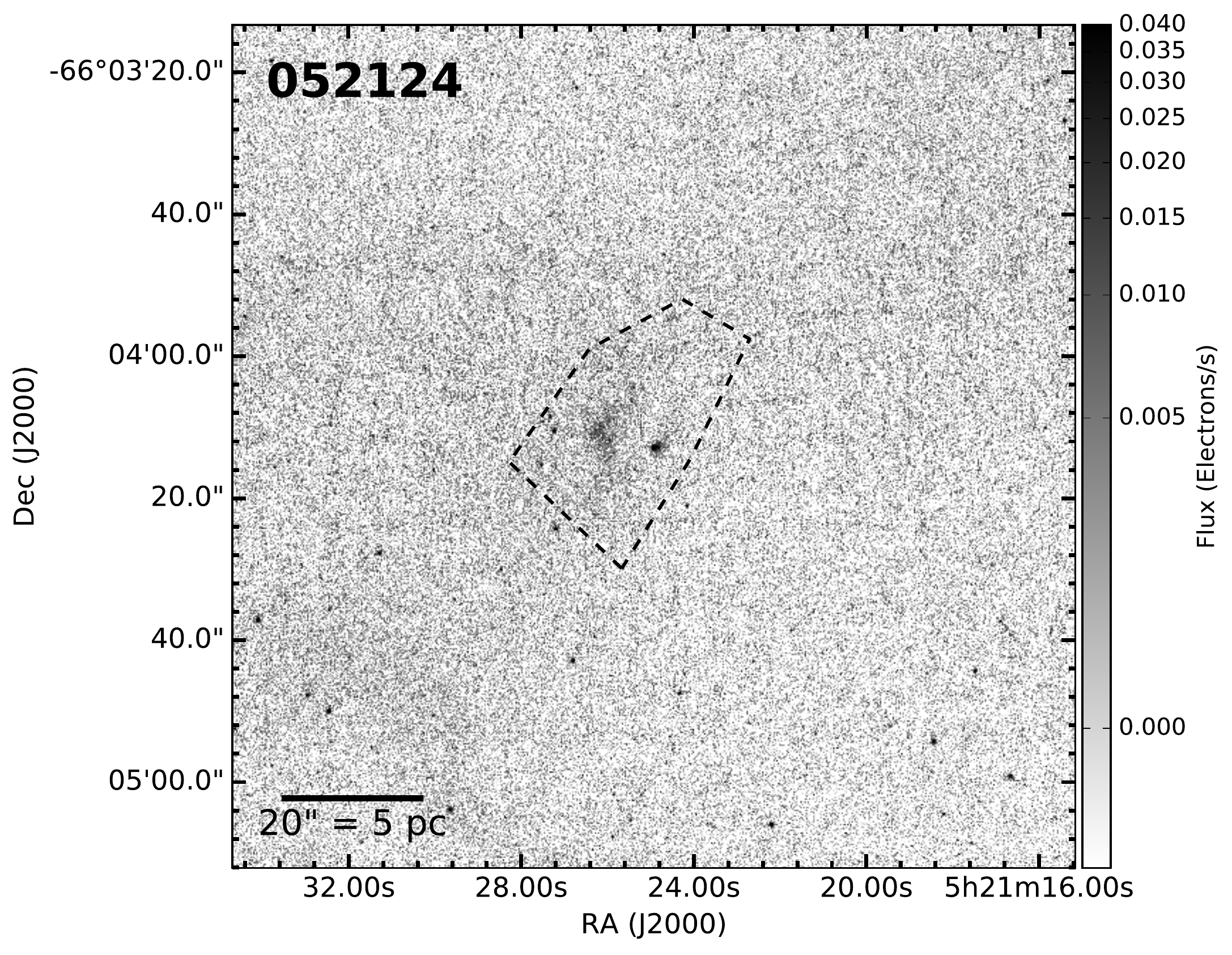}
\includegraphics[clip=true, trim =  0 0 -0.5cm 0, width=0.33\textwidth]{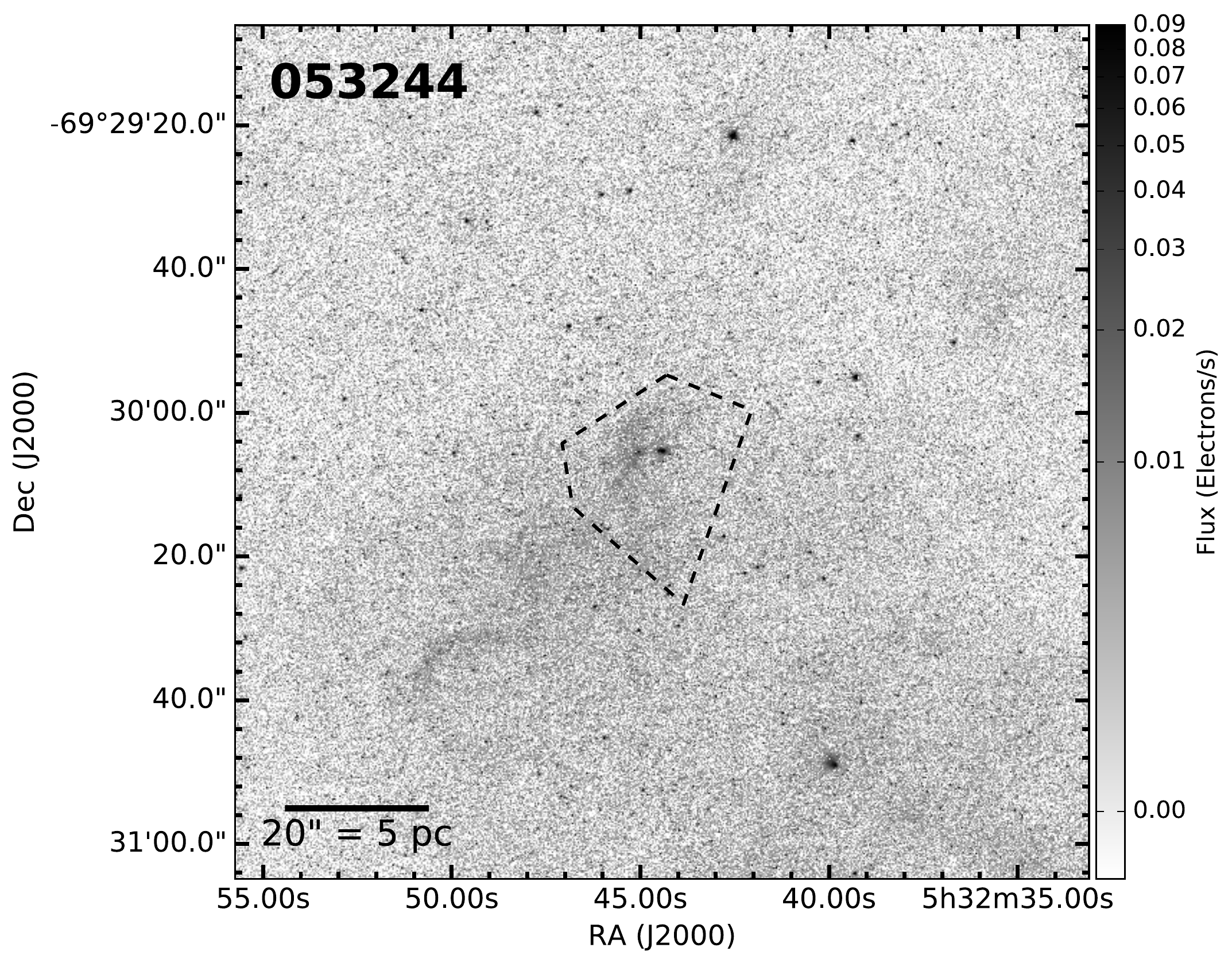}
\includegraphics[clip=true, trim =  0 0 -0.5cm 0, width=0.33\textwidth]{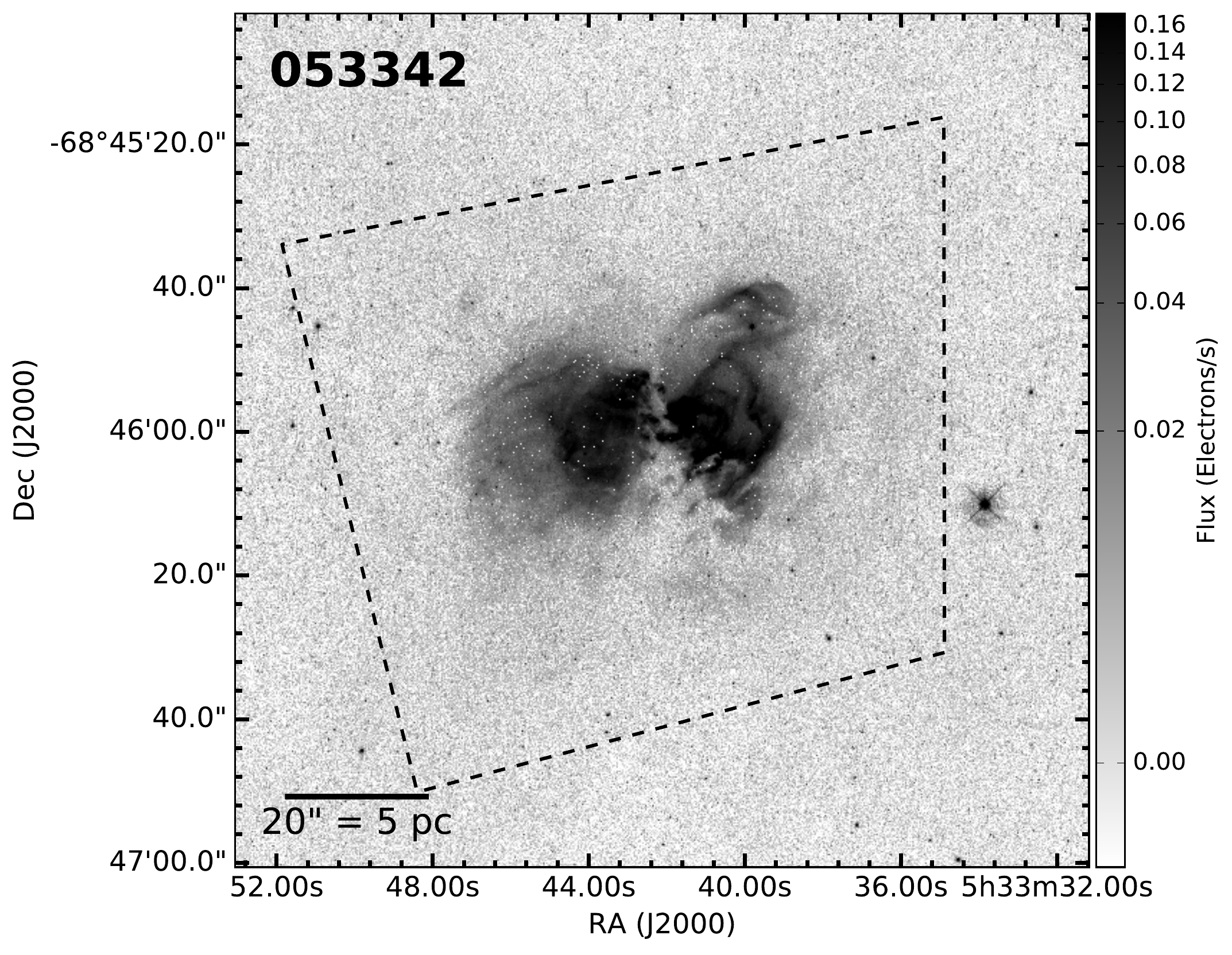}
\includegraphics[clip=true, trim =  0 0 -0.5cm 0, width=0.33\textwidth]{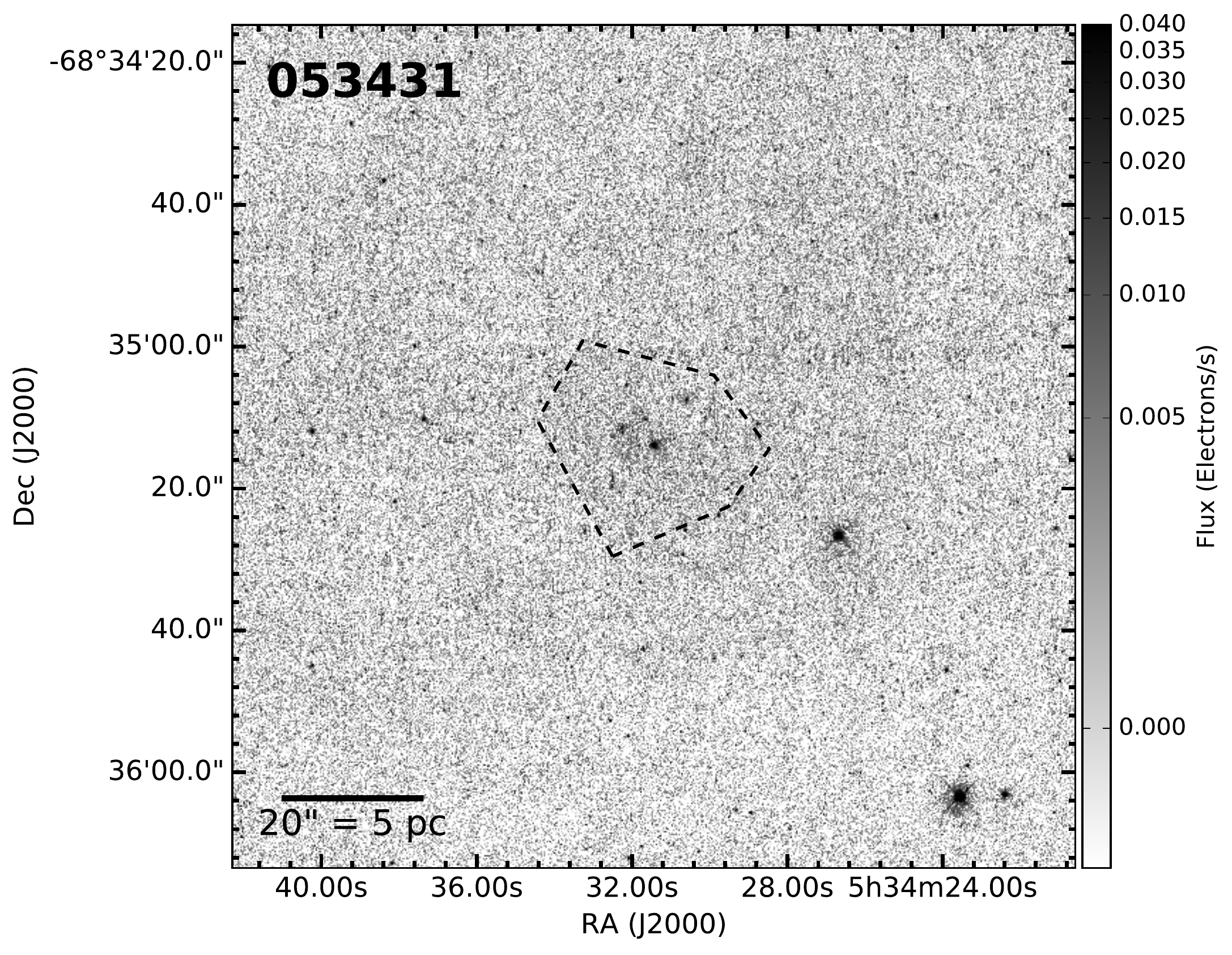}

%}
\end{center}
\caption{$HST$ F656N (H$\alpha$) observations encompassing all 7 MYSOs. The black polygon shows the area in which the flux was summed for determining the ionizing flux (Section \ref{s:spclass}). The images shown here have had their local foreground/background subtracted.}
\label{f:ha}
\end{figure*}
%%%%%%%%%%%%%%%%%%%%%%%%%%%%%%%%%%%%%%%%%%%%%%

We determined the approximate spectral type of each MYSO using its H$\alpha$ luminosity, which is based on the {\em HST} observations in the F656N band (Figure \ref{f:ha}). These images allow for a better determination of the emission associated with the MYSO than the Blanco 4~m telescope (Figure \ref{Halpha}) since $HST$ has superior pointing accuracy and sensitivity. For each F656N image, we draw a polygon around the emission that we judge to be the H$\alpha$ emission most likely associated with the MYSO. The process of picking such emission can be ambiguous since disentangling ionization from the MYSO, other nearby stars, and the general interstellar radiation field can be very difficult (e.g., MYSO 053244). However, we find that selecting slightly larger or smaller areas than the ones shown in Figure \ref{f:ha} will not change the spectral types determined below.

For each polygon shown in in Figure \ref{f:ha}, the positive H$\alpha$ pixels were summed to estimate the total flux. We subtracted the local foreground/background contribution from each of these positive pixels. We determined the local foreground/background value by selecting an empty area in the F656N map and calculated the mean pixel value in this area. For each source, the stellar sources were subtracted from the image in order to avoid stellar continuum contamination. For both \sthree\ and \ssix, the stellar subtraction was negligible compared to the total H$\alpha$ emission. However, the subtraction was more significant for the other sources.

For these \emph{HST} observations, the flux surrounding each YSO is given in Electrons\,s$^{-1}$. This was converted to a flux, $F_{{\rm H}\alpha}$ in erg\,s$^{-1}$\,cm$^{-2}$, by multiplying by the inverse sensitivity given in the WFC3 FITS header, PHOTFLAM ($1.632\times10^{-17}$ erg\,cm$^{-2}$\,\AA$^{-1}$\,Electron$^{-1}$), and the root-mean-square bandwidth of filter plus detector, PHOTBW (41.89\,\AA). The H$\alpha$ luminosity was then calculated via $L_{{\rm H}\alpha} = 4\pi F_{{\rm H}\alpha} D^2$, where $D$ is the distance to the LMC ($\sim$\,50\,kpc).

Spectral types can be determined from the hydrogen ionizing luminosity, $Q_0$. $L_{{\rm H}\alpha}$ and $Q_0$ can be related by 

\begin{equation}
L_{{\rm H}\alpha} = V n_p n_e \alpha_{{\rm eff},{\rm H}\alpha} E_{{\rm H}\alpha} 
\end{equation}
\begin{equation}
%Q_0 = \frac{4}{3}\pi r^3 n_p n_e \alpha_B
Q_0 = V n_p n_e \alpha_B ,
\end{equation}
where $V$ is the volume of the region, $n_p$ and $n_e$ are the proton and electron densities of the region, $E_{{\rm H}\alpha}$ is the energy of an $H\alpha$ photon, $\alpha_{{\rm eff},{\rm H}\alpha}$ is the H$\alpha$ recombination rate, and $\alpha_B$ is the case B hydrogen recombination rate (i.e., optically thick to ionizing radiation; excludes recombinations into the $n=1$ state). Assuming an electron temperature of $10^4$\,K \citep[e.g.,][]{MH05}, we adopt $\alpha_B = 2.59\times10^{-13}$\,cm$^3$\,s$^{-1}$ and $\alpha_{{\rm eff},{\rm H}\alpha} = 1.17\times10^{-13}$\,cm$^3$\,s$^{-1}$ \citep{Draine1992}. Therefore,
\begin{equation}
%Q_0 = L_{{\rm H}\alpha}\times \alpha_B/(\alpha_{{\rm eff},{\rm H}\alpha}\times E_{{\rm H}\alpha}) \approx 1.37\times10^{-12} L_{{\rm H}\alpha}~\rm{erg}\,\rm{s}^{-1}
Q_0 = L_{{\rm H}\alpha}\times \frac{\alpha_B}{\alpha_{{\rm eff},{\rm H}\alpha}\times E_{{\rm H}\alpha}} \approx 1.37\times10^{-12}~\rm{s}^{-1} \left(\frac{L_{{\rm H}\alpha}}{\rm{erg\,s}^{-1}}\right) .
\end{equation}
Note that no extinction correction was made, causing $Q_0$ to be underestimated.

%$L_{{\rm H}\alpha}$ is converted to the number of hydrogen ionizing photons, $Q_0$, via the formula $Q_0 = L_{{\rm H}\alpha}\times \alpha_B/(\alpha_{{\rm eff},{\rm H}\alpha}\times E_{{\rm H}\alpha})$. Assuming an electron temperature of $10^4$\,K \citep[e.g.,][]{MH05}, we adopt the case B hydrogen recombination rate (i.e., optically thick to ionizing radiation; excludes recombinations into the $n=1$ state) $\alpha_B = 2.59\times10^{-13}$\,cm$^3$\,s$^{-1}$ and an H$\alpha$ recombination rate $\alpha_{{\rm eff},{\rm H}\alpha} = 1.17\times10^{-13}$\,cm$^3$\,s$^{-1}$ \citep{Draine1992}. Using these numbers one gets the simple conversion $Q_0 = 1.37\times10^{-12} L_{{\rm H}\alpha}$. Note that no extinction correction was made, causing $Q_0$ to be underestimated.

In Table\,\ref{t:spclass} we match $Q_0$ to the approximate spectral type following the observational effective temperature scales of class\,V stars in \citet{Martins2005}. Note that metallicity does not have a major effect on the spectral type of these OB-stars, affecting the classification by no more than half a spectral type \citep{Smith2002}. Each spectral type also has a corresponding mass \citep{Martins2005} which we provide in the last column of Table\,\ref{t:spclass}. \citet{Martins2005} only estimated stellar parameters for O-stars, and \sone, \sfour, \sfive, and \sseven\ have calculated $Q_0$ values that are notably smaller than an O9.5V star (log~$Q_0 = 47.56$). Interpolating $Q_0$ to later spectral types is difficult since the \citet{Martins2005} model is not well fit by a simple functional form. \cite{Hanson1997}, which calculates very similar values for $Q_0$ as \citet{Martins2005}, suggests that a B0V star has log~$Q_0 = 47.18$, and thus we adopt this spectral type for these four sources. The \citet{Martins2005} spectral types as a function of mass fits well with an exponential function, and we interpolate a B0V star to have a mass of 14\,$M_\odot$.

%\cite{Martins2005} only estimated stellar parameters for O-stars, with \stwo's $Q_0$ value (log~$Q_0 = 46.9$) notably smaller than an O9.5V star (log~$Q_0 = 47.56$). Interpolating $Q_0$ to later spectral types is difficult since the model is not well fit by a simple functional form. \cite{Hanson1997}, which calculates very similar values for $Q_0$ as \cite{Martins2005}, shows a large drop in $Q_0$ between O9.5V and B0V spectral types, suggesting \stwo\ has a spectral type between a B0V and B0.5V. We adopt a spectral type of B0.5V, corresponding to a mass of 11~$M_\odot$ in \cite{Hanson1997}. %The flux, $Q_0$, and spectral types for the three MYSOs are shown in Table 3.

%calculations are in blanking_pixels.py
\begin{deluxetable}{ccccc}
\tablecaption{Determination of the Spectral Type of the MYSOs.\label{t:spclass}}
\tablewidth{0pc}
\tablehead{
\colhead{MYSO} & \colhead{Flux}  & \colhead{log($Q_0$)} & \colhead{Spectral} & \colhead{Mass} \\
\colhead{} &  \colhead{(10$^3$ Electrons\,s$^{-1}$)} & \colhead{(Photons s$^{-1}$)} & \colhead{Type} & \colhead{($M_\odot$)}
 }
\startdata
\sone & 1.1 & 47.2 & B0V & 14 \\
%\stwo  & 0.59$^*$ & 46.9 &  B0.5V & 11 \\
\stwo  & 4.7 & 47.8 &  O9.5V & 16 \\
\sthree  & 16 & 48.4 & O8V & 21 \\
\sfour & 1.1 & 47.2 & B0V & 14 \\
\sfive & 1.6 & 47.4 & B0V & 14 \\
\ssix & 57 & 48.9 & O6V & 31 \\
\sseven & 1.0 & 47.2 & B0V & 14
\enddata
%\small
%$^*$If we use a larger area to integrate the H$\alpha$ emission (see text), this value would be $2.6\times10^3$\,Electrons\,s$^{-1}$ which would give a spectral type of $\sim$O9.5V.
\end{deluxetable}

For this method of calculating spectral types, the ionizing flux is underestimated since the H$\alpha$ emission is not corrected for extinction, and we over-subtract the stellar sources. However, our assumptions thus far assume only one main ionizing source, i.e., we do not account for multiplicity. Indeed, high-mass stars are expected to be binary systems. Since the relationship between ionizing luminosity and spectral type is far from linear, we do not drastically overestimate the mass of the highest mass YSO due to multiplicity in the system. For example, the ionizing luminosity of 2.4 O7.5V stars (each $\sim$25\,M$_\odot$) is the same as an O6V star ($\sim$31\,M$_\odot$); similarly, the ionizing luminosity of 2.4 O9V stars (each $\sim$17\,M$_\odot$) is the same as an O8V star of $\sim$21\,M$_\odot$ \citep{Martins2005}. It is unclear if extinction of the H$\alpha$ band (suggesting bias toward later spectral types) or multiplicity of the YSOs (bias toward earlier spectral types) has a greater impact for the estimated spectral types. 
We also note we estimated the {\em current} spectral types of the sources; since these MYSOs are embedded as indicated by Spitzer emission, they may still be accreting and could eventually become earlier (more massive) spectral types \citep{Zinnecker2007}.

%We also note that for \stwo, there exists extended $H_\alpha$ emission toward the west (Fig. 1 left panel, emission toward the right of the central MYSO) that was not summed for the flux calculation. Only the compact flux was summed since this extended flux may be ionized due to other stars to the west (right) of the MYSO. However, if this extended flux is added, the spectral type is earlier as indicated in the note for Table 3.

%
%\clearpage 
%%%%%%%%%%%%%%%%%%%%%%%%%%%% FIGURE %%%%%%%%%%%%%
\begin{figure*}[t!]
\begin{center}
%\centerline{
\includegraphics[clip=true, trim =  0 0 -0.5cm 0, width=0.247\textwidth]{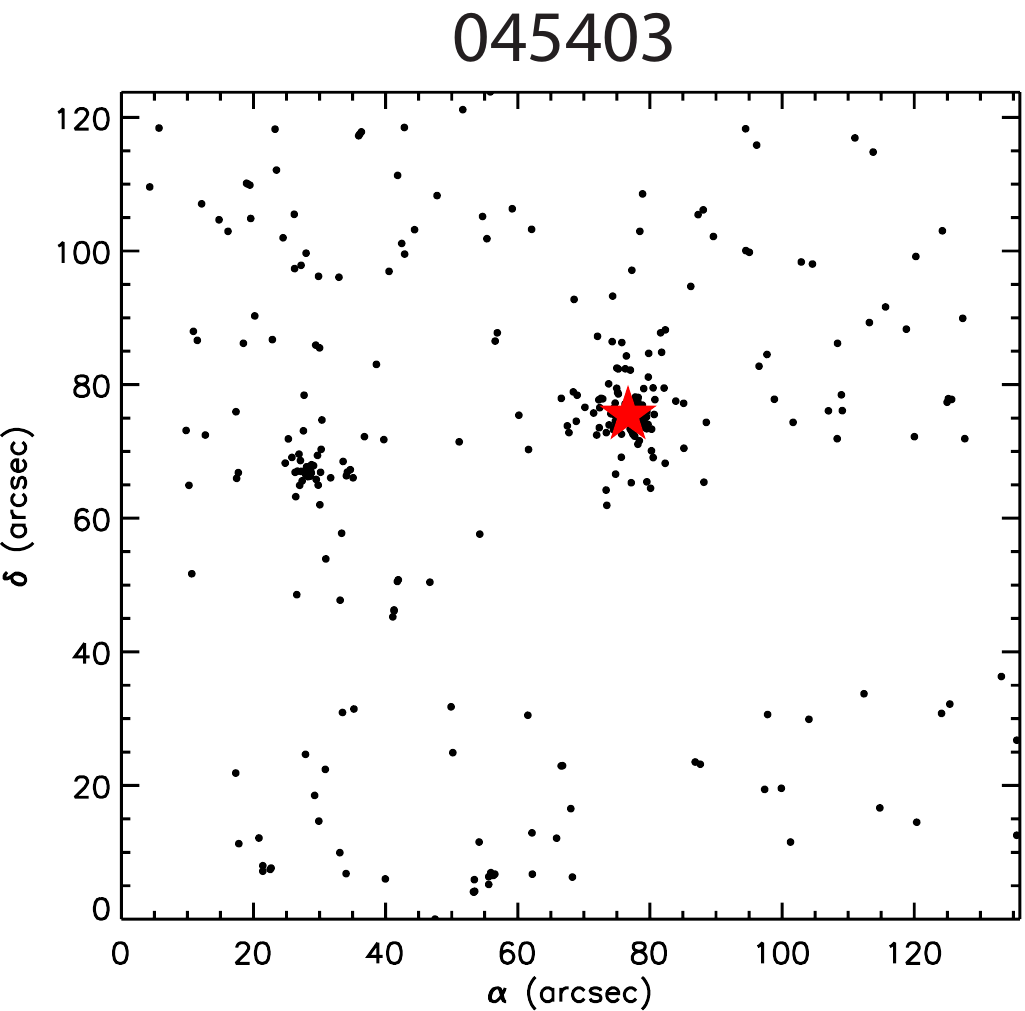}
\includegraphics[clip=true, trim =  0 0 -0.5cm 0, width=0.247\textwidth]{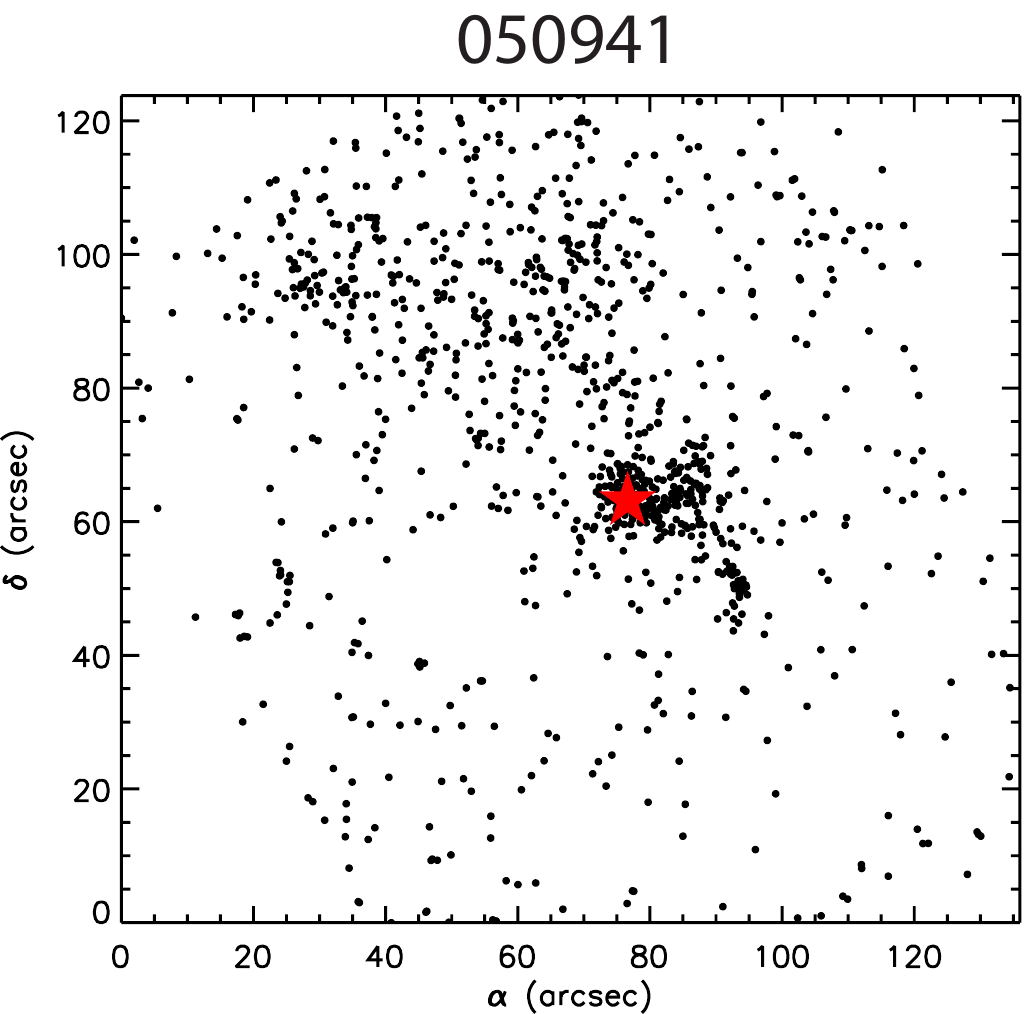}
\includegraphics[clip=true, trim =  0 0 -0.5cm 0, width=0.247\textwidth]{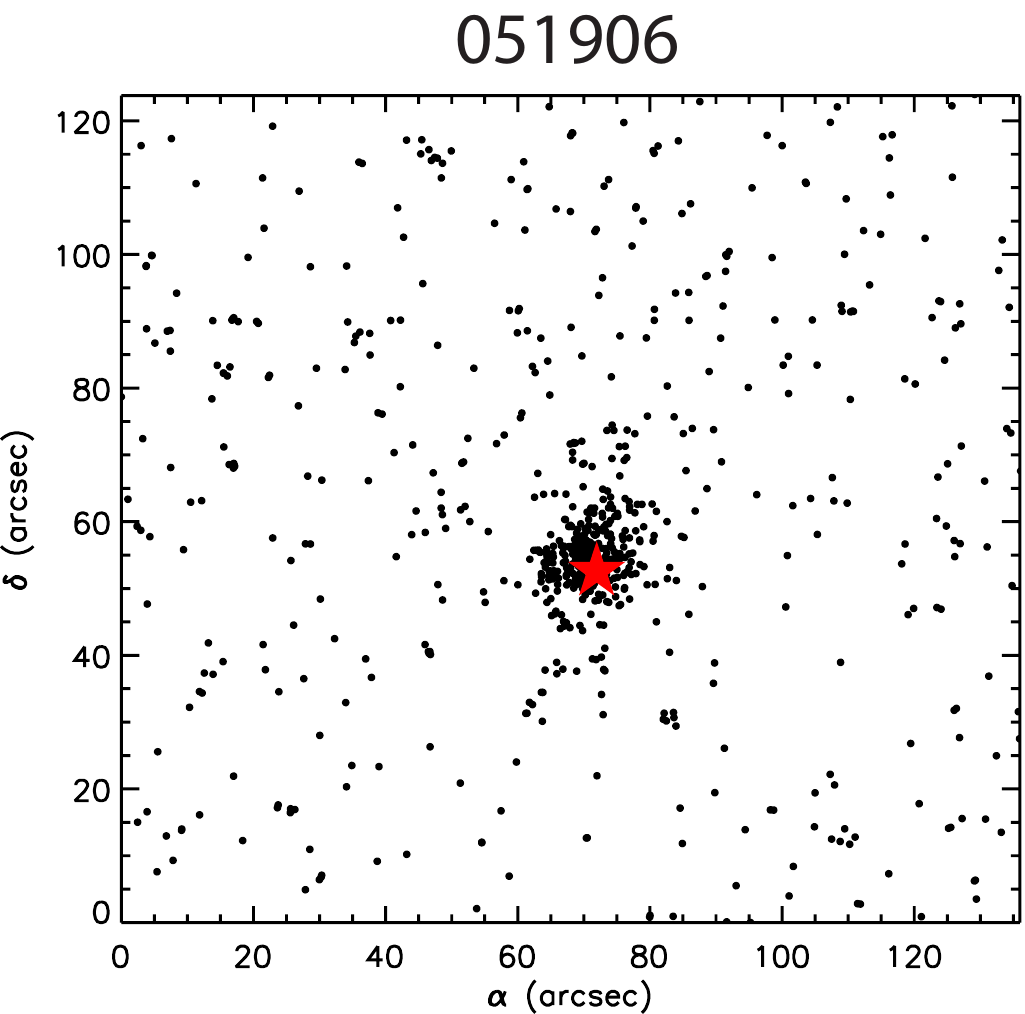}
\includegraphics[clip=true, trim =  0 0 -0.5cm 0, width=0.247\textwidth]{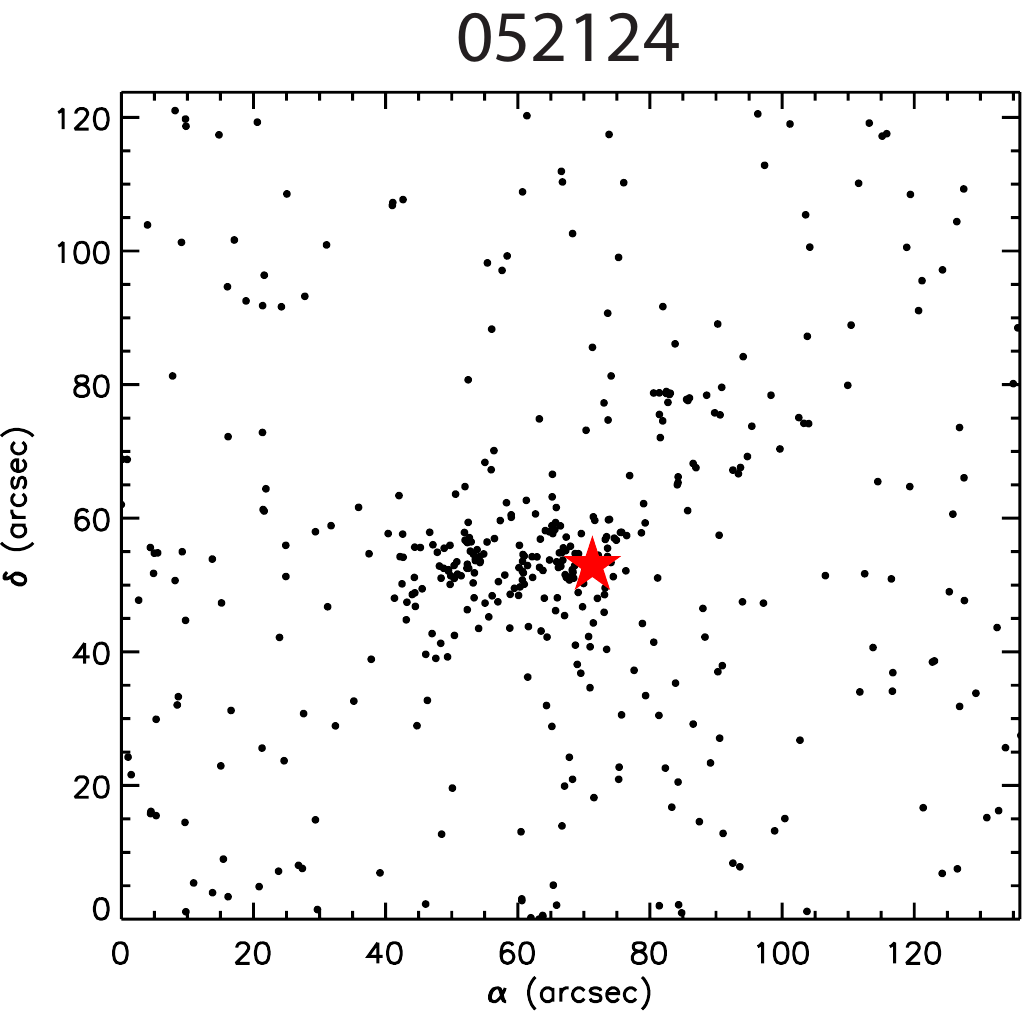}

\includegraphics[clip=true, trim =  0 0 -0.5cm 0, width=0.247\textwidth]{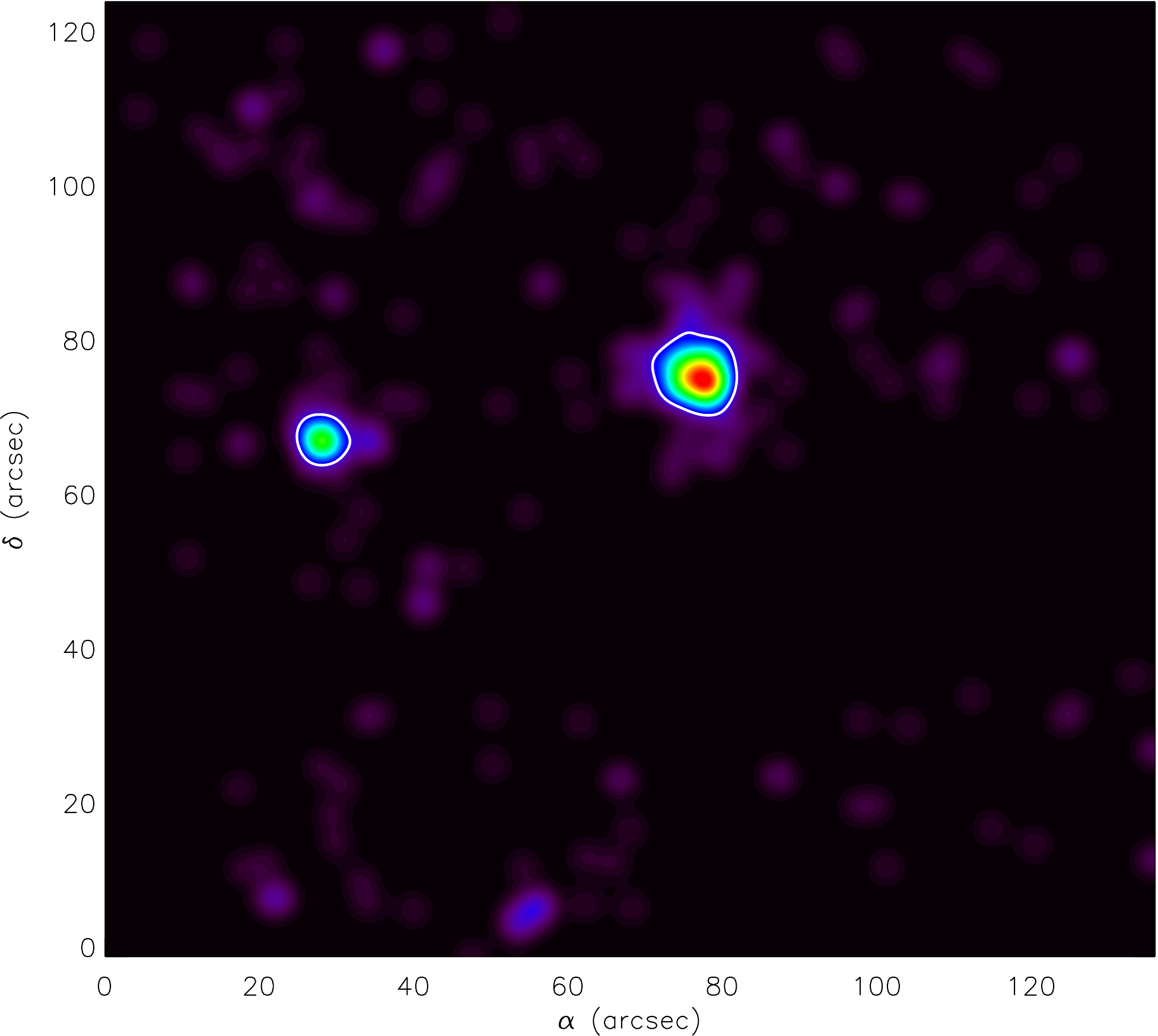}
\includegraphics[clip=true, trim =  0 0 -0.5cm 0, width=0.247\textwidth]{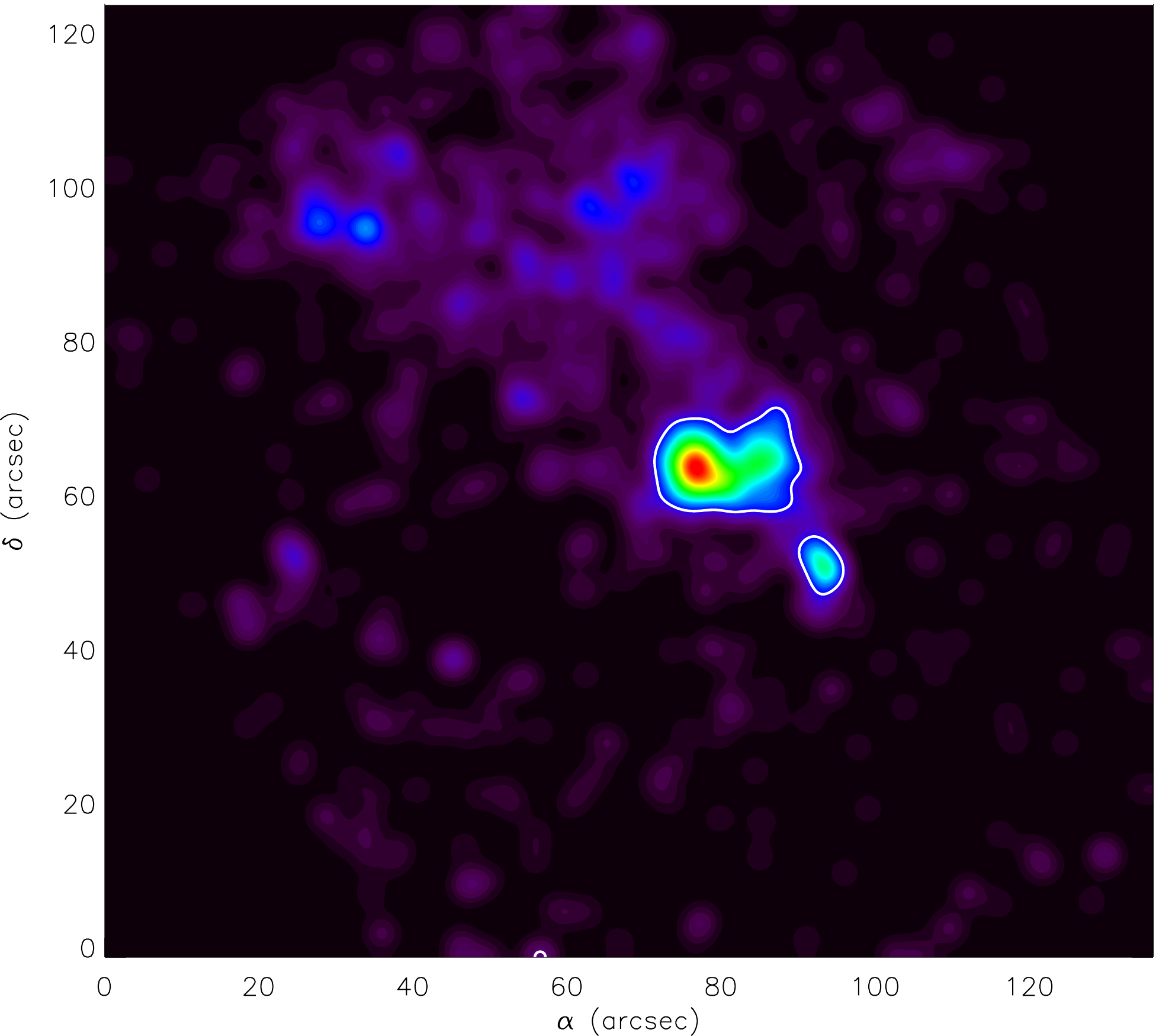}
\includegraphics[clip=true, trim =  0 0 -0.5cm 0, width=0.247\textwidth]{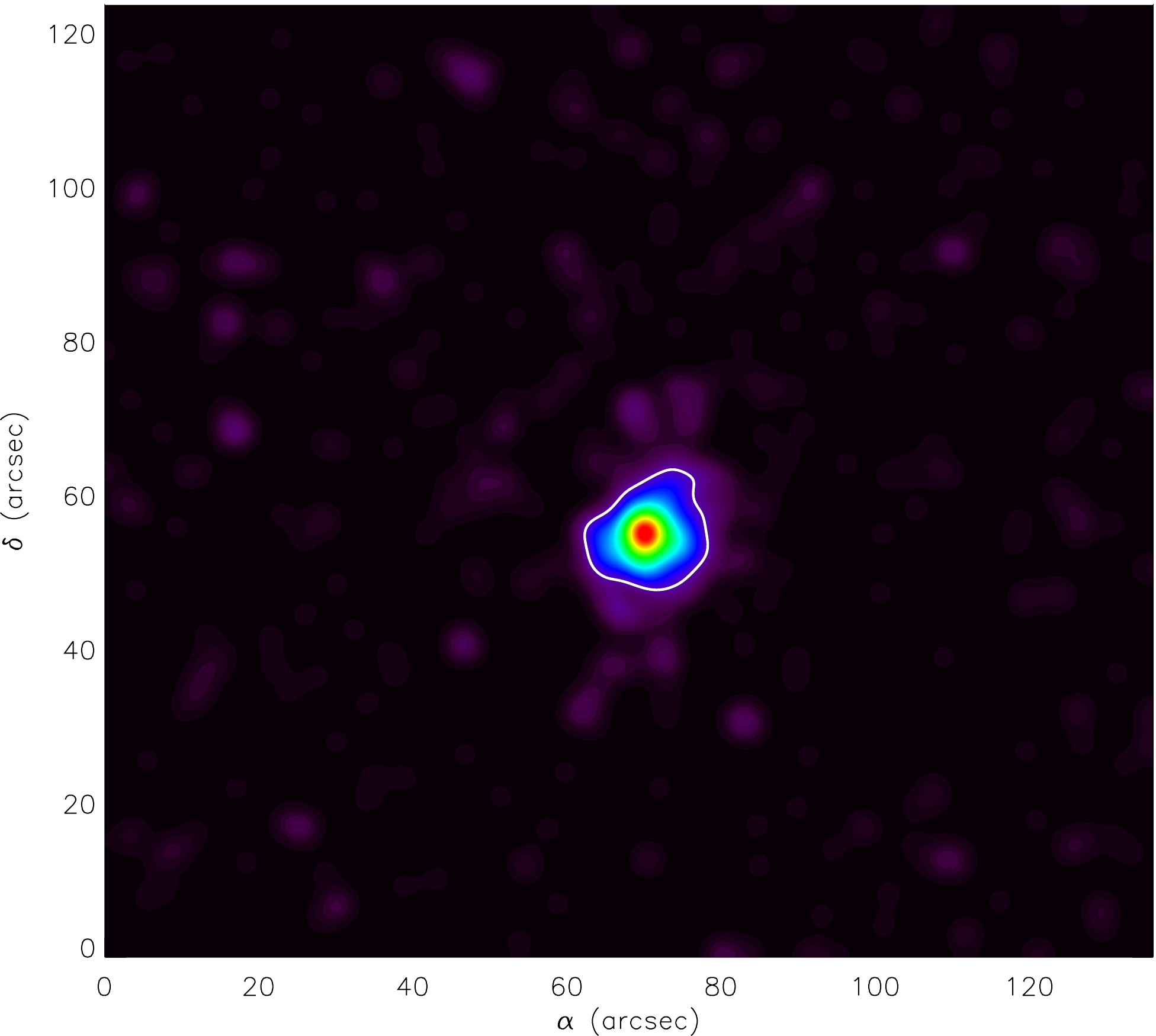}
\includegraphics[clip=true, trim =  0 0 -0.5cm 0, width=0.247\textwidth]{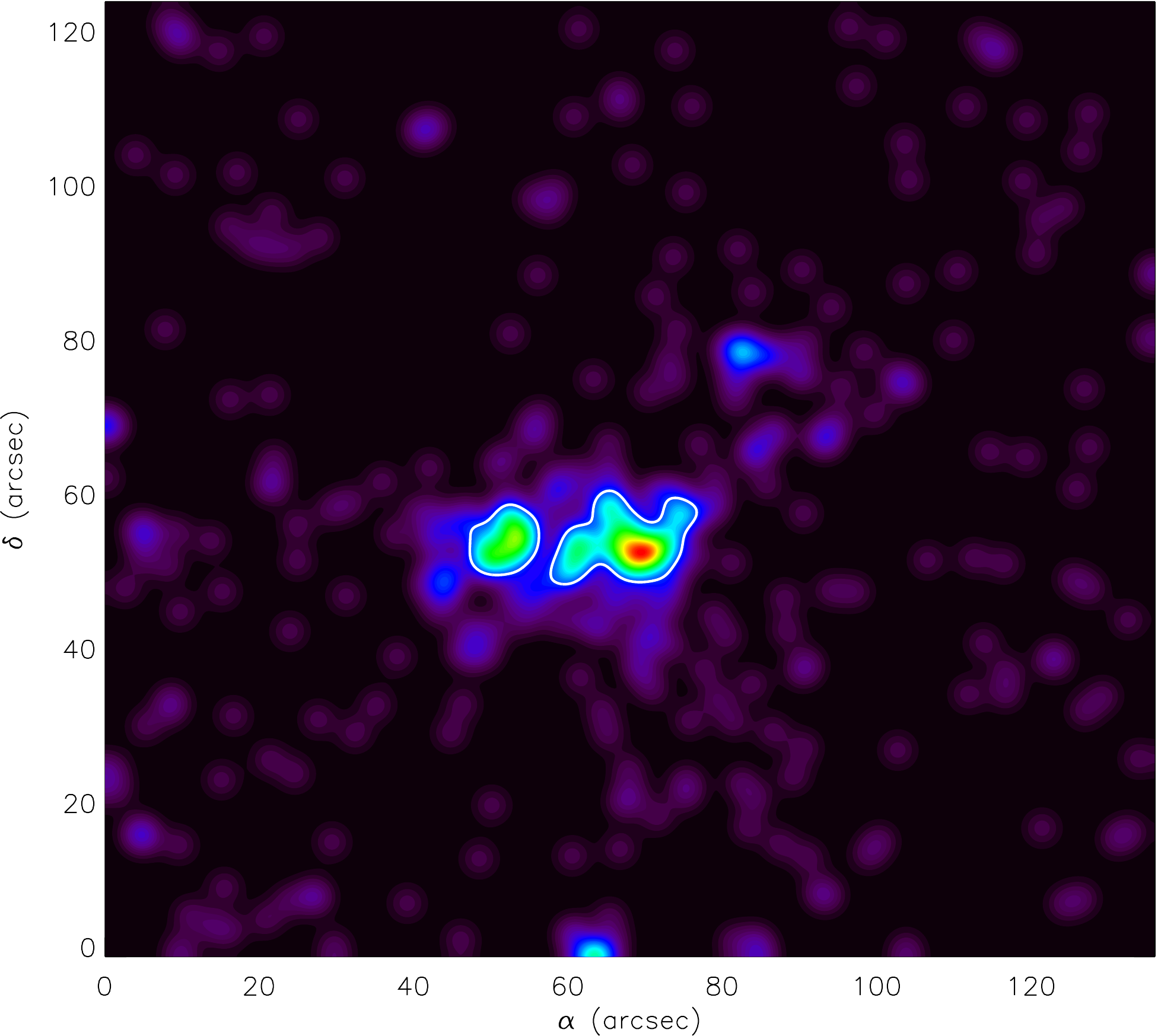}
%\LARGE 045403 \normalsize \\

\includegraphics[clip=true, trim =  0 0 -0.5cm 0, width=0.247\textwidth]{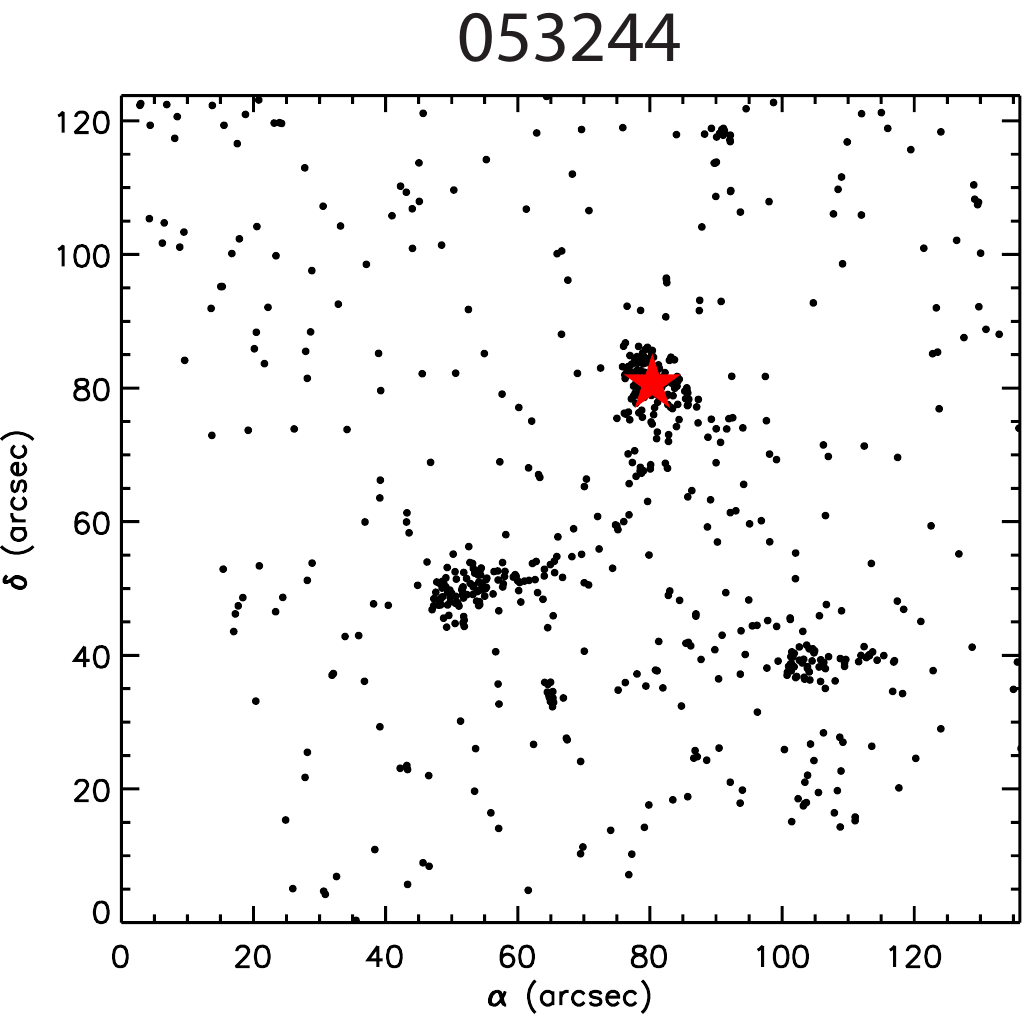}
\includegraphics[clip=true, trim =  0 0 -0.5cm 0, width=0.247\textwidth]{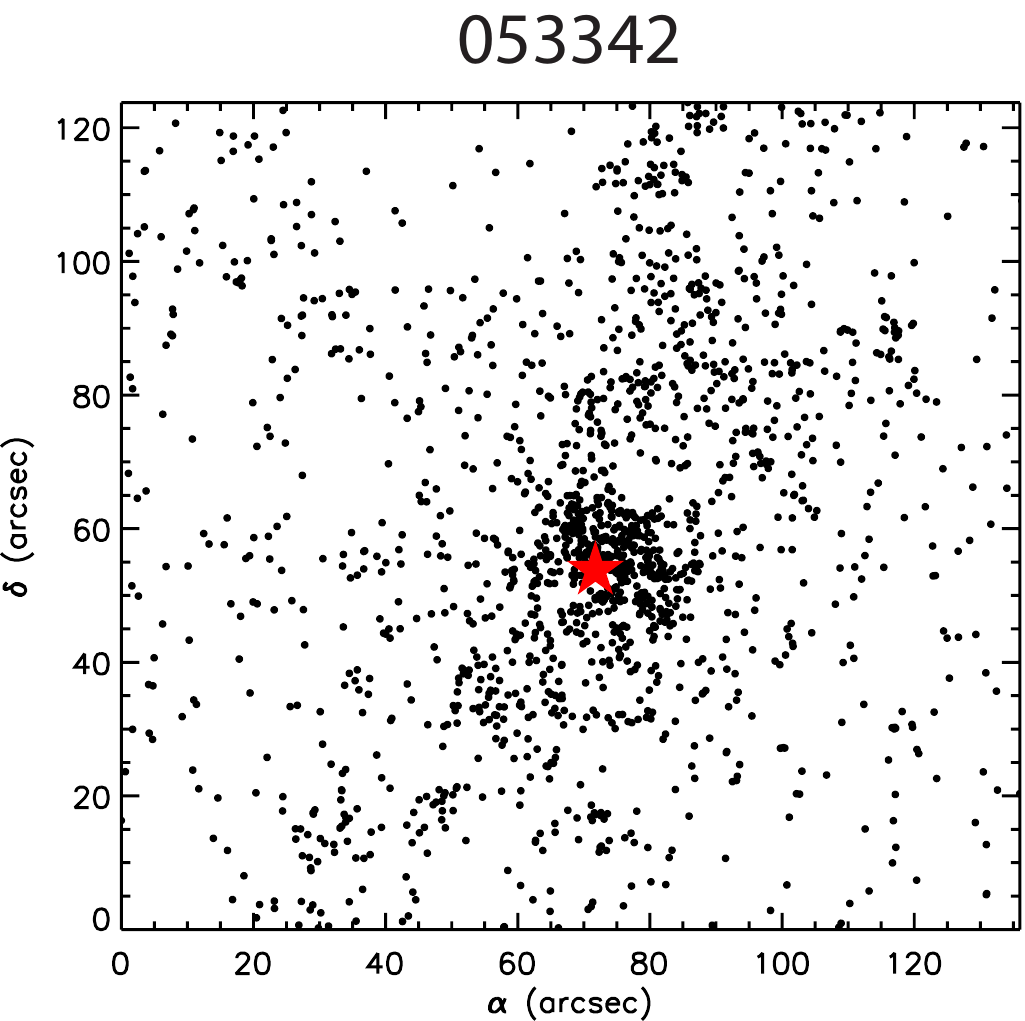}
\includegraphics[clip=true, trim =  0 0 -0.5cm 0, width=0.247\textwidth]{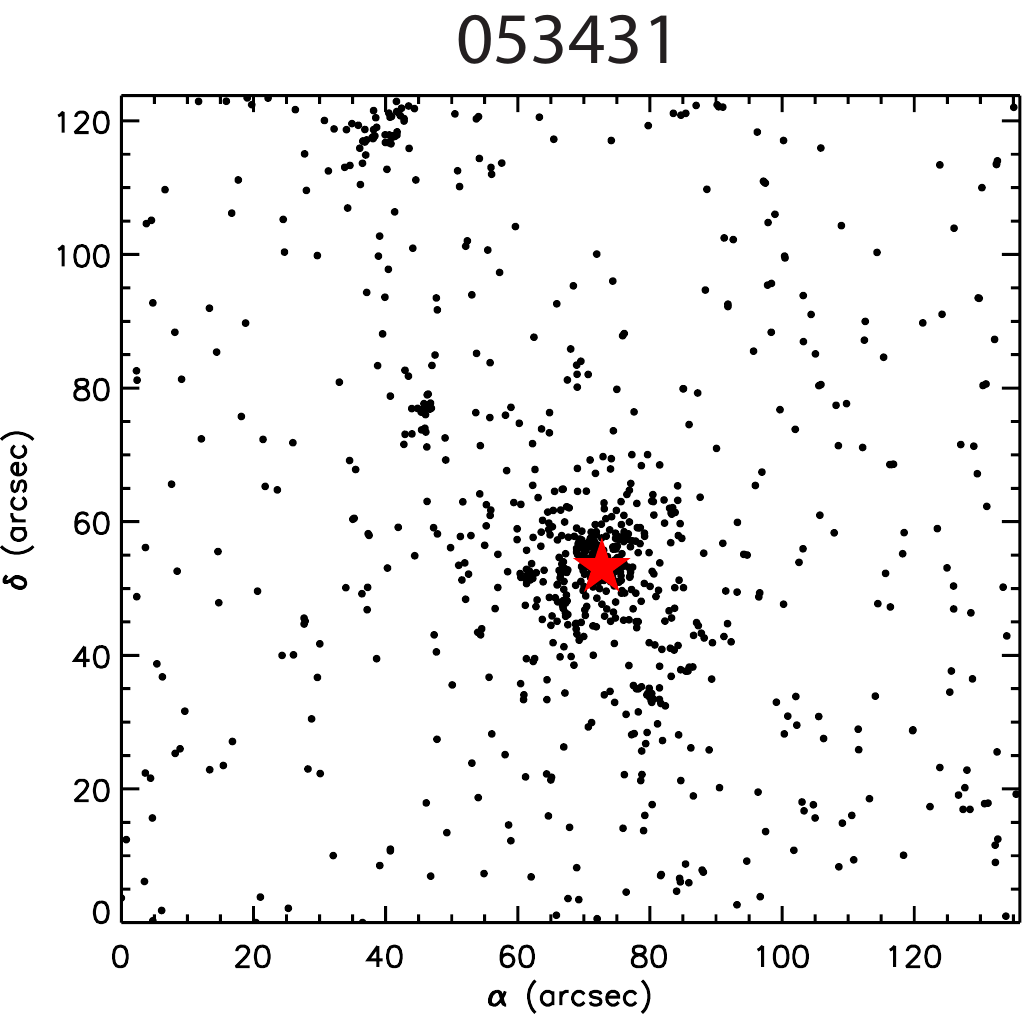}

\includegraphics[clip=true, trim =  0 0 -0.5cm 0, width=0.247\textwidth]{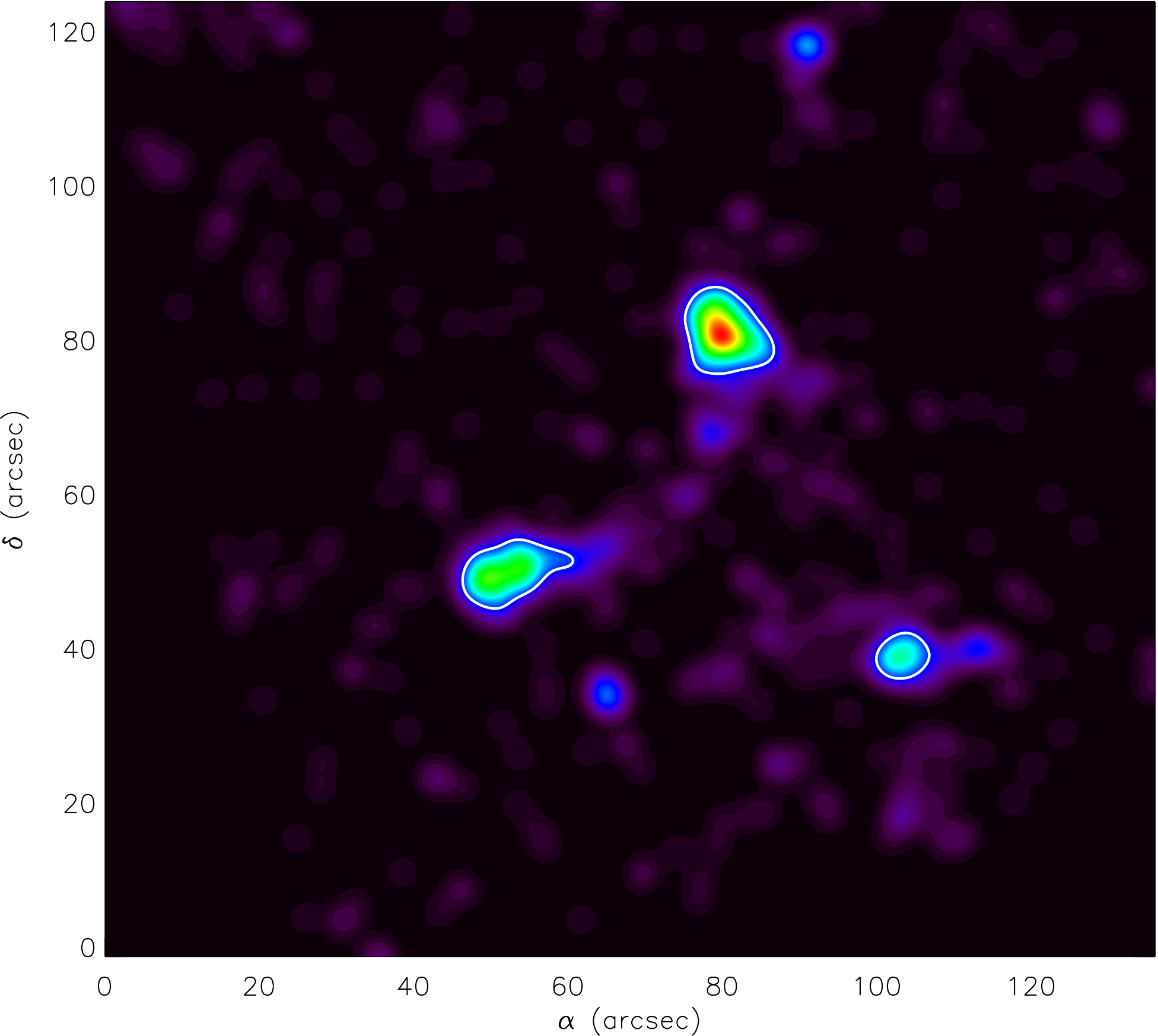}
\includegraphics[clip=true, trim =  0 0 -0.5cm 0, width=0.247\textwidth]{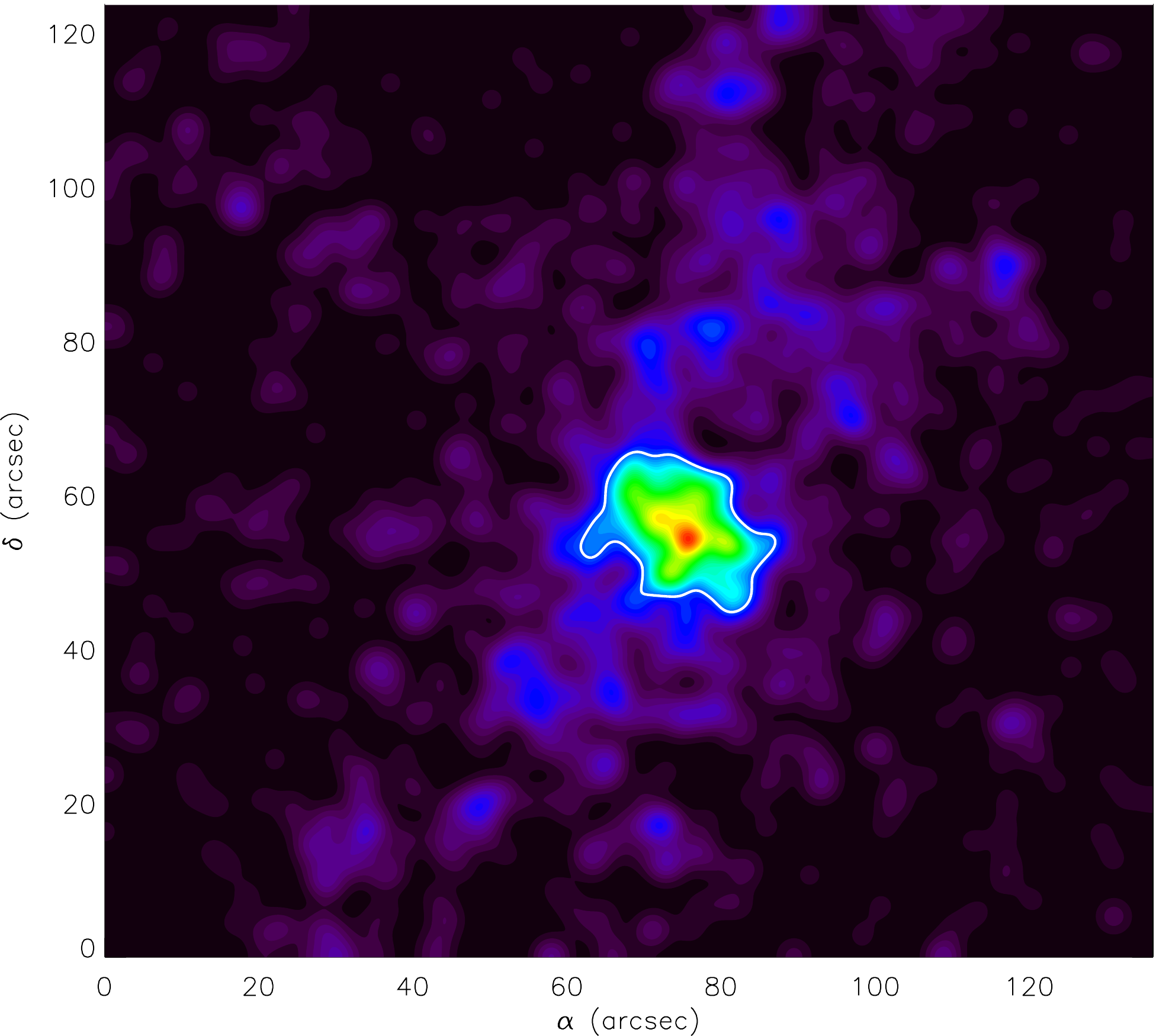}
\includegraphics[clip=true, trim =  0 0 -0.5cm 0, width=0.247\textwidth]{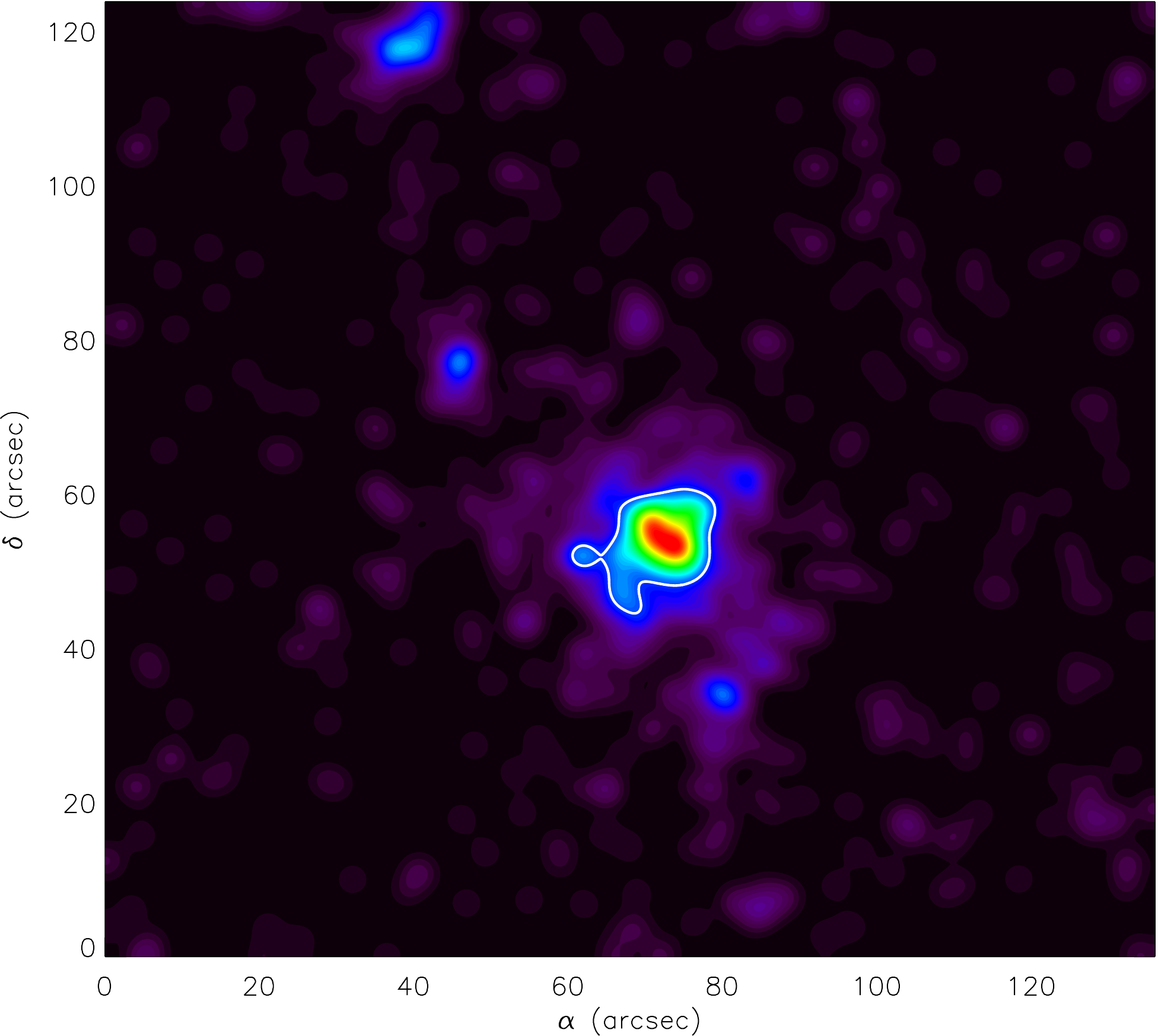}

\end{center}
\caption{\noindent Distribution of pre--main-sequence stars in regions encompassing the seven MYSOs. All maps 
have the same scale and orientation (north is up, east is left). {\em Top Panels:}
Charts of the young stellar populations detected across the observed field-of-views. The young stars in every region are 
plotted with black points. The positions of the target MYSOs are indicated by the large red star symbols.
{\em Bottom Color Scale Panels:}
The surface stellar density maps of the observed fields-of-view constructed with the Kernel Density Estimation from the detected  pre--main-sequence population. Each color in the maps shows the same clustering significance above the local pre--main-sequence population. 
White contours show the 3$\sigma$ isopleth surrounding each MYSO, where 1$\sigma$ is 1.20, 2.17, 2.63, 0.99, 1.91, 2.70, and 2.06\,PMS-stars\,pc$^{-2}$ for \sone, \stwo, \sthree, \sfour, \sfive, \ssix, and \sseven, respectively.
These maps show that the seven sources are far from isolated, in the sense that
their immediate environments are populated by a large sample of young PMS stars, which are distributed 
in a clustered fashion.
}
\label{f:kdemaps}
\end{figure*}
%Sensitivities in pms stars/pc^2
%045403      1.20238
%050941      2.16559
%051906      2.63420
%052124     0.994311
%053244      1.90858
%053342      2.70357
%053431      2.05979

%%%%%%%%%%%%%%%%%%%%%%%%%%%%%%%%%%%%%%%%%%%%%%

\section{Cluster Analysis}\label{s:clusanl}

The stellar charts of the selected young stellar populations around each of the MYSOs are shown in Figure\,\ref{f:kdemaps} (top panel). These maps show that the considered MYSOs are far from being formed in an isolated environment because there is a large number of young stars around each of the MYSOs. Note that, as discussed in Section\,\ref{PMSstars}, the field encompassing \sone\ has many photometric sources removed due to the large diffraction spike in the field, making the cluster appear more isolated than it actually is. There are two facts derived from these maps: 1) the vast majority of PMS stars are assembled around the MYSOs, apparently in star-forming clusters, and 2)~these clusters are not entirely isolated themselves. Specifically, while in the region \sthree\ there is only
one compact stellar over-density  around the MYSO,  the other regions have clear evidence of young stars 
in the field surrounding the MYSOs, which are loosely distributed across almost the entire observed field. This indicates that the stellar 
clusters around the MYSOs for all regions \emph{except} \sthree\ are themselves not isolated either, but they probably belong to a larger stellar constellation, related to a larger-scale star formation event and to their parental molecular clouds.

\subsection{Stellar Surface Density Maps}

We investigate the clustering behavior of the PMS stars in the observed regions by first identifying and characterizing the stellar clusters in the regions. To this aim we build surface stellar density maps with the use of our stellar samples and the application of the {\em Kernel Density Estimation} (KDE) method \citep{Silverman1992}. Density maps are constructed by convolving the stellar catalog with a Gaussian kernel. The main input parameter is  the full-width at half-maximum (FWHM) of the kernel, which specifies the minimum size of any stellar clustering that can be identified. There is no concrete method to define the optimal KDE kernel for smoothing the stellar maps, which is thus determined through experimentation. The minimum permitted FWHM size, corresponding to the typical PSF size of  $\sim$\,2.5\,WFC3 pixels is about 0.1\arcsec, which at the distance of the LMC corresponds to $\sim$\,0.025~pc. However, this limit is essentially the {\em resolution} in our photometry (depending on waveband), and therefore a KDE map of lower resolution, i.e., built with a larger FWHM, should be used for identifying statistically important stellar over-densities. 

A reasonable minimum size for the identified stellar clusterings is $\sim$\,1\,pc, which corresponds to a FWHM of $\sim$\,4\arcsec\ ($\sim$100 pixels). Our experiments showed that kernels smaller than this limit produce very noisy maps in which density fluctuations do not allow any concrete identification. On the other hand, kernels larger than 100 pixels begin to over-smooth the data so that the derived size-scale for the detected over-densities are overestimated. Therefore, we use for our cluster analysis the kernel size of $\sim$\,4\arcsec. The constructed KDE maps are also shown in Figure\,\ref{f:kdemaps} (bottom panel).

%These maps visualize more clearly the results derived above from the stellar charts of the regions, i.e., that there are conspicuous stellar concentrations around the selected MYSOs, which are highly clustered around the YSOs, while also being part of larger loose concentrations for two of the regions. The KDE maps are moreover used for the determination of the borders of the main stellar clusterings in the regions and their characterization. Two measures of the size of each cluster are given in column 5 of Table\,1; its {\em equivalent radius}, defined as the radius of a circle with the same area, $A_{\rm cl}$, as that covered by the 3$\sigma$ isopleth ($r_{\rm equiv} = \sqrt{A_{\rm cl}/\pi}$), and the  {\em maximum radius}, $r_{\rm max}$, defined by the area enclosed by the largest circle that encompass the entire cluster, equivalent to the half of the largest separation between the PMS stars in the cluster.  The ratio of these radii, $r_{\rm max} / r_{\rm equiv}$, provides a characterization of the {\em elongation} of each cluster, given in column 8 (a circular cluster has an elongation of unity). 

%%%%%%%%%%%%%%%%  TABLE  %%%%%%%%%%%%%%%%%%%%%%%%%
\begin{deluxetable*}{ccccccccccccc}
%\tablewidth{0pc}
\tablecaption{Characteristics of young stellar clusters detected around the seven MYSOs.\label{t:cluschar}} 
\tablehead{
\colhead{MYSO} & 
\colhead{Spectral\tablenotemark{a}} &
\colhead{\mmax \tablenotemark{a}} & 
\multicolumn{2}{c}{Position (J2000)}  & 
\colhead{$N_\star$\tablenotemark{b}} & 
\colhead{$r_{\rm equiv}$\tablenotemark{c}} & 
\colhead{$r_{\rm max}$\tablenotemark{c}} &  
\colhead{Elong.\tablenotemark{c}}  & 
\multicolumn{2}{c}{\mecl\ ($M_\odot$) Estimated\tablenotemark{d}} &
\multicolumn{2}{c}{\mecl\ ($M_\odot$) Extrapolated}  \\ 
\colhead{} &
\colhead{Type} & 
\colhead{($M_\odot$)} & 
\colhead{Right Ascension} & 
\colhead{Declination} & 
\colhead{} & 
\colhead{(pc)} & 
\colhead{(pc)} & 
\colhead{} & 
\colhead{1\,Myr} &
\colhead{2.5\,Myr} & 
\colhead{IMF\tablenotemark{e}} &
\colhead{\mmax\,--\,\mecl\,Relation\tablenotemark{f}} 
}%\colhead{\multicolumn{2}{c}{Cluster Density Peak}} 
\startdata 
\sone & B0V & 14 & 4$^{\rm h}$54$^{\rm m}$03.6$^{\rm s}$& $-$67$^{\circ}$16$^{\prime}$18.3$^{\prime\prime}$ & 86 & 1.9 & 2.2 & 1.16 & 110 & 170 & 210 & 200 \\
\stwo & O9.5V & 16 & 5$^{\rm h}$09$^{\rm m}$41.9$^{\rm s}$& $-$71$^{\circ}$27$^{\prime}$41.7$^{\prime\prime}$ & 397 &   2.7 & 3.4 &  1.28  & 250 & 360 & 240 & 280 \\
\sthree & O8V & 21 & 5$^{\rm h}$19$^{\rm m}$07.0$^{\rm s}$& $-$68$^{\circ}$21$^{\prime}$35.1$^{\prime\prime}$ & 375 &   2.4 & 2.7 &  1.14  & 350 & 510 & 360 & 490 \\
\sfour & B0V & 14 & 5$^{\rm h}$21$^{\rm m}$25.6$^{\rm s}$& $-$66$^{\circ}$04$^{\prime}$12.3$^{\prime\prime}$ & 86 & 2.4 & 4.2 & 1.74 & 90 & 140 & 210 & 200 \\
\sfive & B0V & 14 & 5$^{\rm h}$32$^{\rm m}$44.3$^{\rm s}$& $-$69$^{\circ}$30$^{\prime}$05.5$^{\prime\prime}$ & 122 & 1.9 & 2.0 & 1.06 & 170 & 250 & 210 & 200 \\
\ssix & O6V & 31 & 5$^{\rm h}$33$^{\rm m}$41.4$^{\rm s}$& $-$68$^{\circ}$46$^{\prime}$02.6$^{\prime\prime}$ & 517 &   3.1 & 4.0 &  1.29  & 400 & 610 & 670 & 1220 \\
\sseven & B0V & 14 & 5$^{\rm h}$34$^{\rm m}$31.7$^{\rm s}$& $-$68$^{\circ}$35$^{\prime}$13.6$^{\prime\prime}$ & 222 & 2.7 & 3.4 & 1.29 & 220 & 350 & 210 & 200 \enddata
%\tablecomments{A detailed description of the parameters is given in Section\,\ref{s:cluscat}.}

\tablenotetext{a}{Spectral types and masses are derived using the \emph{HST} H$\alpha$ observations according to the effective temperature scales by \citet{Martins2005}. Spectral type estimates do not take into account multiplicity or extinction. See Section\,\ref{s:spclass} for more information.} %Hanson1997 for B-stars... similar Q_0 as Martins for O-stars
\tablenotetext{b}{Number of stars in the cluster from the complete clean photometric sample.} 
\tablenotetext{c}{Two radii are given; the equivalent radius (defined as the radius of a circle with the same area) and the maximum radius of the cluster. Elongation is the ratio of the two radii, $r_{\rm max} / r_{\rm equiv}$.}
\tablenotetext{d}{Approximate embedded cluster mass (\mecl) assuming isochrone ages of 1~Myr and 2.5\,Myr for all stellar sources within the cluster. Cluster mass is likely underestimated due to differential spatial extinction and high extinction of sources at the envelope scale.}
\tablenotetext{e}{\mecl\ expected from analytically based on the IMF \citep{Weidner2004} for the observed \mmax\ given in column 3.}
\tablenotetext{f}{\mecl\ expected from the typical \mmax -- \mecl\ relation \citep{Weidner2013} for the observed \mmax\ given in column 3.}
\end{deluxetable*}
%all 1 Myr
%045403 : (87.4+81.5)/2*1.28=108.096
%050941 :  (185.9 + 202.4)/2*1.28=248.512
%051906 : (279.7 + 263.8)/2*1.28=347.84
%052124 : (67.7+75.0)/2*1.28=91.328
%053244 : (133.8+124.5)/2*1.28=165.312
%053342 : (293.6 + 331.9)/2*1.28=400.32
%053431 : (173.8+171.7)/2*1.28=221.12

%all 2.5\,Myr
%045403 : (123.4+136.4)/2*1.28=166.272
%050941 : (262.6+293.6)/2*1.28=355.968
%051906 : (401.2+402.0)/2*1.28=514.048
%052124 : (100.7+115.5)/2*1.28=138.368
%053244 : (203.1+208.2)/2*1.28=263.232
%053342 : (446.7+512.4)/2*1.28=613.824
%053431 : (262.8+286.2)/2*1.28=351.36

%just PMS
%045403 : (42.6+46.9)/2*1.28=57.28
%052124 : (42.2+54.2)/2*1.28=61.696
%053244 : (57.5+59.2)/2*1.28=74.688
%053431 : (86.9+91.9)/2*1.28=114.432

\subsection{Stellar Clusters around MYSOs}\label{s:cluscat}

The identification of stellar clusterings in the KDE maps as statistically important over-densities, was made for those having density above a certain threshold in the maps. This threshold is given in $\sigma$ above the local density background, where $\sigma$ is the standard deviation of each map. Regions in the KDE that appear at a minimum level of 3$\sigma$ and persist at higher levels are considered as bonafide stellar clusters. The 3$\sigma$ identification threshold for each map is shown with the white isopleth line in the KDE maps of Figure\,\ref{f:kdemaps}. With this method we identified a single important concentration in each region. These concentrations correspond to the compact clusters seen around the MYSOs in both the stellar and KDE maps of Figure\,\ref{f:kdemaps}. For \sone, \stwo\, \sfour, and \sfive, secondary smaller clusters  above the 3$\sigma$ threshold (and a tertiary for \sfive) are found.
%In the region 050941 there is secondary smaller cluster located to the south-west of the main cluster, identified in Section\,\ref{sec:stwo} as BSDL770. We do not consider this small cluster, which is likely part of the same star-forming region, in our analysis.

The characteristics of the MYSO clusterings, as defined within the 3$\sigma$ isopleth of the KDE maps, are given in Table\,\ref{t:cluschar}.  Cols.\,2 and 3 show the spectral type and maximum mass, \mmax\,, for each source (see Section\,\ref{s:spclass}). Coordinates of the clusters' centers, which correspond to their KDE density peaks are given in Cols.\,4 and 5. Col.\,6 shows the number of the stars in the clean photometric sample within the borders of every cluster. The approximate size of each cluster is given by the so-called {\em equivalent radius}  \citep[e.g.,][]{RomanZuniga2008} in Col.\,7. This radius is defined as the radius of a circle with the same area, $A_{\rm cl}$, as the area covered by the cluster ($r_{\rm equiv} = \sqrt{A_{\rm cl}/\pi}$). We also give in Col.\,8 the radius, $r_{\rm max}$, defined by the area enclosed by the smallest circle that encompass the entire cluster, equivalent to the half of the distance between the two farthest PMS stars in the system. These radii measurements imply that all seven clusters are compact.  The ratio of these radii, $r_{\rm max} / r_{\rm equiv}$, provides a characterization of the {\em elongation} of each cluster which we provide in Col.\,9, \citep[e.g.,][]{Schmeja2006}. A circular distribution, with axis ratio equal to unity, has an elongation parameter of 1, while an elongated distribution with an axis ratio of 10 has an elongation parameter of $\sim$3 \citep{Schmeja2006}.  The measurements of this parameter of the detected clusters show that most of the main clusters are slightly elongated. Col.\,10 and 11 gives the estimated cluster mass based on observations, while Cols.\,12 and 13 give predicted masses based on the estimated mass of the MYSO and studies by \citet{Weidner2004} and \citet{Weidner2013}. These cluster masses are discussed in more detail in the next section.

\begin{figure*}[t!]
\begin{center}
%\centerline{
\includegraphics[clip=true, trim =  0 0 -0.5cm 0, width=0.247\textwidth]{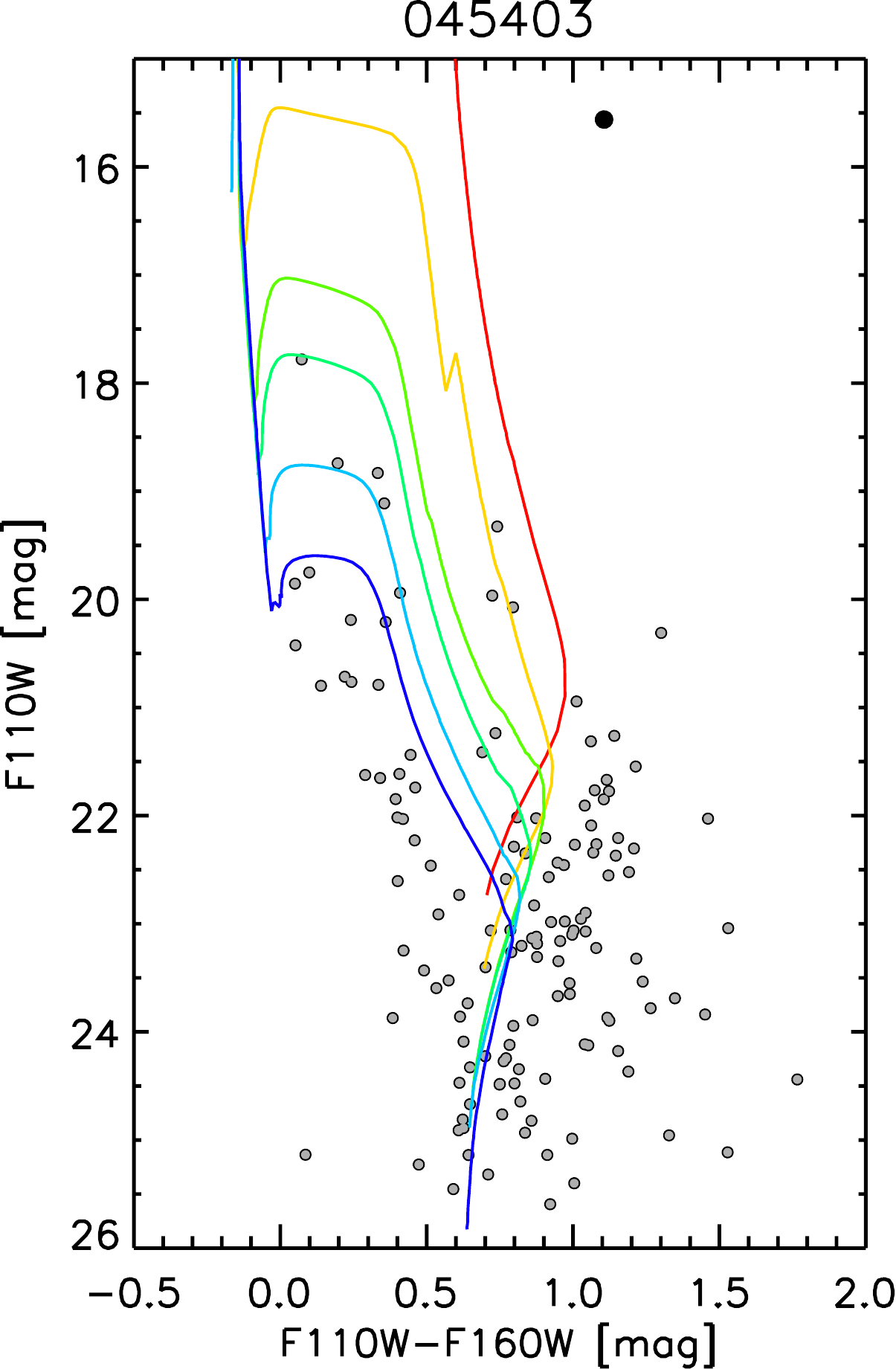}
\includegraphics[clip=true, trim =  0 0 -0.5cm 0, width=0.247\textwidth]{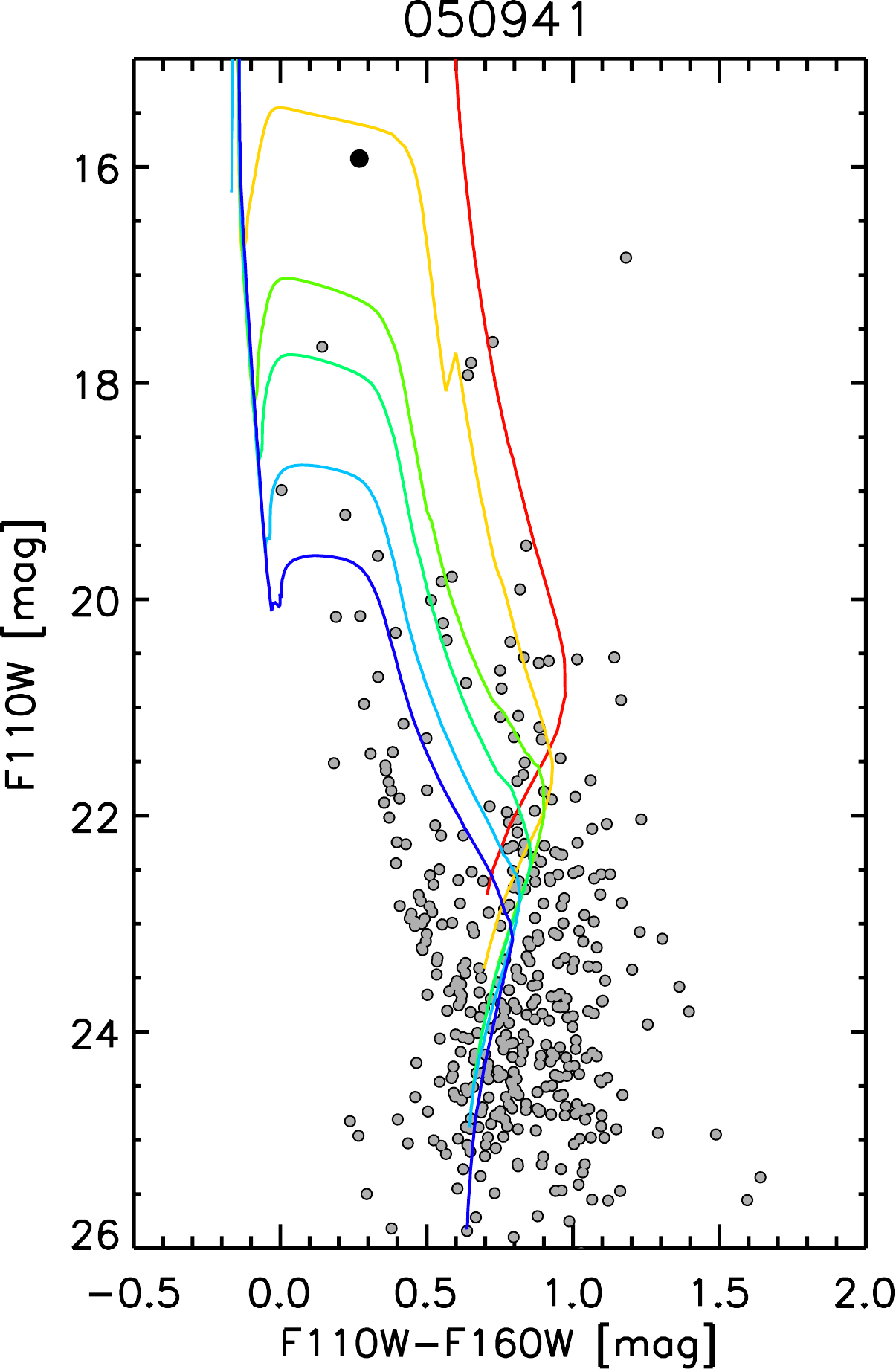}
\includegraphics[clip=true, trim =  0 0 -0.5cm 0, width=0.247\textwidth]{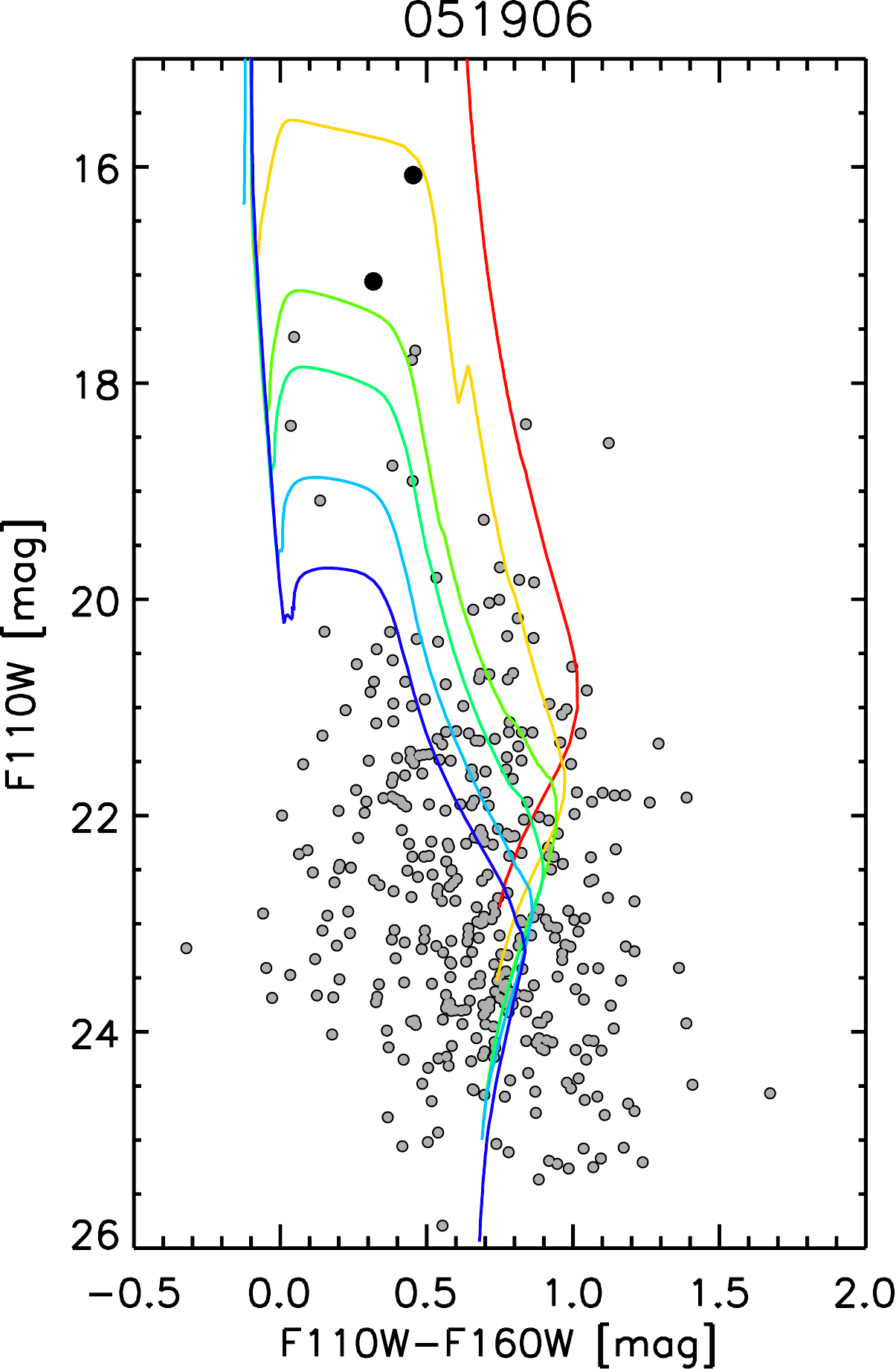}
\includegraphics[clip=true, trim =  0 0 -0.5cm 0, width=0.247\textwidth]{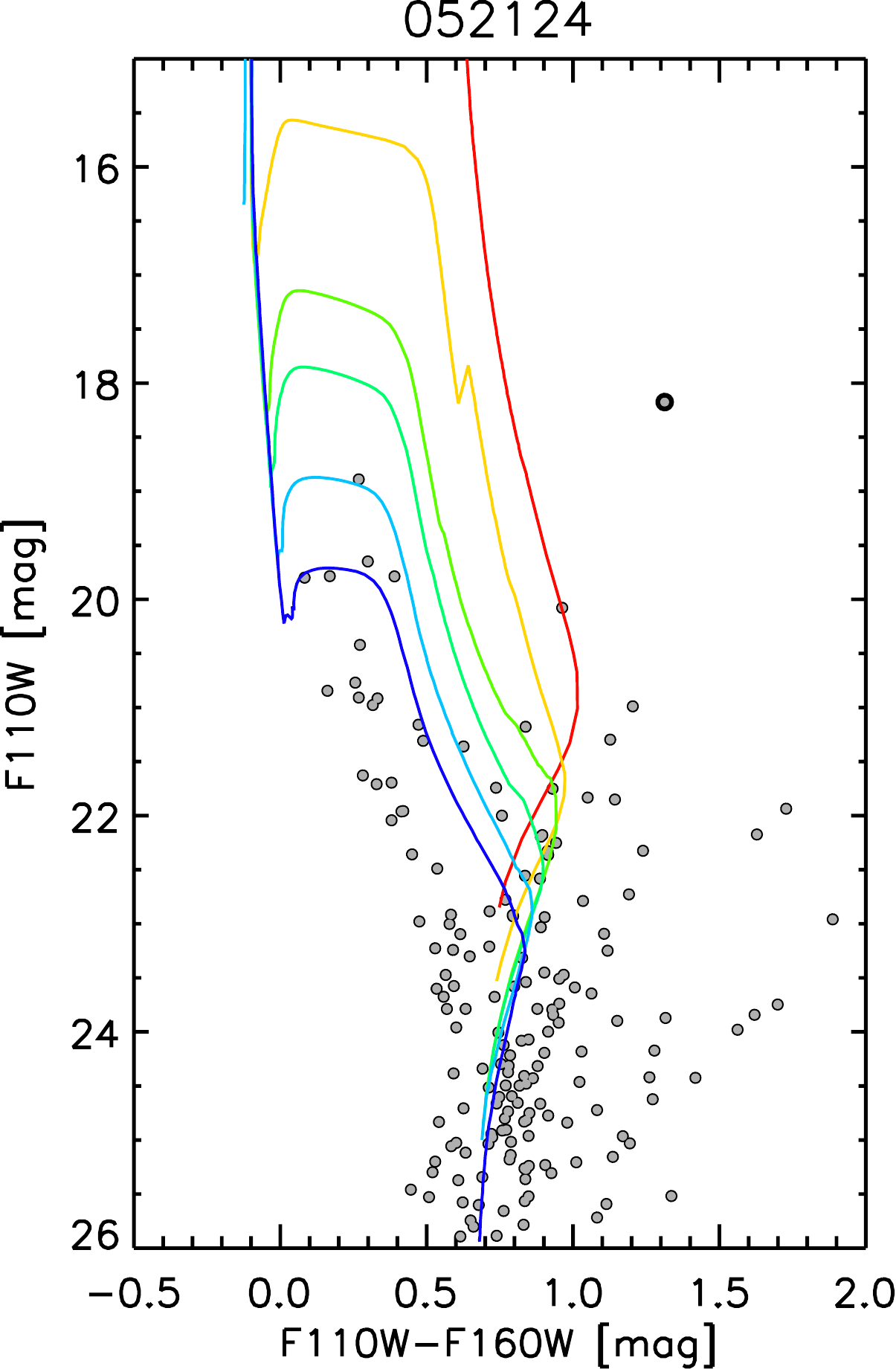}
\includegraphics[clip=true, trim =  0 0 -0.5cm 0, width=0.247\textwidth]{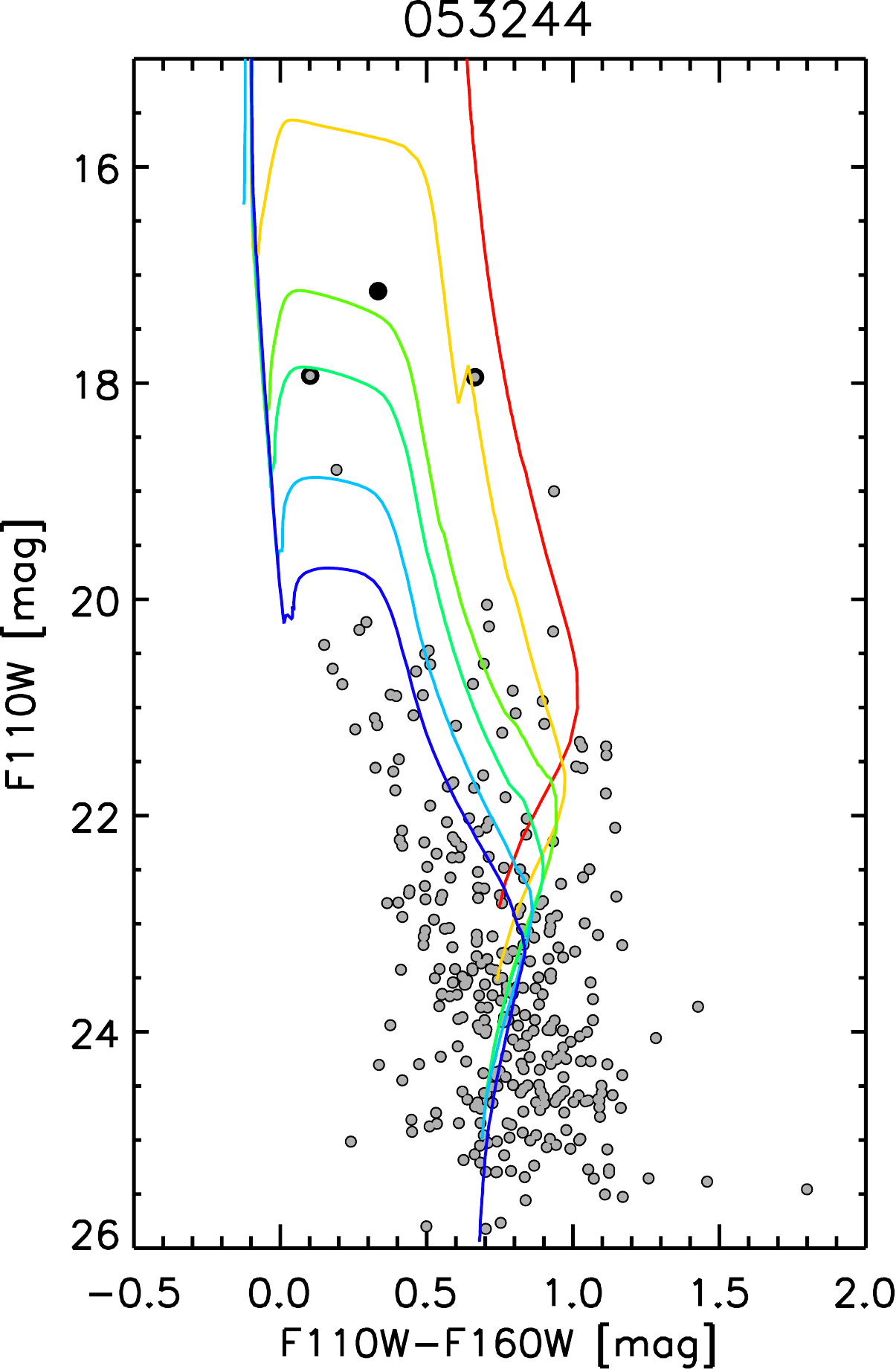}
\includegraphics[clip=true, trim =  0 0 -0.5cm 0, width=0.247\textwidth]{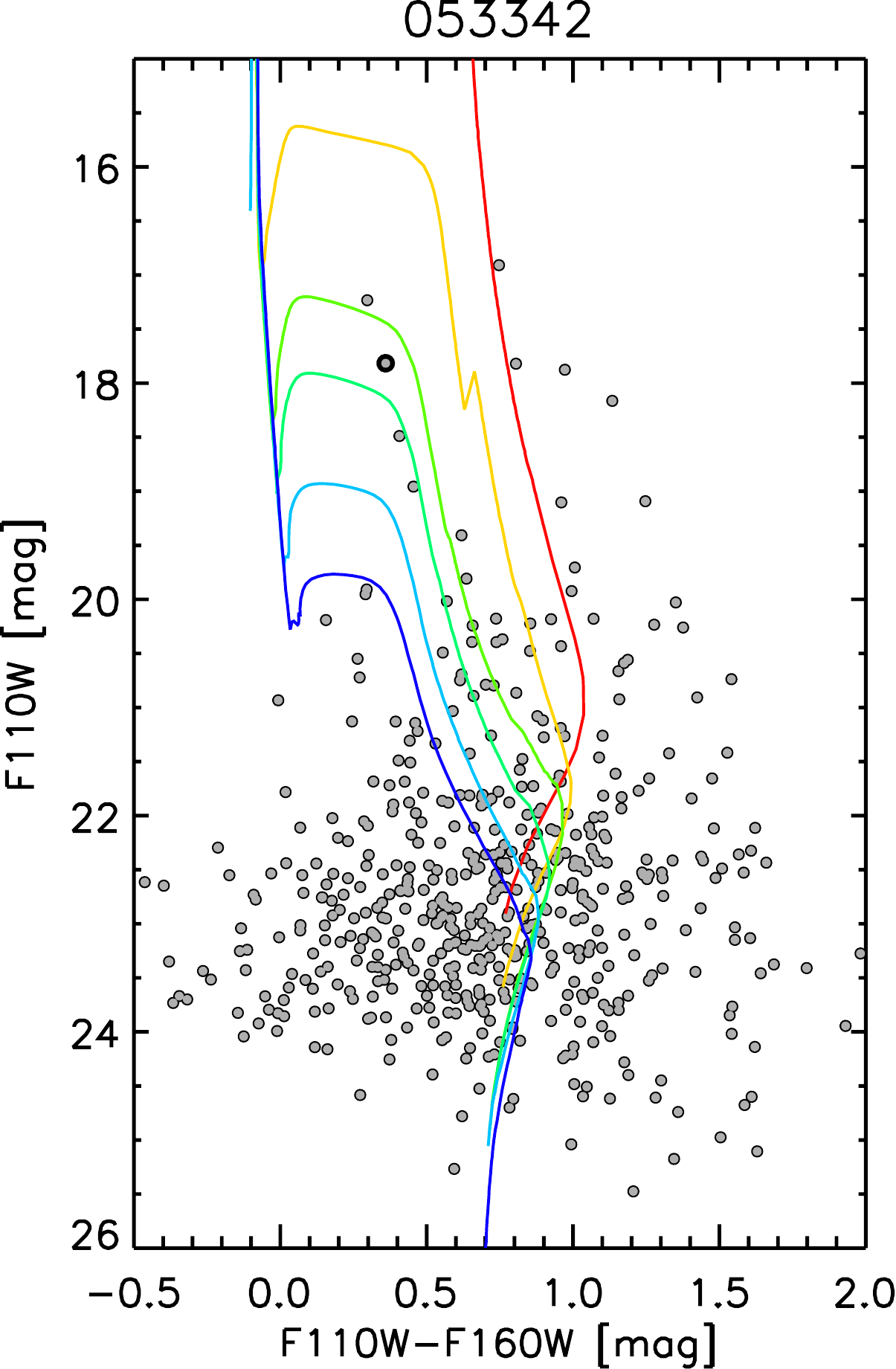}
\includegraphics[clip=true, trim =  0 0 -0.5cm 0, width=0.247\textwidth]{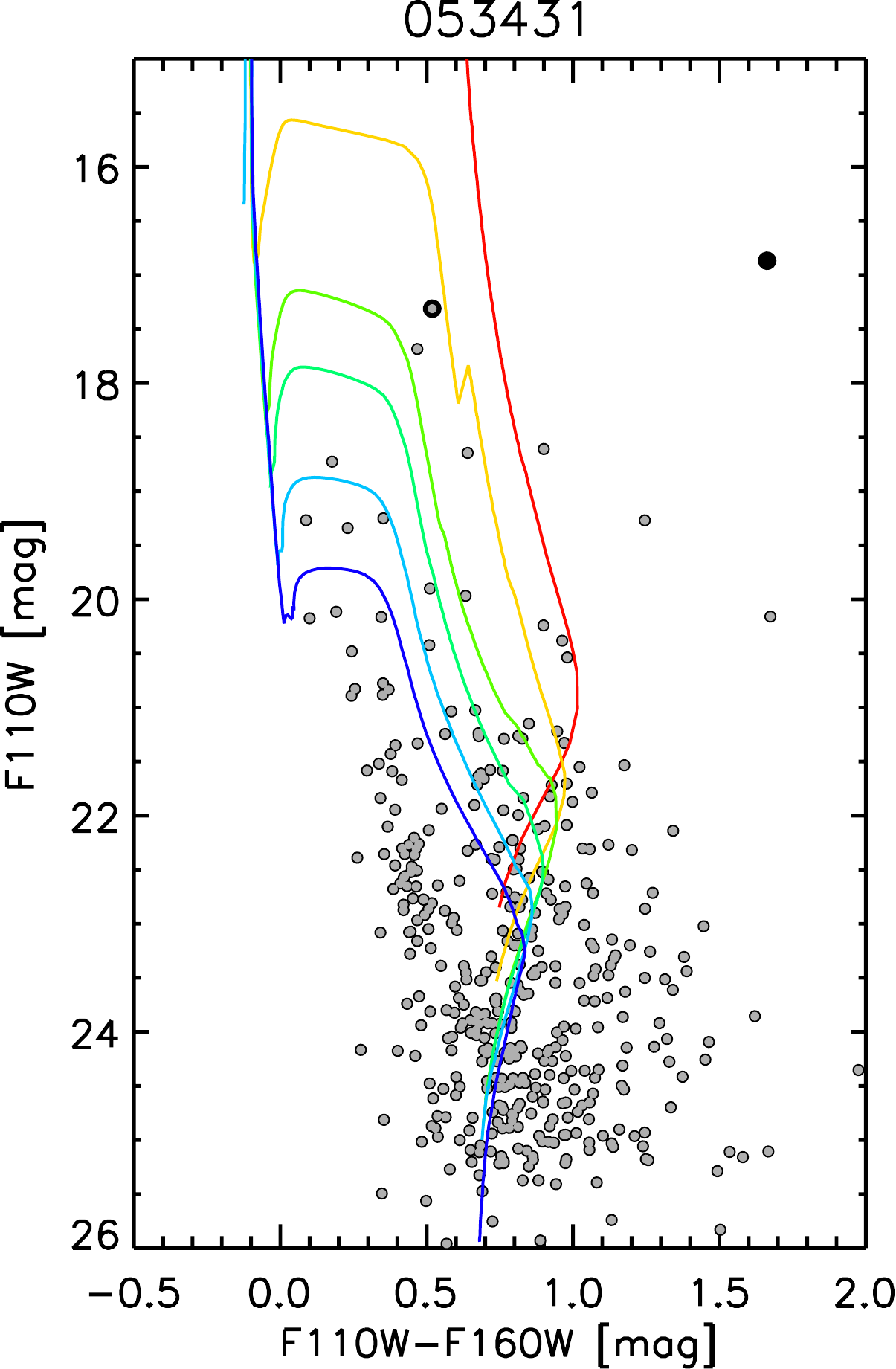}
%}
\end{center}
\caption{\noindent F110W, F160W color-magnitude diagrams of the all stars (gray circles) in the clean photometric sample within the 3$\sigma$ isopleth of the KDE maps. Black circles show the locations the brightest sources (F160W magnitudes less than 18 mag) within 1$\arcsec$ of the central MYSO. Black circles without a corresponding gray circle were sources that did not have a valid photometric fit in the F555W/F814W band and therefore were not included in our clustering analysis. Isochrones from the Padova family of models are shown corresponding to ages 0.01 (red), 0.1 (yellow), 0.5 and 1.0 (green), 2.5 (blue), and 5.0~Myr (violet). 
}
\label{f:clustercmd}
\end{figure*}

\subsection{Estimates of Cluster Masses}\label{s:clusterest}
For the estimation of the total mass of the cluster around each MYSO, we extract all stellar sources from the clean photometry sample within the 3$\sigma$ isopleth of the density maps. These stars within the stellar clustering are shown in Figure\,\ref{f:clustercmd}. We provide two estimates of the observed cluster masses: one using the 1\,Myr isochrone and another using the 2.5\,Myr isochrone (see Section \ref{s:pmss} for justification). The masses of each star are based on the luminosities of the F110W and F160W filters and are interpolated from the Padova grid of models \citep{Bressan2012, Chen2014, Tang2014}, corrected for the average extinction in the field (see Section\,\ref{PMSstars}). The masses of the stars were summed to calculate the {\em detected} mass of the embedded cluster, \mecl. The total cluster masses estimated by each of the F110W and F160W filters typically differed by less than 10\%; we adopted the average of these two measurements as the ``detected" cluster mass. 

In order to add the undetected mass to the total cluster mass, we determined the mass of our photometric detection limit. The detection limit for field stars in the F814W filter ($I$-band) has an apparent magnitude $m_I \approx 27$\,mag (Figure\,\ref{f:fs.cmds}). Using a distance modulus of 18.5\,mag, the absolute magnitude detection limit is $M_I = 8.5$\,mag, which is approximately a K4 star of 0.7~$M_\odot$ \citep{Cox2004}. We assume that the IMF of the cluster behaves like a \cite{Kroupa2001} IMF, which suggests that the actual cluster mass should be 28\% higher than the detected mass, calculated above. We add the missing mass to the detected cluster masses to find the final estimated cluster masses for ages of both 1\,Myr and 2.5\,Myr, which are reported as the ``Estimated \mecl'' in Col.\,10 and 11 of Table\,\ref{t:cluschar}. The number of stars, radii, and masses of the clusters are comparable to typical embedded Galactic clusters \citep{LadaLada2003}.

A protostar (later than O-type) of a constant mass will have its luminosity decrease as it evolves from the pre--main-sequence to the main sequence. Therefore, using different age isochrones can drastically change the mass of the cluster. As seen in Table\,\ref{t:cluschar}, the estimated cluster masses for the 2.5\,Myr isochrones are approximately 50\% higher than that of the 1\,Myr isochrones. In order to have a better idea of the uncertainty of the tabulated cluster masses, we also fit Padova isochrones for protostellar ages of 0.5 and 5\,Myr. For 0.5\,Myr isochrones, cluster masses were $\sim$25\,--\,30\% lower than those estimated using 1\,Myr isochrones. For 5\,Myr isochrones, cluster masses were typically $\sim$15-20\% higher than those estimated using the 2.5\,Myr isochrones. Moreover, extinction, which is both differential within the cluster and local due to protostellar envelopes at small scales, makes the determination of accurate masses for the PMS stars quite difficult, causing our analysis to probably underestimate the total mass of the clusters. Although there is a considerable amount of uncertainties, we consider the masses estimated from the 1\,Myr and 2.5\,Myr isochrones to be the best estimates of the cluster masses.

%The Padova 1\,Myr isochrone model suggests a mean mass for the stars in the clusters of about 0.65\,$M_\odot$ and a minimum mass between 0.03 and 0.11\,$M_\odot$.  As discussed above, the detection limit of the \emph{HST} observations is $\sim$0.7\,$M_\odot$ which is significantly higher than these limits, suggesting that our assumed extinctions for the clusters do not work perfectly for this isochrone. 

%Given that these MYSOs form in much more desolate environments than other LMC MYSOs, it may be expected that they have different clustering characteristics. We investigate whether these MYSOs sample the universal stellar initial mass function (IMF) by first estimating the mass of the most massive YSO in each region (\mmax) and the surrounding embedded cluster mass (\mecl). For each MYSO, \mmax\  was calculated based on the HST narrow-band H$\alpha$ images. 

%The estimated \mecl\ for the three MYSOs are between 250 and 400~$M_\odot$. 

We consider the MYSOs mass estimates \mmax\ (Section\,\ref{s:spclass}) and we extrapolate them to the total mass of each cluster, as it is expected analytically from the IMF \citep{Weidner2004} and from the typical relation between the mass of the most massive star in a cluster and its cluster mass  \citep{Weidner2013}. These estimates for the clusters masses are also given in Table\,\ref{t:cluschar} (Cols. 12 and 13 respectively). The estimated and extrapolated cluster masses agree well with each other; the estimated cluster masses for both the 1\,Myr and 2.5\,Myr isochrones are almost all within a factor of two of the extrapolated masses using both the IMF and the \mmax\,--\,\mecl\ relation. There are only two exceptions: 1) the mass estimated for 1\,Myr isochrones for \sfour\ is over a factor of two different from both extrapolated masses; and 2) the mass estimated for 1\,Myr isochrones for \ssix\ is over a factor of two different from that mass extrapolated from the \mmax\,--\,\mecl\ relation. Based on Figure \ref{f:s.cmds}, the PMS population of the stars within both of the clusters encompassing \sfour\ and \ssix\ tend to be closer to the 2.5\,Myr isochrone than the 1\,Myr isochrone. Therefore, the higher masses estimated from the 2.5\,Myr isochrones are probably more accurate estimates of the cluster masses for these two MYSOs. Given that our estimated cluster masses are similar to both of the extrapolated cluster mass estimates, we cannot declare that any of these 7 isolated MYSOs have significantly unique clustering properties.

%Although the \mecl\ estimated for \stwo\ and \sthree\ based on the observed stars does match well that calculated from both the IMF and the \mmax\,--\,\mecl\  relation, there is a high discrepancy for the values derived for \mecl\ of MYSO \ssix.  This suggests that -- given that both assumptions about the form of the IMF and the \mmax\,--\,\mecl\ relation are valid -- this MYSO should belong to a more massive cluster than what we measure, i.e., that we underestimate the cluster mass due to high incompleteness in our stellar sample or we underestimate the extent of this cluster. Indeed, the surrounding dispersed stellar distribution implies that this MYSO may be a member of a larger more extended stellar concentration. On the other hand, the apparent lack of a GMC around this MYSO may indicate a smaller gas reservoir that is incapable of forming a higher \mecl, and therefore, given the uncertainties of the estimated parameters \mmax\ and \mecl\ (primarily due to uncertainties that arise from stellar models, extinction, and multiplicity), we cannot rule out the possibility that this cluster is entirely inconsistent with both the `typical' IMF and the \mmax\,--\,\mecl\ relation.

The investigated regions show differences in their stellar clustering behavior. The region around MYSO \sthree\ encompasses one single centrally condensed star cluster with no other apparent stellar concentration around it (Figure\,\ref{f:kdemaps}). Approximately 35\% of all PMS stars observed in the region belong to the cluster, with the remaining stars being uniformly distributed in the field. On the other hand, the regions for the rest of the MYSOs clearly show signatures of multiple clustered environments, hosting additional sparse but not uniform stellar distributions. The clusters around the MYSOs enclose in these regions vary from 15--25\% of the complete observed PMS sample, with the remaining forming the surrounding stellar distributions. %{\bf The other three MYSOs, \sone, \sfive, and \sseven, have PMS populations that are a bit more centrally condensed about the MYSO than \stwo, \sfour, and \ssix, but still show signatures of nearby low-mass clusters that are not found for \sthree.}

With the exception of \sthree, the distributed populations about the regions can account for stars that may have been formed in the same star formation event as the MYSO itself but in a less clustered fashion. Most of these MYOs have a PMS stellar distribution in the region of MYSO that is loose and somewhat remote from the cluster. However, that in the region of MYSO\,\ssix\ appears denser and directly related to the cluster which it encompasses. These dispersed distributions, especially the high extinction in the region of MYSO\,\ssix, clearly imply the existence of molecular clouds (apparently the parental), which were not detected in our ancillary ISM data, but revealed through their faint PMS stars in our \emph{HST} data.

\section{Discussion}\label{discussion}

We investigate the environments of seven apparently isolated MYSOs in the LMC in order to characterize -- and eventually parametrize -- the phenomenon of isolated high-mass star formation at its earliest \emph{HST} observable stages. In the study described in the previous sections, the lack of isolation is apparent for all seven of the MYSOs. The unparalleled resolution of {\em HST} allowed 
for the direct detection of a plethora of faint PMS stars clustered around the MYSOs, which were undetected from previous low-resolution {\em Spitzer} 
and ground-based imaging. This discovery showed that these MYSOs are not isolated, at least not as far as their immediate environments are concerned. The observational contradiction between high- and low-resolution imaging about the isolation of high-mass stellar sources in the Magellanic Clouds is an issue that is already discussed in the literature \citep[e.g.,][]{gouliermis07, carlson11}. Including our dataset, every MYSO  resolved with {\em HST} is found to be surrounded by lower-mass red sources.

While the selected targets meet some strict criteria for isolation, such as e.g., being  at least 80\,pc away from known GMCs or OB associations, this discovery introduces additional constraints to the interpretation of available observations, used as evidence for isolated high-mass star formation. If indeed it is the normal for such ``isolated'' high-mass stars to host compact clusters around them, one may ask the obvious question: {\em ``Does the clustering of stars around a high-mass star under formation, with no other high-mass stars in its vicinity, still account for isolated high-mass star formation?''} {\em HST} reveals populous distributions of faint sources (both clustered and dispersed) around our selected 
MYSOs, which, based on previous low-resolution (and low-sensitivity) observations, are apparently isolated. {\em Does the existence of such distributions in the vicinity of a forming high-mass star challenge its supposed isolation?}

Observations suggest that roughly 4 ($\pm$\,2) percent of all O-type stars in the field of the Milky Way may have formed in isolation \citep{deWit2005}.  This fraction was successfully reproduced by random sampling from a typical stellar IMF and by selecting clusters from a power law cluster mass function (CMF) of slope $\beta=1.7$. A study by \citet{Parker2007} showed that selecting clusters from a standard CMF with  $\beta=2$ \citep[see, e.g.,][]{LadaLada2003} increases the fraction of isolated O-stars (defined in \citealt{Parker2007} as a star with a mass $>$17.5\,$M_\odot$) even more to about 17 percent. However, if \citet{Parker2007} restrict their definition of an ``isolated'' O-star as those from stellar clusters of mass less than 100\,$M_{\odot}$ that do not contain any other stars $>$10~$M_\odot$, the fraction of apparently isolated O-stars drops dramatically to between 1\,and\,5 percent. This result suggests that isolated O-stars {\em ``are low-mass clusters in which massive stars have been able to form''} \citep{Parker2007}. %However, the cluster masses about our sample of MYSOs are much larger than 100\,$M_{\odot}$.

On the other hand, as pointed out by \citet[][and references therein]{Weidner2013}, the IMF might not necessarily be randomly sampled, as was assumed by \citet{Parker2007}. Instead the IMF could be optimally sampled in accordance to the \mmax\,--\,\mecl\ relation \citep{Kroupa2013}. For this optimal sampling, the IMF is scale-free and the upper mass limit \mmax\ on which the IMF is sampled changes based on the cluster mass \mecl. Our observations provide a basis for testing both optimal and random sampling of the IMF. 

%The evidence presented with our analysis suggests that the most likely ``isolated'' massive stars do not form in complete isolation from their environments, but in compact clusters with faint stars. 
By estimating the mass of the clusters around the investigated MYSOs according to the proposed empirical relation between the mass of the most massive star and the cluster mass \citep[e.g.,][optimal sampling]{Weidner2013}, we find masses similar to those estimated from the data (Table\,\ref{t:cluschar}). In the instance of MYSO \ssix, the estimated mass is quite smaller than that estimated by the  \mmax\,--\,\mecl\ relation. This difference may simply be due to the uncertainties for \mecl\ about this MYSO are especially large due to extinction and the method used in determining the area encompassing the cluster. We therefore cannot rule out the \mmax\,--\,\mecl\ relation.

%which are larger than both those measured from the data and calculated from extrapolation of a typical stellar IMF (Table\,\ref{t:cluschar}). The difference is not surprising because we specifically selected MYSOs in unique, isolated star-forming environments. Moreover, this difference is only significant for only one MYSO, \ssix, and the uncertainties for \mecl\ about this MYSO are especially large due to extinction and the method used in determining the area encompassing the cluster. We therefore cannot entirely rule out the \mmax\,--\,\mecl\ relation.

Our analysis, which is focused on MYSOs, i.e., high-mass stars {\em that are embedded and may still be accreting}, provides observational evidence that indeed apparently isolated MYSOs do form within clusters. While the cluster masses are small ($\lesssim$600~$M_\odot$; Table\,\ref{t:cluschar}, Col.\,10 and 11), they are \emph{all} larger than 100~$M_\odot$.\footnote{Using the 1\,Myr isochrone, the cluster mass for \sfour\ is only 90\,$M_\odot$ However, as discussed in Section \ref{s:clusterest}, the 2.5\,Myr isochrone is a better indicator of the actual age of this cluster.} Moreover, the O-stars in the sample contain intermediate mass B-stars. Therefore, none of the seven MYSOs in the sample would qualify for the \citet{Parker2007} definition of an ``isolated O-star'' occurring due to random sampling. We note that, although some of the MYSOs in our sample may not satisfy the \citet{Parker2007} definition of an O-star (i.e., we estimate four MYSOs to have masses of 14~$M_\odot$ while \citealt{Parker2007} require 17.5~$M_\odot$), random sampling of the IMF would suggest that these MYSOs are even more likely to be isolated. Based on our strict criteria for isolation, we selected MYSOs that are strong candidates to become isolated field stars. If these are indeed the most likely candidates to become field stars, then these data do not support the random sampling scenario suggested by studies of in situ formation of field O-stars. This may suggest that 1) our criteria poorly selects MYSOs that will become part of the isolated field population, and/or 2) observations of isolated ``evolved'' (i.e., unembedded main-sequence) O-stars \citep[e.g.,][]{deWit2004,deWit2005,Lamb2010} do not properly characterize their initial star-forming environments. 

%probability_selecting.nb
If 1 to 5\% of MYSOs form in isolation, then of the 248 GC09 MYSOs, $\sim$ 2 to 12 MYSOs in the LMC should be isolated.\footnote{The GC09 MYSO sample contains some early B-stars. Randomly sampling would predict that an even higher fraction of these 248 MYSOs will be isolated.} Our selection criteria certainly eliminates many of the definite non-isolated MYSOs in the LMC; indeed half of the GC09 MYSOs have another MYSO within 25~pc (Figure\,\ref{nearestneighbor}). In other words, if our selection criteria for isolation at least work in part, we are not randomly selecting seven MYSOs from the 248 GC09 MYSOs; instead, one could imagine that we are ``randomly'' selecting seven MYSOs from a smaller subset. Let us arbitrarily assume that we are randomly selecting 7 MYSOs from a subset of 50 MYSOs rather than 248. If only 2.48 or 12.4 (corresponding to 1 or 5\% of all GC09 MYSOs) of these 50 sources are isolated, the probability of \emph{not} random selecting an isolated MYSO from this subset is 68\% or 12\%, respectively. If we assume that our selection criteria does even better and we are randomly selecting from a subset of 25 MYSOs, these probabilities are now 43\% or 0.2\%, respectively. Given these scenarios, it is unlikely that 5\% of all MYSOs in the LMC are isolated; however, it is certainly possible that 1\% of the sources are isolated. In summary, if our selection criteria for MYSOs increase our chances of selecting isolated MYSOs and the IMF is randomly sampled, the LMC likely has significantly fewer than 5\% of its MYSOs forming in isolation.

%that were in the most isolated environments, and if the IMF is randomly sampled, the expectation would be that some of these seven MYSOs would qualify for the \citet{Parker2007} definition of isolation. We therefore focus on the second possibility for the discrepancy, i.e., isolated field ``evolved'' O-stars poorly characterize their initial star-forming environment, which is discussed in detail in \citet{Weidner2013}.

%The small masses of the clusters are further verified from extrapolation of a typical stellar IMF with the most massive star having a mass equal to that of the MYSO for objects \sthree\ and \stwo, while for object \ssix\ a slightly larger cluster mass was derived (Table\,\ref{t:cluschar}; Col.\,11). These results are consistent with single OB stars being the most massive member of a group of smaller stars, as expected by a universal clustering law \citep[e.g.,][]{Hunter2003}, implying that the star-forming process is continuous from rich clusters to poor groups \citep{Oey2004}.

Although we cannot rule out random sampling, we also cannot rule out the possibility that absolutely no high-mass stars form in isolation. The search for isolated high-mass star formation based on populations of ``evolved''  O-type stars certainly imposes limitations in identifying clusters about O-stars because ionization and winds from the high-mass star may have erased any signature of the original gas in the cluster, and clusters are subject to ejections and dissolution \citep[e.g.,][]{Gvaramadze2012}. \citet{Weidner2013} specifically cautioned interpreting clusters as isolated if they are known to be old ($\gtrsim$4 Myr) or gas-free since these clusters can lose a considerable amount of stars. Moreover, \citet{Pflamm2010} showed that after formation in a cluster, O-stars can be expelled via a binary ejection event coupled with a subsequent sling-shot due to a supernova explosion, making it impossible to trace them back to clusters. Furthermore, observations of the more evolved O-type stars may have been limited by the dynamic range of the telescope since the brightness of an O-star may outshine the surrounding faint sources. In other words, the field O-stars may not represent their initial environment and may not supply any evidence of how the IMF is sampled.

%This inconsistency is not surprising, because if we assume that the observed isolated massive stars do actually form in low-mass clusters then the \mmax\,--\,\mecl\ relation is challenged. Indeed, according to the findings by \citet{Parker2007}, the relationship between the mass of a cluster and the mass of the most massive star in that cluster cannot be fundamental, but {\em ``a typical result of star formation in clusters.''} 

%\emph{If these MYSOs do not form in isolation or in clusters with masses less than 100~$M_\odot$, then which MYSOs do?} Our study would benefit from an increased sample, but \emph{perhaps such isolated MYSOs simply do not exist}. Based on our findings on the best (to our knowledge) candidates for on-going isolated massive star formation in the \emph{entire} LMC, it is unlikely that high-mass stars can form in isolation or in clusters with masses less than 100~$M_\odot$. Therefore, the observations do not support (though do not completely rule out) that the isolated field population is due to random sampling of the IMF.

Another result presented in this study is that all but one MYSO (\sthree) has an unambiguous detection of \mbox{CO(1--0)} in its immediate vicinity. Assuming that molecular gas in the LMC is reliably traced by CO emission, the reservoir of molecular gas associated with the MYSOs is small ($M_{\rm{CO}} \lesssim 2\times 10^4~M_\sun$), and well below the mass threshold that is usually adopted for GMCs. Our results may therefore indicate that ``isolated'' high-mass star formation can occur in low-mass gas clouds, contrary to the usual assumption that high-mass stars only form in GMCs. Alternatively, previous studies of the YSO population in the LMC have suggested that lower luminosity YSOs may outlive their natal GMCs \citep{Wong2011} and this may also be the case for MYSOs. The latter scenario suggests that GMCs in the LMC can be efficiently disrupted on $\sim$Myr timescales, which is in moderate tension with empirical arguments for GMC lifetimes of 20--30~Myr \citep[e.g.,][]{Kawamura2009}. Alternatively, such observations could suggest multiple epochs of high-mass star formation in a GMC.

%, but it does not necessarily require a completely-populated stellar IMF \citep[see also the discussion by][]{Lamb2015}. The mass of the formed star is then set not by the upper IMF, as is statistically confirmed for normal clusters \citep{OeyClarke2005}, but by the parental gas reservoir. This scenario is in line with both the competitive accretion model of star formation which requires a significant population of lower mass stars \citep[e.g.,][]{Bonnell2004}, and the core accretion model, which suggests that the accreted gas is controlled by the mass of the fragmented core alone \citep[e.g.,][]{Krumholz2009}.

With the exception of \sthree, the MYSOs in our sample show prevalence of multiple clusters in the region, clearly indicating that the high-mass stars are not forming in isolation across the extent of $\sim$\,few\,10\,pc. On the other hand, in the case of MYSO\,\sthree, apart from its own surrounding low-mass cluster, there are no additional clusters in its vicinity within a distance $\gtrsim$\,60\,pc,  suggesting that this object is an {\em isolated compact cluster}. We chose the most isolated MYSOs in the entire LMC, and only one source has been confirmed as an isolated cluster. Therefore, an isolated compact cluster about an O-star appears to be rare phenomenon. Searching for in situ isolated high-mass stars may be instead a search for isolated compact clusters that contain an O-star.

%Talk about the newly discovered star cluster well below scale height of Galaxy.

%Say something about triggered star formation?

%This study does not exclude O-stars forming in unbound environments, as discussed in \citet{Bressert2012}. However, since no early O-stars are isolated during formation, it shows that a GMC that will create early O-stars must fragment into multiple O-stars.

%Velocity lowered and/or redirected by dynamical friction or supernova slingshot

\section{Summary and Conclusions}

A galaxy-wide search throughout the entire LMC shows that there are very few MYSOs that are forming outside of GMCs and not near other MYSOs or OB associations, i.e., they form in apparent isolation. Based on an ancillary set of imaging data from both {\em Spitzer} and ground-based telescopes, we constructed from typical star formation indicators a dataset of MYSOs that are considered to be the best candidates for forming in isolation. These sources are confirmed MYSOs with {\em Spitzer} IRS spectroscopy, and they emit enough ionizing photons to produce \ion{H}{2} regions around them, confirmed with H$\alpha$ imaging. They are also more than 80\,pc away from any other MYSO \citep{GC09}, OB association star \citep{Lucke1970}, or GMC \citep{Fukui2008,Wong2011}. 

Our \emph{HST} follow-up observations clearly demonstrate that while these MYSOs appear to be in isolated environments, they are actually surrounded by a plethora of PMS stars. Our clustering analysis of these stars shows that all MYSOs are members of compact clusters. Six of the regions have significant sub-structure, with the PMS stars being both sparsely distributed and in the compact clusters. These stellar alignments appear to be the signatures of the parental molecular cloud, which is presently undetected by CO surveys. A seventh analyzed MYSO (\sthree) was found to be surrounded by a single isolated compact low-mass stellar cluster with no other stellar distribution being associated with it, indicating that the parental cloud of this object did not produce stars in a dispersed fashion. Moreover, \sthree\ contains no known clusters within 60~pc \citep{Bica2008}. Such an isolated cluster containing an O-star is a rare occurrence in the context of high-mass star formation.
%Such a bimodal stellar assembling in both clustered and dispersed patterns is being observed across larger scales in the giant SMC \ion{H}{2} region N66 \citep{Gouliermis2014}.

The observed population of isolated field O-stars that are expected to form in situ \citep[e.g.,][]{deWit2004,Lamb2010,Oey2013,Lamb2015} are often considered to be a phenomenon of random sampling of the IMF, which allows O-stars to form in relative isolation \citep[i.e., in clusters $<$100~$M_\odot$ with no other star $>$10~$M_\odot$][]{Parker2007}. In other words, in situ O-stars forming in a cluster of mass $<$100~$M_\odot$ is rare but not impossible. However, while the previous confirmations of isolated high-mass star formation among field main-sequence O-type stars (after correcting for runaways) provide evidence of {\em in situ} formation, they do not provide information on the {\em environment} where formation took place; radiation and winds from the high-mass star and dynamical events may have erased the signatures of the parental gas and the clustering around the O-star. 

We investigate isolated high-mass star formation at a much earlier stage, i.e., the embedded MYSO stage. Based on our selection criteria, we have selected the best candidates for in situ, isolated high-mass star formation. We find cluster masses about these MYSOs to be larger than 100~$M_\odot$, suggesting that these MYSOs are not as isolated as typical field O-stars. While we cannot entirely rule out random or optimal sampling of the IMF, we suggest that a randomly sampled IMF should find that significantly less than 5\% of LMC MYSOs are isolated.

With the present study we demonstrate that the investigation of the phenomenon referred to as ``isolated high-mass star formation'' requires the investigation of sources at earlier stages of their formation, such as MYSOs, which should still be embedded in their natal environments. Our investigation is the only observational study \citep[apart from that presented by][]{Selier2011} that approaches the issue strictly from this perspective. Based on our findings we argue that panchromatic high-resolution observations in the vicinity of apparently isolated MYSOs (and not main-sequence stars) will allow a better understanding of the conditions and the parameters that set the stage for high-mass stars to form in isolation. %The results on the remaining four MYSOs in our sample are qualitatively indistinguishable from those presented here. In our follow-up study, we repeat our detailed analysis for these MYSOs and provide parameters for the clusters surrounding them in the context of a statistical investigation of the phenomenon.

\acknowledgements
I.W.S. and L.W.L. acknowledges NASA grant HST-GO-12941 06-A. D.A.G. acknowledges the German Research Foundation (Deu\-tsche For\-schungs\-ge\-mein\-schaft, DFG)  grant GO\,1659/3-2. D.R.W. is supported by NASA through Hubble Fellowship grant HST-HF-51331.01 awarded by the Space Telescope Science Institute. A.H. acknowledges support from the Centre National d'\'Etudes Spatiales (CNES). Based on observations made with the NASA/ESA {\em Hubble Space Telescope}, obtained from the data archive at the Space Telescope Science Institute (STScI). STScI is operated by the Association of Universities for Research in Astronomy, Inc.\ under NASA contract NAS 5-26555. The Mopra radio telescope is part of the Australia Telescope National Facility which is funded by the Commonwealth of Australia for operation as a National Facility managed by CSIRO.  The University of New South Wales Digital Filter Bank used for the observations with the Mopra Telescope was provided with support from the Australian Research Council. The National Radio Astronomy Observatory is a facility of the National Science Foundation operated under cooperative agreement by Associated Universities, Inc. This research has made use of the SIMBAD database, operated at CDS, Strasbourg, France, and APLpy, an open-source plotting package for Python hosted at http://aplpy.github.com.

\appendix

\section{Isolation Characterization of each MYSO}
Below is the characterization of the isolation of each MYSO studied in this paper. These summaries are primarily based on \citet{Bica2008} and \citet{Seale2014}, which we will henceforth refer to BBDS08 and S14, respectively. Nebular complexes below that are labeled as "NXXX" are from \citet{Henize1956} survey while those labeled with DEM\,LXXX are from the \citet{Davies1976} survey. The summaries primarily concentrate on possible (recent) star formation activity within 60~pc.
 
\subsection{MYSO\,\sone}\label{sec:sone}
MYSO\,\sone\ (full GC09 name: \lsone; Figure\,\ref{045403_2panel.png}) is located toward the northwest of the LMC and lies between the two OB associations LH\,3 and LH\,6 \citep{Lucke1970}. \sone\ is located within the very faint nebular association DEM\,L19, 60$\arcsec$ from the center, which is located inside the Shapley-VI star-forming region \citep{van1981}. Of the 7 MYSOs observed with $HST$, \sone\ is the closest to any OB association; while it is 100~pc from the center of any \citet{Lucke1970} OB association, more dispersed stars are found as close as 43~pc to the MYSO. Other than being within the DEM\,L19 nebular association, \sone\ has no nearby ($<$60~pc) clusters or associations.
%The closest BBDS08 source to \sone\ is the nebular association DEM~L19 \citep{Davies1976} and is located 60~pc away.

S14 identified MYSO \sone\ as {\em Herschel} source HSOBMHERICC J73.515422--67.271269 (henceforth in the appendix, we will drop the \citealt{Seale2014} nomenclature ``HSOBMHERICC'' for brevity). The closest YSO to \sone\ is located 12~pc away to the east and has an FIR luminosity of 260\,$L_\sun$. In GC09, this YSO was identified as 045412.05--671627.1 and was considered a probable galaxy. However, this is likely a YSO based on the S14 {\em Herschel} analysis and clustering of young stellar sources about 045412.05--671627.1, as identified with the $HST$ photometry from this study. Between 20 and 60\,pc, there are six additional unclassified {\em Herschel} sources. In addition, a possible YSO (FIR luminosity of 450\,$L_\sun$) is located at a distance of 41~pc, but according to GC09, this source (045350.52--671354.7) is more likely to be a galaxy.

\subsection{MYSO\,\stwo}\label{sec:stwo}
MYSO\,\stwo\ (full GC09 name: \lstwo; Figure\,\ref{050941_2panel.png}) is located toward the edge of the LMC and found in the very faint H$\alpha$ region DEM~L91. The extended emission emanating southwest of the MYSO was classified as a small \ion{H}{2} region with an embedded cluster in \citet{Bica1999} and given the name BSDL770. The closest BBDS08 source to \stwo\ that is not associated with the \stwo\ star-forming complex is a star cluster 4\farcm3 (64\,pc) away.

MYSO\,\stwo\ was identified as J77.423753--71.461699 in S14 and has two nearby dim {\em Herschel} sources in the star formation complex. These two sources are located in the extended emission toward the southwest of \stwo\ as seen in Figure\,\ref{050941_2panel.png}, with one identified as a YSO and the other without an identification. Two possible dust clumps are located at 17 and 24\,pc away. Between 25 and 60\,pc, there are three additional unclassified {\em Herschel} sources.

%dust clumps > 1000 L_sun prob have high mass stars

\subsection{MYSO\,\sthree}
MYSO\,\sthree\ (full GC09 name: \lsthree; Figure\,\ref{051906_2panel.png}) is located toward the center of the LMC and was classified as N118. This MYSO has the least amount of CO(1--0) emission out of seven MYSOs observed with HST (see Figure\,\ref{COspec}). The closest source to \sthree\ in the BBDS08 catalog is a cluster located 4\farcm4 (66\,pc) away.

MYSO\,\sthree\ was identified as J79.777027--68.360242 in S14 and has two unclassified {\em Herschel} sources, J79.73786--68.358743 and J79.72038--68.350657, approximately 13 and 21\,pc away, respectively. These two objects are unclassified because they are very dim, and in the \emph{HST} observations, they are present in diffuse regions. These sources are likely low-mass YSOs or dust clumps. The next closest {\em Herschel} source is the YSO J79.724858--68.381149, which was identified in GC09 as the probable YSO 051854.14--682251.9. This source is located approximately 26\,pc away from the MYSO and has a FIR luminosity of 1200\,$L_\sun$, indicating that it is likely an intermediate mass YSO. There are ten more lower mass (220\,$L_\sun$ or less) YSOs or unclassified $Herschel$ sources in the range of 30 to 60\,pc.

\subsection{MYSO\,\sfour}\label{sec:sfour}
MYSO\,\sfour\ (full GC09 name: \lsfour; Figure\,\ref{052124_2panel.png}) is located in the north central part of the LMC within the large (24$\arcmin \times$20$\arcmin$) nebular association DEM\,L154. \citet{Bica1999} identified a stellar association (BSDL1324) that is 4.8\,pc east of the central protostar and appears to be part of the same star-forming complex, but with a much lower stellar density than the near vicinity of the MYSO (based on our cluster analysis; Section\,\ref{s:clusanl} and Figure\,\ref{f:kdemaps}). The closest cluster/association not associated with \sfour\ is located 27~pc away. Five more clusters or associations are located within 27-60~pc.

MYSO\,\sfour\ was identified as J80.353044--66.069699 in S14. The star-forming complex also includes two unclassified $Herschel$ sources, J80.348375--66.067789 and J80.356315--66.075533, within the same star formation region. GC09 identifies an intermediate mass ``probable'' YSO (052123.03--660346.9) that is located only 4.8\,pc northwest from \sfour. This source was not directly identified in S14, possibly due to confusion of emission with the large $Herschel$ beam relative to $Spitzer$, but given the minor PMS clustering about this source (Figure\,\ref{f:kdemaps}, northwest of MYSO\,\sfour), it is likely a YSO. Two more unclassified S14 sources are within the WFC3/UVIS field of view, J80.356315--66.075533 and J80.369139--66.083579, but no obvious stellar clustering exists amongst the former source and the latter source lies outside the WFC3/IR field of view. Between 23 and 60~pc, there exist 3 more probable intermediate or low-mass YSOs, 9 unclassified sources, and 1 possible dust clump.

\subsection{MYSO\,\sfive}\label{sec:sfive}
MYSO\,\sfive\ (full GC09 name: \lsfive; Figure\,\ref{053244_2panel.png}) is located just west of 30-Doradus, near DEM\,L224. There are two stellar clusters located within 60~pc, which are known as BSDL2278 \citep{Bica1999} and SL558 \citep{Shapley1963} and are located 41 and 43~pc away from \sfive, respectively.

MYSO\,\sfive\ was identified as J83.183507--69.501648 in S14. GC09 identified two nearby YSOs, 053249.44--693037.0 and 053239.66--693049.5 (S14 names J83.205424--69.510268 and J83.164597--69.51341) at distances of 10 and 12~pc, respectively. In the Figure\,\ref{f:kdemaps} maps about MYSO\,\sfive, 053249.44--693037.0 is located in the contour toward the southeast of the MYSO and 053239.66--693049.5 is located in the contour southwest of the MYSO. Within 60~pc of the MYSO, there are 12 unclassified sources, a probable dust clump at 28~pc away (J83.168656--69.531878), and a definite YSO located 28~pc away (GC09 name 053239.22--693153.9).

\subsection{MYSO\,\ssix}
MYSO\,\ssix\ (full GC09 name: \lssix; Figure\,\ref{053342_2panel.png}) is located northwest of 30-Doradus. As seen in Figure\,\ref{053342_2panel.png}, there is a filamentary nebular feature across the YSO region, following roughly the north-south orientation (seen by the brown/red colors in the figure) and bubbles of ionized gas (seen by the blue colors, as well as the H$\alpha$ emission seen in Figure \ref{f:ha}) that seem to expand toward both east and west directions. \ssix\ is classified as the nebula N150 and DEM\,L233. The east and west bubbles are classified as separate sources in BBDS08. Specifically, the bright stellar-like feature seen at the top right of the right panel of Figure\,\ref{053342_2panel.png} is located 24$\arcsec$ (6\,pc) from the MYSO and was classified by \citet{Bica1999} as a small \ion{H}{2} region with an embedded cluster and given the name BSDL2316.  Our {\em HST} images show that BSDL2316 is probably part of the same star-forming region as \ssix.

S14 identifies two {\em Herschel} sources (J83.426728--68.767114 and J83.427082--68.765922, only resolved with PACS 100~$\mu$m) along the filamentary structure in the region of MYSO \ssix. The emission is dominated by J83.426728--68.767114, with a FIR luminosity of 21000\,$L_\sun$. S14 classified both of the {\em Herschel} sources as probable dust clumps rather than YSOs, which is likely due to the filamentary dust lane containing the objects. The environment surrounding this MYSO has many more {\em Herschel} sources nearby, with three unclassified sources between 16 and 30\,pc, and another 21 {\em Herschel} sources between 30 and 60\,pc. These 21 sources are a mix of unclassified sources, probable YSOs, and probable/possible dust clumps. None are high-mass because their FIR luminosities are about 240\,$L_\sun$ or less, and all are undetected with PACS 100\,$\mu$m. Only one of these sources (58\,pc away from the MYSO) was identified in GC09.
%Considering these findings one deduces that \ssix\ is far less isolated than the other sources.

\subsection{MYSO\,\sseven}\label{sec:sseven}
MYSO\,\sseven\ (full GC09 name: \lsseven; Figure\,\ref{053431_2panel.png}) is also located northwest of 30-Doradus and was identified as BSDL2408 in \citet{Bica1999}. The only BBDS08 sources within 60~pc are the clusters SL580 and SL583 \citep{Shapley1963}, which are located 28 and 45~pc away, respectively.

MYSO\,\sfive\ was identified as J83.631354--68.587125 in S14. Within the same star-forming region of the MYSO, S14 identified two more lower-mass probable YSOs (J83.634501--68.588075 and J83.628613--68.585888) that are each located 1.4~pc from \sfive. The S14 source J83.634501--68.588075 is coincident with the red, cometary-like source seen just southeast of the MYSO in Figure\,\ref{053431_2panel.png}. There are three more S14 sources within the WFC3/UVIS field of view. One of these sources is an unclassified source (J83.644506--68.574154) located 13~pc away from the MYSO, and the other two are YSOs (J83.653678--68.568566 and J83.672002--68.566178) which are located 18 and 23~pc away. These three sources all lie upon the faint extinction feature that extends north/northeast from the MYSO. Within 60~pc, S14 identifies 10 unclassified sources, 1 probable YSO, and 1 probable dust clump.

\bibliography{isolated_msf}

\end{document}